\documentclass[twocolumn 
,showpacs,superscriptaddress,prb,amsmath,amssymb,floatfix,eqsecnum]{revtex4}
\usepackage{amsmath,amssymb}
\usepackage{bm}
\usepackage{epsfig}
\usepackage{psfrag}
\usepackage{color}

\newcommand{\bd}{\bm}

\begin{document}

\title{Dynamic structure factor of Luttinger liquids with quadratic energy dispersion and
long-range interactions}

\author{Peyman Pirooznia}  
\affiliation{Institut f\"{u}r Theoretische Physik, Universit\"{a}t
  Frankfurt,  Max-von-Laue Strasse 1, 60438 Frankfurt, Germany}

\author{Florian Sch\"{u}tz} 
\affiliation{Department of Physics, Brown University, Providence,
RI 02912-1843, USA}

\author{Peter Kopietz}  
\affiliation{Institut f\"{u}r Theoretische Physik, Universit\"{a}t
  Frankfurt,  Max-von-Laue Strasse 1, 60438 Frankfurt, Germany}

\date{February 7, 2008}

\begin{abstract}
We calculate the dynamic structure factor $S ( \omega , q )$
of spinless fermions in one dimension with quadratic energy 
dispersion $k^2/2m$ and long range density-density interaction whose 
Fourier transform $f_q$  is dominated by  small
momentum-transfers
 $ q  \lesssim q_0 \ll k_F$.  Here $q_0$ is a 
momentum-transfer cutoff
and $k_F$ is the Fermi momentum.
Using functional bosonization and  
the known properties 
of symmetrized closed fermion loops, we
obtain an expansion of the {\it{inverse}}
irreducible polarization to 
second order in the small parameter $ q_0 / k_F $.
In contrast to perturbation theory based on conventional bosonization, 
our functional bosonization approach is not plagued by
mass-shell singularities.
For interactions which can be expanded as 
$f_q = f_0 + f_0^{\prime \prime} q^2/2 + O ( q^4)$
with $f_0^{\prime \prime } \neq 0$ we show that the 
momentum scale $q_c =   1/ | m f_0^{\prime \prime} |$
separates two regimes characterized by a different
$q$-dependence of the width $\gamma_q$ of the collective zero sound mode
and other features of $S ( \omega , q )$.
For  $q_c \ll q \ll k_F$ all integrations in our functional bosonization result for 
$S ( \omega , q )$ can be evaluated analytically; we find that
the line-shape in this regime is non-Lorentzian
with an overall width $\gamma_q \propto q^3/(m q_c )$  and  a
threshold singularity $[ ( \omega - \omega_q^-) \ln^{2} ( \omega - \omega_q^- )]^{-1}$
at the lower edge  $ \omega \rightarrow \omega_q^- = v q - \gamma_q$,
where $v$ is the velocity of the zero sound mode.
Assuming that higher orders in perturbation theory
transform the logarithmic singularity into an 
algebraic one, we find for  the corresponding 
threshold exponent  $\mu_q = 1 - 2 \eta_q$ with
$\eta_q \propto q_c^2/ q^2$.
Although for $q \lesssim q_c$
we have not succeeded to explicitly
evaluate  our functional bosonization result for $S ( \omega , q )$,  
we argue  that for any one-dimensional model belonging to  the Luttinger
liquid universality class  the width of the zero sound mode scales as
$q^2/m$ for $q \rightarrow 0$.
\end{abstract}


\maketitle

\section{Introduction}
\label{sec:model}
Recently several authors have calculated the dynamic structure factor $ S ( \omega , q )$
in the Luttinger liquid phase of model systems for interacting fermions with non-linear 
energy dispersion in one spatial 
dimension~\cite{Samokhin98,Capurro03,Pustilnik03,Pustilnik06,Pustilnik06b,Cheianov07,Teber06,Teber07,Pereira06,Pereira07,Pereira07b,Pirooznia07,Schoenhammer07,Aristov07,Ploetz07}. 
Mathematically, $S ( \omega , q )$ 
is defined as the spectral density of the density-density correlation function,
\begin{eqnarray}
 S (  \omega , q )
 & = & \int\!dt\!\int\!dx\,
 e^{ i ( \omega t - q x)} \langle
 \delta \hat{\rho} ( x , t ) \delta \hat{\rho} ( 0,0) \rangle\,,
 \nonumber
 \\ & & 
\label{eq:dyndef}
 \end{eqnarray}
where $\delta \hat{\rho} ( x , t )$ is the operator
representing the deviation of the density from its average.
The dynamic structure factor can be directly measured via scattering experiments probing
density-density correlations of the system. 
It is therefore important to have quantitatively accurate
theoretical predictions for the line-shape of $S ( \omega , q )$.

Although there is general agreement that in the Luttinger liquid regime
of one-dimensional interacting fermions $S ( \omega , q )$ exhibits
for small frequencies $\omega$ and wave-vectors $q$ a narrow peak
associated with the collective zero sound (ZS) mode~\cite{Pines89,footnote1},
a quantitative understanding of the precise line-shape of the ZS resonance 
in generic non-integrable models is still lacking. The spectral 
line-shape is expected to depend 
on non-universal parameters of the model under consideration, such as 
the non-linear terms in the expansion of the energy dispersion $\epsilon_k$ around the
Fermi momentum $k_F$, or the coefficients  in  the expansion of the Fourier transform  
$f_q$ of the interaction for small momentum-transfers $q$. 
Because these parameters correspond to couplings 
which are irrelevant  (in the renormalization
group sense)  at the Luttinger liquid fixed point, the  line-shape 
of $S ( \omega , q )$ cannot be obtained using standard 
field-theoretical methods, such as field-theoretical bosonization, which has otherwise been
very successful to obtain the infrared  properties of Luttinger 
liquids~\cite{Solyom79,Haldane81,Giamarchi04,Schoenhammer05}.
Recall that the crucial step in the bosonization approach is the linearization of the 
energy dispersion around the Fermi points, $ \epsilon_{  k_F + q} - \epsilon_{ k_F} \approx  v_F q$,
where $v_F$ is the Fermi velocity. If in addition the Fourier transform $f_q$ of the 
interaction is non-zero only for momentum-transfers $q \ll k_F$, we arrive at the 
exactly solvable Tomonaga-Luttinger model (TLM), whose bosonized hamiltonian is 
non-interacting \cite{Solyom79,Haldane81,Giamarchi04,Schoenhammer05} . 
As a consequence, the dynamic structure factor of the  TLM has only a single 
$\delta$-function peak corresponding to a  collective  ZS mode with infinite lifetime.
For spinless fermions with long-range density-density interaction $f_q$ one obtains 
for small $q$,  
\begin{eqnarray}
  S_{\rm TLM} (\omega , q )  = Z_q \delta ( \omega - v_0 | q |  )\,,
 \label{eq:STLM}
\end{eqnarray}
where the velocity $v_0$ and the weight $Z_q$ of the collective ZS mode
can be written as
\begin{equation}
 v_0 / v_F = \sqrt{ 1 + g_0}\,,
 \label{eq:v0def}
\end{equation} 
\begin{equation}
 Z_q =   \frac{ v_F q^2}{ 2 \pi v_0 | q | } = \frac{| q | }{  2 \pi   \sqrt{ 1 + g_0}    }  .
 \label{eq:Zqdef}
\end{equation}
For later convenience we have
introduced the relevant dimensionless interaction at vanishing momentum-transfer,
\begin{equation}
 g_0 = \nu_0 f_0,
 \label{eq:g0def}
\end{equation}
where $ \nu_0 = 1 /( \pi v_F)$ is the non-interacting density of states at 
the Fermi energy. 

The question is now
how  the line-shape of $S (  \omega , q ) $ changes if we do not linearize the 
energy dispersion. There have been many recent attempts to find an answer to this 
question. Roughly, the proposed methods can be divided into four different categories:

{\it{1. Conventional bosonization.}} The established
machinery of conventional 
bosonization~\cite{Solyom79,Haldane81,Giamarchi04,Schoenhammer05} has been used
in Refs.~[\onlinecite{Samokhin98,Teber07,Aristov07}]
to calculate the dynamic structure factor of Luttinger liquids.
Expanding the energy dispersion around $k = k_F$ beyond linear order,
$ \epsilon_{ k_F + q} \approx \epsilon_{ k_F} + v_F q + q^2/(2m) $,
the  quadratic term $q^2/(2m)$ gives rise to cubic interaction vertices 
proportional to $1/m$ in the bosonized model \cite{Schick68}.
Hence,  bosonization maps the original (unsolvable) fermionic many-body 
problem onto another unsolvable  problem involving bosonic degrees 
of freedom. The hope is that perturbation theory for the effective 
boson model is well-defined and more convenient to carry out in practice 
than in the original fermion model \cite{Haldane81}. Unfortunately, this 
strategy fails for the calculation of $S ( \omega , q )$, because already to
second order in $1/m$ one encounters singular terms proportional to 
$1 / ( \omega \pm v_0 q )$, which become arbitrarily large as the frequency 
approaches the  mass-shell $ \omega \rightarrow \pm v_0 q $. Some time ago 
Samokhin \cite{Samokhin98} proposed a simple regularization procedure of 
these mass-shell singularities which we shall review in Sec.~\ref{sec:RPA}. 
Assuming a Lorentzian line-shape, he found that for $q \rightarrow 0$ most 
of the spectral weight is smeared out over an interval of width $ q^2/m$. 
Although this estimate for the width of the ZS resonance was later confirmed 
by various other calculations 
\cite{Pustilnik06,Pereira06,Pereira07,Pereira07b,Aristov07}, the assumption
of a Lorentzian line-shape turns out to be incorrect. It would certainly be 
more desirable to have a controlled method of re-summing the interaction in 
the bosonized hamiltonian to infinite orders such that the unphysical mass-shell 
singularities are properly regularized; apparently this problem has not 
been solved so far.  We shall further elaborate on these mass-shell singularities 
in Secs.~\ref{sec:RPA} and \ref{sec:pert}.

2. {\it{Re-summing fermionic perturbation theory via an effective hamiltonian}}.
Because of the above mentioned problems inherent in standard bosonization,
it seems better to  set up the perturbation expansion in terms of the original 
fermionic degrees of freedom using diagrammatic techniques. In this approach, it is
convenient to first  calculate the polarization function $ \Pi ( i \omega , q) $ for imaginary 
frequencies and then use the  fluctuation-dissipation theorem to obtain the
dynamic structure factor,
\begin{eqnarray}
 S (  \omega , q ) = \pi^{-1} {\rm Im } \Pi (  \omega + i 0 , q )\,.
\end{eqnarray}
For simplicity, we shall focus on the limit of vanishing temperature
throughout this work. For long-range interactions whose Fourier transforms $f_q$
are dominated by small wave-vectors $q  \ll k_F$, one usually avoids the direct 
expansion $\Pi ( \omega , q )$ in powers of the bare interaction, but instead expands 
its irreducible part $\Pi_{\ast} ( \omega , q )$ which is defined via
\begin{equation}
 \Pi^{-1} ( \omega , q  ) = f_q +   \Pi_{\ast}^{-1} ( \omega , q  )\,.
 \label{eq:Piirrdef}  
\end{equation}
In a recent paper, Pustilnik {\it{et al.}} \cite{Pustilnik06} did not follow this 
standard approach, but expanded the full (i.e., reducible) polarization 
$\Pi (  \omega  , q ) $ in powers of the bare interaction. They found 
already at the first order in the bare interaction that the correction to $S ( \omega , q )$
diverges logarithmically if $\omega$ approaches a certain threshold edge $\omega^{-}_q$ 
from above. The authors of Ref.~[\onlinecite{Pustilnik06}] then proposed a re-summation 
procedure of the most singular terms in the perturbation series to all orders using an 
effective hamiltonian constructed in analogy with 
the $X$-ray problem. In this way, they succeeded to
transform the logarithmic threshold singularity into an algebraic one, characterized by 
a certain momentum-dependent threshold exponent. The spectral line-shape can therefore 
not be approximated by a Lorentzian as implicitly assumed by Samokhin \cite{Samokhin98};
on the other hand, Samokhin's  result that the overall width of the ZS resonance scales 
as $q^2/m$ was confirmed by Ref.~[\onlinecite{Pustilnik06}]. However, Pustilnik 
{\it{et al.}}~\cite{Pustilnik06} did not explicitly analyze the higher-order 
terms in the perturbation series to demonstrate that the logarithmic singularity 
encountered at the first order can really be re-summed to all orders 
to yield an algebraic singularity. Moreover, they did  not keep track of the (finite) 
renormalization of the ZS velocity $v$, which
determines the precise energy scale of the collective ZS resonance and its position relative 
to the energy of the single-pair particle-hole continuum, which a priori need not be identical. 

{\it{3. Integrable models.}}
Another method to calculate the dynamic structure factor of Luttinger liquids is based
on the analysis of exactly solvable models
belonging to the Luttinger liquid universality class, such as the XXZ-chain 
\cite{Pereira06,Pereira07,Pereira07b} or the Calogero-Sutherland 
model \cite{Pustilnik06b,Cheianov07}. These calculations have confirmed the results obtained 
by Pustilnik {\it{et al.}} \cite{Pustilnik06} for generic (not necessarily integrable) 
one-dimensional Luttinger liquids: The spectral line-shape is non-Lorentzian, exhibits 
algebraic threshold singularities, and the weight is smeared over a frequency interval 
proportional to  $q^2/m$ for $q \rightarrow 0$. But since the re-summation procedure of 
Ref.~[\onlinecite{Pustilnik06}] is not rigorous and higher-order terms in the
perturbation series have not been explicitly analyzed, one cannot exclude
the possibility that the algebraic threshold singularities are a special feature of 
integrable models, and that in generic non-integrable models the higher order terms in 
the perturbation series do not conspire to transform logarithmic singularities into 
algebraic ones. Note also that the effective two-body  interaction in the spinless fermion 
model obtained from the XXZ-chain via the usual Jordan-Wigner transformation involves also 
momentum-transfers of the order of $k_F$. This model is therefore different from the 
forward scattering model with quadratic dispersion considered here, where the Fourier 
transform of the density-density interaction $f_q$ is only finite for $q \ll  k_F$.
Apparently, an exactly solvable model with non-linear energy dispersion and
density-density interaction $f_q$  involving only small momentum 
transfers and $f_{q=0} >0$ does not exist. 
Although the momentum dependence of the interaction is 
irrelevant in the renormalization group sense, the line-shape of $S ( \omega , q )$ is 
essentially determined by irrelevant couplings, so that models with different sets of 
irrelevant couplings might also exhibit different spectral line-shapes.

{\it{4. Functional bosonization.}}
This is an alternative method of describing fermionic many-body systems with dominant 
forward scattering in terms of bosonic degrees of freedom. In the context of the TLM,
the functional bosonization idea has been introduced by Fogedby \cite{Fogedby76} and 
by Lee and Chen \cite{Lee88}. Later this technique has been used to bosonize interacting 
fermions  with dominant forward scattering in arbitrary dimensions \cite{Kopietz95}, and
to estimate the effect of the non-linear energy dispersion on the single-particle Green 
function \cite{Kopietz96}. For a review of this approach see Ref.~[\onlinecite{Kopietz97}],
where the advantages of this method for calculating the dynamic structure factor have 
already been advocated. Like in conventional bosonization, in the functional bosonization
approach the non-linear terms in the energy dispersion give rise to interaction vertices
in the effective bosonized action of the system. However, the interaction vertices in
functional bosonization are rather different from the vertices due to the non-linear 
dispersion in conventional bosonization. In fact, the interaction 
vertices in functional bosonization can be identified diagrammatically with symmetrized 
closed fermion loops, which can be calculated exactly for quadratic  dispersion
in one dimension \cite{Neumayr98,Neumayr99,Kopper01}.  While in conventional 
bosonization a quadratic energy dispersion gives rise to cubic vertices 
in the bosonized Hamiltonian \cite{Haldane81,Schick68},
within functional bosonization a quadratic dispersion leads to infinitely many
vertices involving an arbitrary number of boson fields. The fact that perturbation 
theory for $S ( \omega , q )$ based on functional bosonization is different from perturbation 
theory based on conventional bosonization is obvious if one considers the non-interacting 
limit: while functional bosonization yields the exact free polarization $\Pi_0 ( \omega , q )$,
conventional bosonization produces an expansion of $\Pi_0 ( \omega , q )$ in powers of 
$1/m$, which in practice has to be truncated at some low order, leading to unphysical 
mass-shell singularities.

\begin{figure}[tb]    
\centering   
  \vspace{4mm}
\includegraphics[scale=0.3]{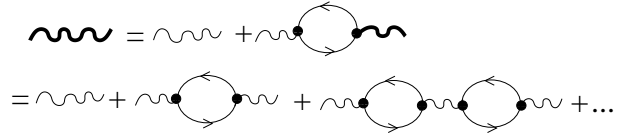} 
\caption{Diagrammatic definition of the screened interaction within 
random phase approximation. The thin wavy line denotes the bare interaction 
and the solid arrows represent non-interacting fermionic single-particle Green 
functions. 
}
 \label{fig:rpainteraction}
\end{figure}

\begin{figure}[tb]    
\centering   
  \vspace{4mm}
\includegraphics[scale=0.3]{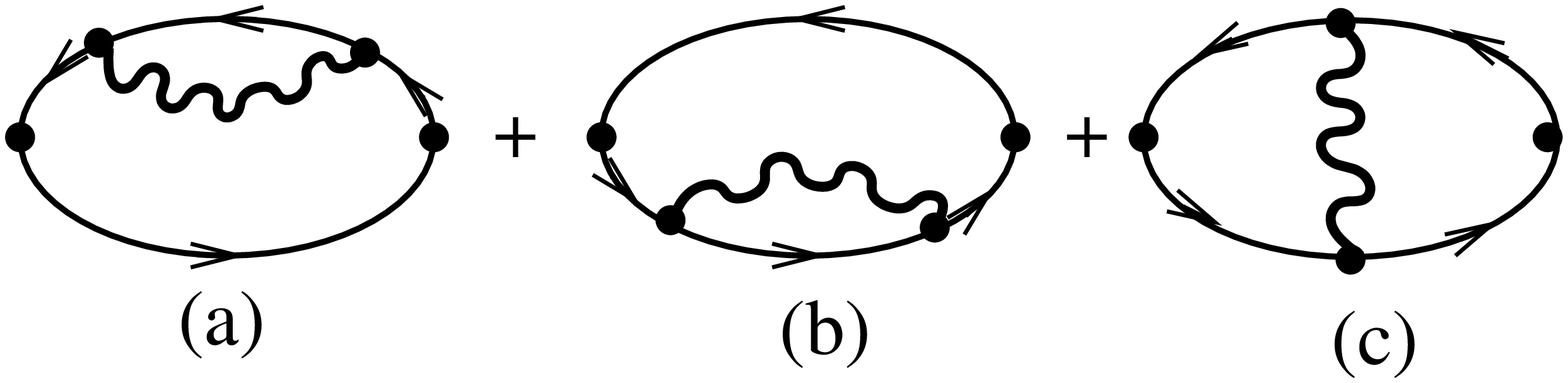} 
\caption{Corrections to the irreducible polarization in an expansion to first order 
in powers of the RPA interaction.
}
    \label{fig:feynman}
\end{figure}

In Ref.~[\onlinecite{Pirooznia07}] two of us have used the functional 
bosonization approach to calculate the width $\gamma_q$ of the ZS mode
in a generalized Tomonaga model with quadratic energy dispersion.
To estimate the effect of non-linear energy dispersion on the dynamic
structure factor, we have truncated the expansion of the inverse irreducible 
polarization at the first order in an expansion in powers of the Gaussian propagator
of the boson fields, which can be identified with the effective screened interaction 
within  random phase approximation (RPA) defined in Fig.~\ref{fig:rpainteraction}.
To this order, the simple first-order Hartree contribution to the bosonic self-energy
in the functional bosonization approach (the corresponding Feynman diagram 
is shown  in Fig.~\ref{fig:pertdiagrams} (a)  in Sec.~\ref{sec:pert})
is in fermionic language equivalent to the sum of the three first-order interaction 
corrections to the  irreducible polarization shown in Fig.~\ref{fig:feynman}.
Functional bosonization thus consistently sums self-energy corrections 
(diagrams (a) and (b) in Fig.~\ref{fig:feynman}) and vertex corrections 
(diagram (c) in Fig.~\ref{fig:feynman}) of the underlying fermion problem.
Actually, the interpretation of the {\it{inverse}} irreducible polarization
as the self-energy of the effective boson theory obtained via functional
bosonization suggests that one should always expand
the {\it{inverse}} $\Pi_{\ast}^{-1} ( \omega , q )$ in powers of the relevant small
parameter \cite{Kopietz97,Pirooznia07}.

Unfortunately, it is not consistent to truncate 
the expansion of  $\Pi_{\ast}^{-1} ( \omega , q )$ at the first order
in the RPA interaction, so that the result $\gamma_q \propto q^3$
for the ZS damping obtained in Ref.~[\onlinecite{Pirooznia07}]  
cannot be trusted. We shall explain this in more detail in 
Sec.~\ref{sec:pert}, where we construct a systematic 
expansion  of $\Pi_{\ast}^{-1} ( \omega , q )$ in powers of bosonic 
loops using functional bosonization and show that for our forward 
scattering model there is a large
intermediate regime $ q_c \lesssim q \ll k_F$ where indeed
$\gamma_q \propto q^3/(m q_c)$. 
The momentum scale $q_c$ is determined by the 
momentum dependence of the interaction $f_q$,
see Eq.~(\ref{eq:qcdef}) below.
Due to the complexity 
of the integrations, in the regime $ q \ll q_c$ we have not been able 
to evaluate our functional bosonization result for $S ( \omega , q )$. 
However, at $ q \approx q_c$ our expression for $\gamma_q$ matches
the result $\gamma_q \propto q^2/m$ obtained by several other
authors for different  model systems for
Luttinger liquids \cite{Samokhin98,Pustilnik03,Pustilnik06,Pereira06}.
We therefore believe that quite generally for 
any model  belonging to the Luttinger liquid universality class
the width of the ZS resonance asymptotically scales as~$q^2$ for $q \rightarrow 0$.

Let us now briefly define our model and introduce some useful notation.
We consider non-relativistic spinless fermions interacting with 
long-range density-density forces in one spatial dimension. The 
Euclidean action is
\begin{equation}
 S [ \bar{c} , c ] = 
 S_0 [ \bar{c} , c ] + \frac{1}{2} \int_Q f_{q} \rho_{-Q} \rho_Q\,,
 \label{eq:Sdef}
\end{equation}
where the non-interacting part can be written in terms of Grassmann fields 
${c}_K$ and $\bar{c}_K$ representing the spinless fermions
as follows,
\begin{equation}
 S_0 [ \bar{c} , c ] = - \int_K ( i \omega - \epsilon_k + \mu ) \bar{c}_K c_K.
 \label{eq:S0def}
\end{equation}
Here, $\mu$ is the chemical potential and the energy dispersion is assumed 
to be quadratic,
\begin{equation}
 \epsilon_k = \frac{k^2 }{ 2m}\,.
\end{equation}
The composite field 
\begin{equation}
 \rho_Q = \int_K \bar{c}_K c_{K+Q}
\end{equation}
represents the Fourier components of the density. The collective label 
$K = (  i \omega , k   )$ denotes fermionic Matsubara frequencies $i {\omega}$ 
and wave-vectors $k$, while $Q = (  i \bar{\omega} , q )$ depends on bosonic
Matsubara frequencies $i\bar{\omega}$. The corresponding integration symbols are
$\int_K = (\beta V)^{-1} \sum_{\omega , k}$, 
and $\int_Q = (\beta V)^{-1} \sum_{\bar{\omega},q}$,
where $\beta$ is the inverse temperature and $V$ is the volume of the system.
Eventually, we shall take  the limit of infinite volume $V \rightarrow \infty$ and
zero  temperature $\beta \rightarrow \infty$, where 
$\int_K = \int \frac{ d \omega dk }{(2 \pi)^2}$ and
$\int_Q = \int \frac{d \bar{\omega} d q }{(2 \pi)^2}$.
We assume that the Fourier transform $f_q$ of the interaction is suppressed
for momentum-transfers $q$ exceeding a certain cutoff $q_0 \ll k_F$. For explicit 
calculations it is sometimes convenient to use a sharp cutoff \cite{Pirooznia07},
\begin{equation}
 f_q = f_0 \Theta ( q_0 - | q | )\,.
 \label{eq:ftheta}
\end{equation}
However, as will be discussed in detail
in Sec.~\ref{sec:sharp}, 
the vanishing of all derivatives of $f_q$ at $q=0$ eliminates an important 
damping mechanism, so that it is better to work with a more realistic smooth cutoff,
such as a Lorentzian,
\begin{equation}
 f_q =   \frac{f_0 }{ 1  + q^2/ q_0^2 }\,. 
 \label{eq:Lorentzcut}
\end{equation}
Throughout this work we assume that the momentum-transfer cutoff $q_0$ (which for 
Lorentzian interaction can be identified with the Thomas-Fermi screening wave-vector)
satisfies
\begin{equation}
 p_0 \equiv \frac{q_0 }{  2 k_F }   \ll 1 \,.
 \label{eq:kappadef} 
\end{equation}
The precise form of $f_q$ is not important for our purpose, as long as
for small $q$ we may expand
\begin{equation}
 f_q = f_0 + \frac{1}{2} f_0^{\prime \prime} q^2 + O ( q^4 ), \; \; 
 \mbox{with  $f_0^{\prime \prime} \neq 0$}\,.
 \label{eq:fsmall}
\end{equation} 
By dimensional analysis, we may use the second derivative $ f_0^{\prime \prime}$
of the Fourier transform of the interaction to construct a new momentum scale 
\begin{equation}
 q_c = \frac{1}{m | f_{0}^{\prime \prime} | }\,,
 \label{eq:qcdef}
\end{equation}
which will play an important role in this work. Note that for Lorentzian cutoff
$f_0^{\prime \prime} = -2 f_0 /q_0^2 < 0$ and $q_c  =  q_0^2 / ( 2 m f_0)$, 
but in general the momentum scale $ q_c$ is  independent of the ultraviolet cutoff 
$q_0$. We  assume that
\begin{equation}  
 q_c \ll q_0 \ll k_F.
 \label{eq:qckF}
\end{equation}
For simplicity, we shall refer to the model defined above as the 
{\it{forward scattering model}} (FSM). If we further simplify the FSM by
linearizing the energy dispersion around the two Fermi points,
$\epsilon_{ \pm k_F + q} - \epsilon_{ k_F } \approx \pm v_F q$
and by extending the linear dispersion at each Fermi point
to the infinite line $- \infty < q < \infty$, then
the FSM reduces to the spinless TLM with dimensionless forward scattering 
interactions $\tilde{g}_2 = \tilde{g}_4 = g_0 $ in ``g-ology''-notation 
\cite{Solyom79}. In contrast to the TLM, the FSM does not require ultraviolet 
regularization, because the quadratic energy dispersion in one dimension renders 
all loop integrations ultraviolet convergent. Hence the usual problems 
associated with the removal of ultraviolet cutoffs and the associated 
anomalies \cite{Chubukov07} simply do not arise in the FSM.

To conclude this section, let us give a brief outline of the rest of 
this work. In Sec.~\ref{sec:RPA} we shall discuss the dynamic structure 
factor of the FSM within the RPA; although in this approximation the ZS mode 
is not damped, it is still instructive to start from the RPA because
it allows us to understand the origin of the mass-shell singularities
encountered in conventional bosonization. In Sec.~\ref{sec:Functionalbos},
we outline the functional bosonization approach to the FSM, which we then 
use in Sec.~\ref{sec:pert} to derive a self-consistency equation for 
$\Pi^{-1}_{\ast} ( \omega , q )$ which does not exhibit any mass-shell 
singularities. In Sec.~\ref{sec:sharp}, we present an evaluation
of this expression for sharp momentum-transfer cutoff (\ref{eq:ftheta}),
while in Sec.~\ref{sec:smooth} we consider a general  interaction $f_q$
of the type (\ref{eq:fsmall}). We also present explicit results for the
spectral line-shape of $S ( \omega , q )$ and the ZS damping.
In Sec.~\ref{sec:summary}, we briefly summarize our main results and
point out some open problems. There are four appendices with technical details:
In appendix \ref{sec:reduction}, we derive explicit expressions for the 
symmetrized closed fermion loops of the FSM. 
In the following two
appendices \ref{sec:threeloop} and \ref{sec:fourloop}
we carefully discuss the symmetrized three-loop and the four-loop 
which are 
needed for the calculations in the main part of this work. 
Finally, in 
appendix \ref{sec:FRG} we present a non-perturbative functional renormalization 
group flow equation for the irreducible polarization of the FSM. Although in 
this work we shall not attempt to further analyze this rather complicated 
integro-differential equation, we use it in Sec.~\ref{sec:selfcon} to justify our
self-consistency equation for $\Pi_{\ast} ( \omega , q )$.

\section{RPA for the forward scattering model}
\label{sec:RPA}

\begin{figure}[tb]    
   \centering
  \vspace{4mm}
  \includegraphics[scale=0.3]{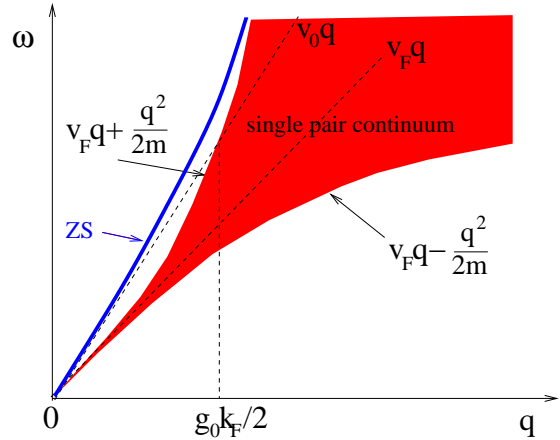} 
  \caption{(Color online) 
Regime in the $\omega$-$q$ plane where $S_{\rm RPA} ( \omega , q ) $ is finite.
The shaded regime represents the single-pair particle-hole continuum, while the thick
line corresponds to the ZS mode. For weak interaction $g_0 \ll 1$ the linear approximation 
$\omega_q \approx v_0 | q |$ to the dispersion of the ZS mode crosses the upper 
boundary of the single-pair continuum at $q  \approx g_0 k_F/2$. However, the  non-linear 
corrections to the ZS dispersion (\ref{eq:dispersion}) are such that it
never intersects the single-pair continuum, so that there is no Landau damping. 
}
    \label{fig:SRPAweight}
\end{figure}

\begin{figure}[tb]    
   \centering
  \vspace{4mm}
\includegraphics[scale=0.3]{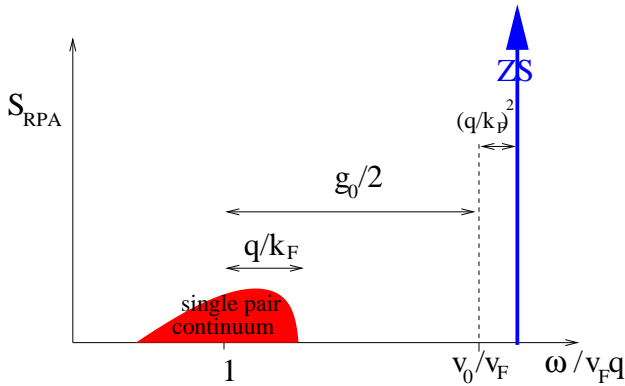} 
  \caption{(Color online)
Schematic behavior of  $S_{\rm RPA} ( \omega , q ) $ for fixed $q \ll k_F$  
as a  function of $\omega$ for $q / k_F \ll g_0 \ll 1$.In this regime, 
the distance between the upper edge of the single-pair particle-hole continuum 
and the position of the ZS peak (indicated by a thick arrow) is much larger than 
the width of the particle-hole continuum.
}
 \label{fig:SRPAcut}
\end{figure}

Because the RPA is exact for the TLM due to the vanishing
of the symmetrized closed fermion loops with more than two external 
legs~\cite{Kopietz95,Kopietz97,Dzyaloshinskii74,Metzner98}, it seems at
the first sight reasonable to use the RPA as a starting point 
of the perturbative calculation of 
the dynamic structure factor of the FSM.
It turns out, however, that the RPA result for $S ( \omega , q )$
exhibits some unphysical features (see below) which are related to 
the fact that the effect of interactions on the energy scale of the 
single-pair particle-hole continuum is not included in the RPA.
A better starting point would be the so-called RPAE or 
``time-dependent Hartree-Fock approximation'', because it takes the 
renormalization of the single-pair particle-hole continuum approximately 
into account \cite{Schoenhammer07,Ploetz07}. On the other hand, the RPA 
is sufficient to understand the relation between the mass-shell 
singularities and the expansion of the free polarization in powers of 
$1/m$, so that in this section, we shall carefully work out the
spectral line-shape of the FSM using the simple RPA. The irreducible 
polarization  is then approximated by
the non-interacting one,
\begin{eqnarray}
 \Pi_{\ast} ( Q ) \approx  
 \Pi_{0} (Q) & = & - \int_K G_0 (K) G_0 (K+Q)\,,
 \label{eq:Pi0}
\end{eqnarray}
where 
 \begin{equation}
 G_0 ( K ) = \frac{1}{ i {\omega} - \xi_k }\,,
\label{eq:G0xidef}
\end{equation}
with 
 \begin{equation}
\xi_k = \frac{k^2}{2m} - \frac{k_F^2}{2m}\,.
 \label{eq:xikdef}
 \end{equation}
For $\beta \rightarrow \infty$ and $V \rightarrow \infty$
the integrations can be performed analytically,
\begin{eqnarray}
 \Pi_{0} (Q)
 &  = &
 -\frac{1}{V} \sum_{k}     
 \frac{\Theta(- \xi_{k} )-\Theta(- \xi_{k+q})}{i\omega -
 \xi_{ k+ q} +\xi_{ k}} 
 \nonumber
 \\
 & = & \frac{m}{\pi q } 
 \ln \left| \frac{  i \bar{\omega} +  v_F q  + \frac{q^2}{2m} }{ i \bar{\omega} + v_F q 
  - \frac{q^2}{2m} } \right| \,.
\label{eq:Pi0res}
\end{eqnarray}
The corresponding RPA structure factor has been discussed in 
Ref.~[\onlinecite{Pirooznia07}].
It consists of two contributions,
\begin{equation}
 S_{\rm RPA} ( \omega , q ) = Z_q \delta ( \omega - \omega_q ) +
 S_{\rm RPA}^{\rm inc} ( \omega , q )\,,
 \label{eq:SRPA2} 
\end{equation}
where the first term represents the undamped ZS mode with weight
\begin{equation}
 Z_q = \frac{ v_F q^2}{ 2 \pi \omega_q} W_q ,
\end{equation}
and energy \cite{footnoteprint}
\begin{eqnarray}
 \omega_q & = & v_F | q | \sqrt{ 1 + \frac{q}{k_F} \coth \left( \frac{q }{k_F g_0}  
 \right) + \left[ \frac{q}{2k_F}\right]^2 }
 \nonumber
 \\
 & = &  v_0 | q |
 \left\{  1 
  + \frac{  g_0 (4  + 3g_0)    }{6 x_0^2} 
 \left[ \frac{ q }{2 k_F g_0} \right]^2  + O ( q^4 )  \right\}\,.
 \nonumber
 \\
 & &
 \label{eq:dispersion}
 \end{eqnarray}
The dimensionless function
\begin{equation}
 W_q = \frac{ \left[ \frac{q }{k_F g_0} \right]^2}{ \sinh^2 \left( \frac{q }{k_F g_0}  \right) }
 \label{eq:Wqdef} 
\end{equation}
can be identified with the relative contribution of the ZS peak to the 
$f$-sum rule \cite{Pirooznia07}
\begin{equation}
 \int_0^{\infty} d \omega \omega S ( \omega , q ) = \frac{v_F q^2}{2 \pi}.
 \label{eq:fsum}
\end{equation}
The second part $ S_{\rm RPA}^{\rm inc} ( \omega , q )$
in Eq.~(\ref{eq:SRPA2}) represents the incoherent continuum due to excitations
involving a single
particle-hole pair (single-pair continuum)~\cite{footnotethreedim}. 
Because the ZS mode never touches the
single-pair continuum, there is no Landau damping and 
 within RPA the ZS mode is undamped.
The damping of the ZS mode is 
due to excitations involving more than a single particle-hole pair 
(multi-pair excitations), which are neglected in  RPA. 
The regime in the 
$\omega$-$q$ plane where $S_{\rm RPA} ( \omega , q ) $ is finite is shown
in Fig.~\ref{fig:SRPAweight}. The corresponding qualitative shape  of  
$S_{\rm RPA} ( \omega , q ) $ for fixed $q \ll k_F$ is shown in Fig.~\ref{fig:SRPAcut}.
In the limit $g_0 \rightarrow 0$ the ZS mode disappears and the incoherent part 
$S_{\rm RPA}^{\rm inc} ( \omega , q )$ reduces to the dynamic structure factor 
of the free Fermi gas, which in one dimension for $q<2k_F$ is simply a box-function 
of width $q^2/m$ centered around $v_F | q |$,
\begin{eqnarray}
 S_0 ( \omega , q ) & = &
 \lim_{ g_0 \rightarrow 0}
 S_{\rm RPA}^{\rm inc} ( \omega , q ) 
 \nonumber
 \\
 &   = & \frac{m}{ 2 \pi | q| } 
 \Theta \left(  \frac{q^2}{2m} - \bigl|   \omega -v_F | q |\bigr|  \right) .
\end{eqnarray}
For finite $g_0$, the shape of $S_{\rm RPA}^{\rm inc} ( \omega , q )$
is modified as shown quantitatively in Fig. 1 of Ref.~[\onlinecite{Pirooznia07}].
The small shaded hump in Fig.~\ref{fig:SRPAcut} represents schematically the
incoherent part of $S_{\rm RPA} ( \omega , q )$ for finite $g_0$. 
For $ |q | / k_F \ll g_0$, the relative weight of the single-pair continuum 
is negligibly small, so that the ZS peak carries most of the spectral weight.
For example, the relative contribution of the single-pair continuum
to the $f$-sum rule vanishes as $ (q / g_0 k_F)^2  \ll 1$.

It is instructive to see which features of $S_{\rm RPA} ( \omega , q )$
are recovered if we expand the inverse non-interacting polarization 
$\Pi_0^{-1} ( Q )$ in powers of the inverse mass $m^{-1}$.
Therefore we introduce the dimensionless variables,
 \begin{eqnarray}
 i y  & = & \frac{ i \omega}{v_F q }  \; , \; p  =  \frac{q  }{ 2 k_F} ,
 \label{eq:ypdef}
\end{eqnarray}
and rewrite Eq.~(\ref{eq:Pi0res}) as
 \begin{equation}
 \Pi_0 (   i \omega ,q ) = \nu_0 \tilde{\Pi}_0 ( iy, p   ) ,
 \label{eq:tildePidef}
 \end{equation}
with the dimensionless function
 \begin{eqnarray}
  \tilde{\Pi}_0 (  i y, p  ) & = & \frac{1}{ 2 p } 
 \ln \left| \frac{ iy +1 +  p }{iy + 1 -  p} 
 \right|
 \nonumber
 \\
 & = & 
 \frac{1}{4  p } 
 \ln \left[ \frac{ y^2 +( 1 +  p)^2 }{y^2 + (1 -  p)^2} 
 \right] .
 \label{eq:Pi0quad}
 \end{eqnarray}
For an interaction with momentum-transfer cutoff $q_0 \ll k_F$ 
the relevant  
dimensionless momenta satisfy $|p|\ll 1$, so that  we expand 
$\tilde{\Pi}^{-1}_0 ( iy , p )$ in powers of $p$. From Eq.~(\ref{eq:Pi0quad}) 
we find
\begin{equation}
 \tilde{\Pi}_0^{-1} (  i y, p  )  = 1 + y^2 
 - \frac{p^2}{3}
 \frac{1- 3 y^2}{1 + y^2}  
 + O ( p^4).
 \label{eq:tildePi0expansion}
 \end{equation}
For later reference, we note that the correction of order
$p^2$ in Eq.~(\ref{eq:tildePi0expansion}) can be written as
\begin{eqnarray}
 - \frac{p^2}{3} \frac{1- 3 y^2}{ (1 + y^2)}  & = &
 p^2 \left[1 - \frac{4}{3 (1 + y^2)} \right]  
 \nonumber
 \\
 &  & \hspace{-10mm} =  p^2   - \frac{2p^2}{3}
 \left[ \frac{1}{1-iy} +  \frac{1}{1+iy} \right].
 \label{eq:laterref} 
\end{eqnarray}
For $p \rightarrow 0$ we recover the result for linearized dispersion,
\begin{equation}
 \lim_{p\rightarrow 0}  \tilde{\Pi}^{-1}_0 (  i y, p  ) 
 \equiv \tilde{\Pi}_0^{-1} ( i y ) = 1 + y^2 ,
 \label{eq:Pilambda0}
\end{equation}
which yields the dynamic structure factor of the TLM given in 
Eq.~(\ref{eq:STLM}). However, after analytic continuation to real frequencies
$iy \rightarrow x + i0 = \frac{\omega}{v_F q } + i 0$, the correction term 
of order $p^2$ in the expansion (\ref{eq:tildePi0expansion}) is singular on 
the mass-shell $ |\omega |  = v_F |q|$.
Although in the non-interacting limit we know that this
mass-shell singularity has been artificially generated by expanding
the logarithm  in Eq.~(\ref{eq:Pi0quad}), it is not at all obvious how to 
regularize  a similar singularity if it is encountered in the interacting system.
This is the reason why a formal expansion in powers of the band curvature 
$1/m$ using either a purely fermionic approach \cite{Teber06} or conventional 
bosonization \cite{Aristov07} is not reliable close to the mass-shell
after analytic continuation. Fortunately, within the functional bosonization 
approach used in this work this problem does not arise, because the effective 
expansion parameter in functional bosonization is not $1/m$, but the combination
$ g_0 q_0/(m v_F)$, see Refs.~[\onlinecite{Kopietz95,Kopietz97}]. In particular, 
in the non-interacting limit functional bosonization yields the exact structure 
factor of the free Fermi gas with quadratic dispersion, containing all orders 
in $1/m$.

It is instructive to examine the RPA dynamic structure factor
if we  nevertheless use the expansion (\ref{eq:tildePi0expansion}) for the 
non-interacting polarization. Then we obtain after analytic continuation 
$iy \rightarrow x  + i 0 = \omega /(v_F q ) + i 0$ for small 
$  |q | \ll g_0 k_F$,
\begin{eqnarray}
 S_{\rm RPA} ( \omega , q ) & \approx & \frac{\nu_0}{\pi} {\rm Im}
 \left[ \frac{1}{ g_0 + \tilde{\Pi}^{-1}_0 ( x + i 0, p ) }\right]
 \nonumber
 \\
 & & \hspace{-10mm} = Z_q^{+} \delta ( \omega - \tilde{\omega}_q^{+} ) +
 Z_q^{-} \delta ( \omega - \tilde{\omega}_q^{-} )\,,
 \label{eq:S2peaks}
 \end{eqnarray}
where $Z_q^{+}$ and $\tilde{\omega}_q^{+}$ reduce for small $q$
to the corresponding expressions $Z_q$ and $ v_0 | q |$ for
linear dispersion [see Eq.~(\ref{eq:Zqdef})], 
and the weight and dispersion of the other mode $\tilde{\omega}_q^{-}$
is for $ |q  | \ll k_F g_0$,
\begin{equation}
 Z_q^{-}  \approx \frac{ 2 | q | }{3 \pi} \left[ \frac{q  }{2 k_F g_0}   \right]^2,
 \label{eq:Zqminus}
 \end{equation}
 \begin{equation}
 \tilde{\omega}_q^{-} \approx  v_F | q| \left[ 1 -  \frac{2}{3g_0} \left(
 \frac{ q }{2 k_F} \right)^2 \right]\,.
\end{equation}
This peak is associated with  the incoherent part 
$S_{\rm RPA}^{\rm inc} (\omega, q)$ of the dynamic structure factor
discussed above,
which in the approximation (\ref{eq:tildePi0expansion}) is replaced by a single 
peak with the same weight. From Eqs.~(\ref{eq:Zqminus}) and (\ref{eq:Zqdef}) 
one easily verifies that for $ | q |  \ll g_0 k_F $ the relative weight of 
the peak associated with the incoherent part is indeed small,
\begin{equation}
 \frac{Z_q^{-}}{Z_q^{+}} = \frac{  4  x_0}{3} \left[ \frac{ q }{ 2  k_F g_0 } \right]^2
 = \frac{  4  \pi^2 x_0 p_0^2 }{3}   \left[     \frac{  v_F q}{f_0 q_0}
 \right]^2\,,
 \label{eq:Zpmratio}
\end{equation}
where we have used $p_0 = q_0 /(2 k_F )$, see Eqs.~(\ref{eq:kappadef}).
Hence, for $ | q |  \ll  g_0 k_F  $ most of the weight of
$S_{\rm RPA} (  \omega , q )$ is carried by the ZS mode 
$\tilde{\omega}_q^+ \approx v_0 | q|$, so that  the incoherent part
corresponding to the mode $\tilde{\omega}_q^{-} \approx v_F | q |$ can be 
neglected \cite{Pirooznia07}. Note that the limits $q \rightarrow 0$ 
and $g_0 \rightarrow 0$ do not commute and that only for 
$ |q | /( 2 k_F ) \ll g_0$ the weight of the mode $\tilde{\omega}_q^{-}$ can be neglected.

Mathematically, the second peak in Eq.~(\ref{eq:S2peaks})
is due to the pole arising from the term
of order $p^2$ in the expansion (\ref{eq:tildePi0expansion})
of the inverse free polarization.
Although after analytic continuation $iy \rightarrow x + i 0$
this term is singular for $x=1$, we know from the
exact result (\ref{eq:Pi0quad}) how this singularity should be regularized:
we simply should smooth out the corresponding
$\delta$-function peak over an interval of width $\text{w}_q \propto q^2/m$.
In fact, we can self-consistently calculate $\text{w}_q$
by noting that after analytic continuation
the singular term in the expansion
(\ref{eq:tildePi0expansion})
gives rise to the following  formally infinite imaginary part 
of the inverse non-interacting polarization,
 \begin{equation}
 {\rm Im } \tilde {\Pi}_0^{-1} ( x + i 0 , p ) = - \Gamma_0 (x , p ) = - \frac{2 \pi}{3} p^2 
[ \delta (1-x ) - \delta (1+x )].
 \label{eq:Imreg}
 \end{equation}
Ignoring the renormalization arising from the (singular)
real part of  $\tilde {\Pi}_0^{-1} ( x + i 0 , p )$ and approximating
the resulting dynamic structure factor in this regime by a Lorentzian
centered at $ \omega = v_F | q |$, we find 
for the full width at half maximum in the limit $g_0 \ll 1$,
 \begin{equation} 
 \text{w}_q  = \frac{ v_F  | q |}{2} \Gamma_0 (1 , p = q /(2 k_F)).
 \label{eq:wq} 
\end{equation}
To obtain a self-consistent estimate for $\text{w}_q$ we follow
Samokhin \cite{Samokhin98} and regularize the
singularity in $\Gamma_0 ( 1, p )$ by replacing
$\delta ( \omega =0)$ 
by the height  of a normalized Lorentzian of width $\text{w}_q$ on resonance,
 \begin{equation}
\left. \delta ( x -1 ) \right|_{x=1} = v_F | q | 
\left. \delta ( \omega - v_F | q | ) \right|_{ \omega = v_F | q | } 
  \rightarrow \frac{ v_F | q |}{ \pi \text{w}_q }.
 \end{equation}
Hence, our self-consistent regularization is
 \begin{equation}
 \Gamma_0 ( 1 , p  ) \rightarrow \frac{ 2 p^2v_F | q |}{3 \text{w}_q }.
 \end{equation}
Substituting this into Eq.~(\ref{eq:wq}) we obtain  the 
self-consistency equation
 \begin{equation}
 \text{w}_q = \frac{1}{3}  \left( \frac{q}{2 k_F} \right)^2 \frac{ (v_F q)^2}{\text{w}_q},
 \end{equation} 
which leads to the
following estimate  
for the width of the single-pair particle-hole continuum,
 \begin{equation}
 \text{w}_q = \frac{1}{2 \sqrt{3}} \frac{q^2}{m}.
 \label{eq:wq0res}
\end{equation}
Of course, it is now known \cite{Pustilnik06,Pereira06,Pereira07}
that the shape of the single-pair continuum
cannot be approximated by a Lorentzian, but the
order of magnitude of its width obtained
within the above regularization is correct for sufficiently small $q$.
Hence, the mass-shell singularity
arising after analytic continuation $iy \rightarrow x + i 0$ 
in the expansion of the inverse non-interacting
polarization (\ref{eq:tildePi0expansion})
in powers of $p = q /(2 k_F )$ is simply related to the
single-pair particle-hole continuum. This singularity can be regularized by  smearing out
the $\delta$-function in the imaginary part over a finite interval of width 
$\text{w}_q \propto q^2/m$.
However, the width $\text{w}_q$ 
should not be confused with the damping of the ZS mode, which
remains sharp within RPA.

In order to obtain the ZS damping, one should calculate
interaction corrections to the irreducible polarization.
The finite overlap between the continuum due to
particle-hole excitations involving more than a single 
particle-hole pair (multi-pair excitations) then determines the ZS damping.  
In three dimensions general phase space arguments \cite{Pines89}
imply that the resulting damping is very small.
In one dimension, an  argument due to Teber \cite{Teber06} suggests  that
the damping of any acoustic collective mode which overlaps
with the two-pair continuum should vanish as $q^3$ for small $q$.
However,  for this argument to be valid,
one should self-consistently calculate the renormalized energy 
of the ZS mode and show that it is immersed in the multi-pair continuum.
This has neither been done in our previous work \cite{Pirooznia07}, nor
in the work by Pustilnik  {\it{et al.}} \cite{Pustilnik06}, where
the renormalization of  the ZS velocity has been ignored.

In this work we shall carefully examine all corrections
to the RPA to second order in an expansion in powers
of the small parameter $p_0 = q_0 / ( 2 k_F )$,
which is naturally generated using the functional bosonization 
approach \cite{Kopietz95,Kopietz97}.
Most importantly, our approach does not suffer from the 
mass-shell singularities discussed above. Moreover, we shall show that
the distinction between the ZS energy  $v_0 |q | $ and
the energy scale $ v_F | q | $ associated with the single-pair
continuum shown schematically in
Figs. \ref{fig:SRPAweight} and \ref{fig:SRPAcut} is an unphysical artefact
of the RPA, which disappears once the corrections to the RPA are self-consistently 
taken into account.

\section{Functional bosonization}
\label{sec:Functionalbos}

In this section we 
outline the functional bosonization approach \cite{Kopietz95,Kopietz96,Kopietz97}
which we then use in  Sec.~\ref{sec:pert} to calculate the dynamic structure factor.
In contrast to Refs.~[\onlinecite{Kopietz95,Kopietz96,Kopietz97}], we shall here
keep track of Hartree corrections to the fermionic self-energy, because these corrections 
contribute to the renormalization of the ZS velocity.

\begin{figure}
  \begin{center}
    \epsfig{file=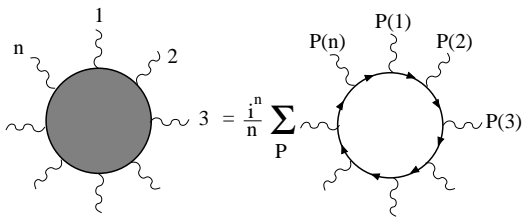,width=0.8\hsize}
  \end{center}
  \caption{
Boson vertex with $n$ external legs in the interaction part $S_{\rm int} [ \delta \phi ]$
of the bosonized effective action, see Eq.~(\ref{eq:Sint}). The arrows denote the 
fermionic Green functions $G_0 ( K )$  within self-consistent Hartree approximation, 
see Eq.~(\ref{eq:G0Hartree}). The sum is taken over the $n!$ permutations of the labels 
of the external legs. For linearized energy dispersion all symmetrized closed fermion 
loops with more than two external legs vanish.  }
  \label{fig:closedloop}
\end{figure}

Decoupling the density-density interaction in Eq.~(\ref{eq:Sdef})
by means of a real Hubbard-Stratonovich field $\phi$,
the ratio of the partition functions with and without 
interaction can  be written as
\begin{equation}
 \frac{\cal{Z}}{{\cal{Z}}_0}
 = \frac{ \int {\cal{D}}[ \bar{c} , c,  \phi] 
 e^{-  S_0 [ \bar{c} , c ] - S_0 [  \phi ]
 - S_1 [ \bar{c} , c , \phi ] } }{
 \int {\cal{D}}[ \bar{c} , c ,  \phi] 
 e^{-  S_0 [ \bar{c} , c ] - S_0[ \phi ] } }
 \; ,
 \label{eq:Zratio}
\end{equation}
where the free fermionic action $S_0 [ \bar{c} , c ]$ is given
in Eq.~(\ref{eq:S0def}),  the free bosonic part is
\begin{equation}
 S_0 [   \phi ]  = \frac{1}{2}  
  \int_Q  f^{-1}_{q} {\phi}_{-Q} \phi_Q
 \; , 
 \label{eq:S0phi}
\end{equation}
and the Fermi-Bose interaction is
\begin{eqnarray}
 S_1 [ \bar{c} , c ,  \phi ]
 & = &  i \int_Q \int_K  \bar{c}_{ K +Q } c_{ K } \phi_Q .
 \label{eq:S1}
 \end{eqnarray}
The fermionic part of the action in the numerator of Eq.~(\ref{eq:Zratio}) 
can be written as
\begin{eqnarray}
 S_0 [ \bar{c} , c  ] + 
 S_1 [ \bar{c} , c ,  \phi ] 
 & = & 
 -  \int_K  \int_{ K^{\prime}} \bar{c}_K [ {\bf{G}}^{-1} ]_{ K K^{\prime}}
  {c}_{ K^{\prime} }  
 , 
 \nonumber
 \\
 & &
\end{eqnarray}
where the infinite matrix ${\bf{G}}^{-1}$ is defined by
\begin{equation}
 [ {\bf{G}}^{-1} ]_{ K K^{\prime}} = \delta_{ K,  K^{\prime}} [ i \omega - \epsilon_k + \mu]
  - i \phi_{ K - K^{\prime}}.
\end{equation}
At finite density, the field $\phi_K$ has a non-zero expectation value,
\begin{equation}
 \phi_Q = - i \delta_{ Q,0} \bar{\phi}  + \delta \phi_Q.
\end{equation}
Here the $\delta$-symbol is for finite $\beta$ and $V$ (where
the components of $Q = ( \bar{\omega} , q )$ are discrete)  given by  
$\delta_{ Q , 0} = \beta V \delta_{ \bar{\omega},0} \delta_{ q , 0}$,
which reduces to $ (2 \pi )^2 \delta ( \bar{\omega} ) \delta (q )$
for $\beta \rightarrow \infty$ and $V \rightarrow \infty$.
We fix the real constant $\bar{\phi}$ from the requirement that 
the effective action $S_{\rm eff} [ \phi ]$
of the $\phi$-field, which is obtained by integrating over the fermionic fields
in Eq.~(\ref{eq:Zratio}), does not contain
a term linear in the fluctuation $\delta \phi_Q$. To do this, we define
the matrix ${\bf{G}}_0^{-1}$ which includes the self-energy correction
due to the vacuum expectation value $\bar{\phi}$,
\begin{equation}
 [ {\bf{G}}_0^{-1} ]_{ K K^{\prime}} = \delta_{ K,  K^{\prime}} [ i \omega - \epsilon_k - \bar{\phi} + \mu],
 \label{eq:G0def}
 \end{equation}
and write
 \begin{equation}
 {\bf{G}}^{-1} = {\bf{G}}_0^{-1} - {\bf{V}},
 \end{equation}
with
 \begin{equation}
 [ {\bf{V}} ]_{ K  K^{\prime} } = i \delta \phi_{ K - K^{\prime}}.
 \end{equation}
Integrating in Eq.~(\ref{eq:Zratio}) over the fermion fields we obtain
the formally exact expression
\begin{equation}
 \frac{\cal{Z}}{{\cal{Z}}_0} =
 e^{ - \beta ( \Omega_1 - \Omega_0 )}
 \frac{ \int {\cal{D}}[ \delta \phi] 
 e^{-  S_{\rm eff} [   \delta \phi ] } }{
 \int {\cal{D}}[  \phi] 
 e^{- S_0[  \phi ] } }
 \; ,
 \label{eq:Zintegrate}
\end{equation}
where $\Omega_1 - \Omega_0$ is the change of the grand canonical potential
due to the vacuum expectation value ignoring fluctuations, 
\begin{equation}
  \Omega_1 - \Omega_0  =  
 \frac{1}{\beta} {\rm Tr}  
 \ln [ \mathbf{G}_{0} ( \bar{\phi} )    \mathbf{G}^{-1}_{0} ( \bar{\phi}=0 ) ]
 - V \frac{ \bar{\phi}^2 }{ 2 f_0} .
 \label{eq:Omega10}
\end{equation}
The effective action for the fluctuations of the bosonic field is
\begin{eqnarray}
 S_{\rm eff} [  \delta \phi   ] & = & 
 S_0 [ \phi_Q \rightarrow -i \delta_{Q,0} \bar{\phi}  + \delta \phi_Q   ] 
 - \beta V   \frac{  \bar{\phi}^2}{2 f_0} 
 \nonumber
 \\ 
 &  & -
 {\rm Tr} \ln [ 1 - {\bf{G}}_0  \bd{V} ]
 \nonumber
 \\
 & = & 
 \frac{1}{2} \int_{Q} 
 f_q^{-1} \delta \phi_{ - Q}  \delta \phi_{ Q}
  - i  f_{0}^{-1} \bar{\phi} \delta \phi_0 
 \nonumber
 \\ 
 &   &  + \sum_{n=1}^{\infty} \frac{
 {\rm Tr} [ {\bf{G}}_0  \bd{V} ]^n}{n}.
 \label{eq:Seff}
\end{eqnarray}
We now fix the vaccum expectation value $\bar{\phi}$ from the saddle point condition
\begin{equation}
  \frac{ \partial \Omega_1}{\partial \bar{\phi}} = - V \frac{ \bar{\phi}}{f_0} + V \rho_0 =0,
 \label{eq:Hselfcon} 
\end{equation}
or equivalently
\begin{equation}
 \bar{\phi} = f_0 \rho_0.
\end{equation}
Here, $\rho_0$ is the density
and $G_0(K)$ is the fermionic Green function
in self-consistent Hartree approximation, where
 \begin{equation}
 \rho_{0} = \int_K G_0 ( K ) = \frac{1}{V} \sum_k \Theta ( \mu - \epsilon_k - f_0 \rho_0 )\,,
 \label{eq:Hartreerho} 
\end{equation}
 \begin{equation}
 G_0 ( K ) = \frac{ 1}{ i \omega - \epsilon_k  - f_0 \rho_0 + \mu}.
 \label{eq:G0Hartree} 
\end{equation}
Note that Eq.~(\ref{eq:G0Hartree}) agrees with  Eq.~(\ref{eq:G0xidef}) if we
take into account that within  self-consistent
Hartree approximation the Fermi momentum $k_F$ is defined via
\begin{equation}
 \frac{k_F^2}{2 m} = \mu - f_0 \rho_0.
\end{equation}
Eq.~(\ref{eq:Hselfcon}) guarantees that the
terms linear in the fluctuations $\delta \phi_Q$
in Eq.~(\ref{eq:Seff}) cancel, so that
our final result for the effective action for the
fluctuations of the Hubbard-Stratonovich field is
\begin{eqnarray}
 S_{\rm eff} [  \delta \phi   ] & = & 
 \frac{1}{2} \int_{Q} 
 f_q^{-1} \delta \phi_{ - Q}  \delta \phi_{ Q}
 + \sum_{n=2}^{\infty} \frac{
 {\rm Tr} [ {\bf{G}}_0  \bd{V} ]^n}{n}
 \nonumber
 \\
 &  = &
 S_2 [ \delta \phi ] + S_{\rm int} [ \delta \phi ],
 \label{eq:Sefffinal}
\end{eqnarray}
with the Gaussian part given by
 \begin{equation}
 S_2 [ \delta \phi ] =
\frac{1}{2} \int_{Q} 
 [ f_q^{-1} + \Pi_0 ( Q ) ] \delta \phi_{ - Q}  \delta \phi_{ Q} ,
 \label{eq:S2}
 \end{equation}
and the interaction part
 \begin{eqnarray}
S_{\rm int} [ \delta \phi ] 
 & = & 
 \sum_{ n=3}^{\infty} \frac{1}{n!} \int_{Q_1} \ldots \int_{ Q_n} 
  \delta_{Q_1 + \ldots + Q_n,0} 
 \nonumber 
\\
 &  \times &
\Gamma_0^{(n)} ( Q_1 , \ldots, Q_n ) \delta \phi_{ Q_1}
 \ldots \delta \phi_{ Q_n}.
 \label{eq:Sint}
 \end{eqnarray}
The vertices $\Gamma_0^{(n)} ( Q_1 , \ldots, Q_n )$ are proportional 
to the symmetrized closed fermion loops $L^{(n)}_S (  Q_1 , \ldots ,  Q_n )$ 
defined in Eq.~(\ref{eq:fermionloop}),
\begin{eqnarray}
 \Gamma^{(n)}_{0} ( Q_1, \ldots , Q_n ) & = &
 i^n (n-1)! L^{(n)}_S ( - Q_1 , \ldots , - Q_n ) .
 \nonumber
 \\
 & &
\end{eqnarray}
A graphical representation of  $\Gamma^{(n)}_{0} ( Q_1, \ldots , Q_n )$ 
is shown in Fig.~\ref{fig:closedloop}.
In appendix~\ref{sec:reduction} we give explicit  expressions for the
symmetrized $n$-loops of the FSM \cite{Neumayr98,Neumayr99,Kopper01} and 
show that $L_S^{(n)} \propto (1/m)^{n-2}$ to leading order in $1/m$.
Moreover, in appendices \ref{sec:threeloop}  and \ref{sec:fourloop}
we carefully discuss  the properties of the 
loops $L_S^{(3)} ( Q_1 , Q_2 , - Q_1 - Q_2)$ and
$L_S^{(4)} (  Q_1 , Q_2 , -Q_1 , - Q_2)$ for the special combinations
of momenta needed in this work. 

The exact irreducible polarization can now be obtained from the
fluctuation propagator  of the Hubbard-Stratonovich field,
\begin{eqnarray}
  \langle \delta \phi_Q \delta \phi_{Q^{\prime} } \rangle & = &
 \frac{ \int {\cal{D}} [ \delta \phi ]  e^{ - S_{\rm eff} [ \delta \phi ] }   
 \delta \phi_Q \delta \phi_{Q^{\prime} } }{
 \int {\cal{D}} [ \delta \phi ]  e^{ - S_{\rm eff} [ \delta \phi ] } }
 \nonumber
 \\
 & = &    
 \delta_{Q + Q^{\prime},0}
 \frac{1}{ f_q^{-1} + \Pi_{\ast} ( Q )},
\end{eqnarray} 
where  the effective 
action  $S_{\rm eff} [ \delta \phi ]$ is defined in
Eq.~(\ref{eq:Sefffinal}). Within the Gaussian approximation this reduces to the 
RPA interaction 
\begin{eqnarray}
  \langle \delta \phi_Q \delta \phi_{Q^{\prime} } \rangle_{S_2}& = & 
\delta_{Q + Q^{\prime},0}
 \frac{1}{ f_q^{-1} + \Pi_{0} ( Q )}
 \nonumber
 \\
 & \equiv &  \delta_{Q + Q^{\prime},0}
 f_{\rm RPA} ( Q ).
 \label{eq:fRPAdef}
\end{eqnarray} 
The corrections to the RPA can now be calculated systematically
in powers of the interaction $S_{\rm int}$
using the Wick theorem. The RPA interaction thereby plays the role of the 
Gaussian propagator, so that we naturally obtain an expansion in powers of the
RPA interaction. As discussed in appendix \ref{sec:reduction},  
the vertices $\Gamma^{(n)}_0 (Q_1 , \ldots , Q_n )$ 
with $n$ external bosonic legs are proportional to $(1/m)^{ n-2}$, so that the 
underlying small parameter of our expansion is the product of the RPA 
interaction and $1/m$. In particular, in the limit of vanishing interaction 
we recover the exact non-interacting polarization $\Pi_0 ( i \omega , q )$
for quadratic energy dispersion.
Our approach based on functional bosonization is therefore fundamentally 
different from conventional bosonization \cite{Samokhin98,Teber06,Aristov07}, where the quadratic term in the energy 
dispersion gives rise to a cubic vertex proportional to $1/m$ which has to be 
re-summed to infinite order to recover the correct  non-interacting polarization.

\section{Calculation of $S ( \omega , q )$ using  functional
bosonization}
\label{sec:pert}

\subsection{One-loop self-consistency equation for
$\Pi_{\ast} ( Q)$}
 \label{sec:selfcon}

\begin{figure}[tb]    
\centering   
  \vspace{4mm}
\includegraphics[scale=0.3]{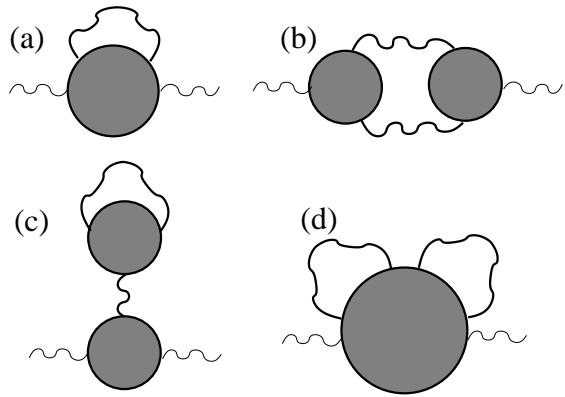}
\caption{
Diagrams arising in the perturbative expansion of the irreducible polarization to
second order in the RPA interaction. The shaded circles represent the vertices of 
$S_{\rm eff} [ \delta \phi ]$, which are related to symmetrized closed fermion 
loops as defined in Fig.~\ref{fig:closedloop}. Diagram (a) is equivalent to the 
three fermionic diagrams shown in Fig.~\ref{fig:feynman}. Diagram (b) is the 
so-called Aslamazov-Larkin diagram, while diagram (c) can be viewed as a higher 
order self-energy correction which renormalizes the relation between density and 
chemical potential. Diagram (d) involving two bosonic loops and the symmetrized fermionic 
six-loop is of fourth order  in $p_0 = q_0 /(2 k_F)$ and can be neglected to order $p_0^2$.
 }
\label{fig:pertdiagrams}
\end{figure}

It is now straightforward to expand the irreducible polarization in powers
of the RPA interaction, which is the Gaussian propagator of our boson field 
$\delta \phi_Q$. The diagrams contributing to $\Pi_{\ast} ( Q )$ up to second 
order in the RPA interaction are shown in Fig.~\ref{fig:pertdiagrams}.
The relevant dimensionless parameter for this expansion
is the ratio $p_0 = q_0 / (2 k_F) $, because
by assumption the range of the interaction 
in momentum space has a cutoff $q_0 \ll k_F$ so that 
each additional bosonic loop integration gives rise to 
a factor of $p_0^2$. Hence, the diagram (d) involving two bosonic loops of
is of order $p_0^4$, because according to Eq.~(\ref{eq:Loopm}) 
the symmetrized fermion loop with six external legs
is proportional to $1/m^4 $ and the two bosonic loop integrations generate
a factor of $q_0^4$. Because the other three diagrams are proportional to $p_0^2$, 
it is consistent to neglect diagram (d) as long as we retain all terms
up to order $p_0^2$.
Evaluating the diagrams (a)--(c) in Fig.~\ref{fig:pertdiagrams} 
we obtain the following expression for the irreducible polarization,
\begin{widetext}
\begin{eqnarray}
  \Pi_\ast ( Q)  & \approx &  \Pi_0 (Q ) -
  \frac{ 1}{2} 
  \int_{{Q^{\prime}}}
  {f}_{\rm RPA}  ({Q}^{\prime})
  \Biggl\{ 
  6 L^{(4)}_S ({Q}^{\prime} ,- {Q}^{\prime} , {Q} ,- Q)
   + 4      {f}_{\rm RPA} (0)     L^{(3)}_{S} ( Q , - Q , 0 )   L^{(3)}_{S} (  {Q}' ,- {Q}' ,0)
  \nonumber
  \\
  &  &    \hspace{40mm} +4   f_{\rm RPA} ( Q+ {Q}') 
  L^{(3)}_{S} (- {Q} , {Q}+ {Q}' ,- {Q}' )
  L^{(3)}_{S} ( {Q}' ,- {Q}- {Q}' , {Q}) \Biggl\}.
  \label{eq:Pipert}
\end{eqnarray}
\end{widetext}
The properties of the symmetrized three- and four-loops
appearing in this expression are discussed in detail in 
appendices~\ref{sec:threeloop} and \ref{sec:fourloop}. 
It turns out, however, that in order to
cure the unphysical features of the RPA
discussed at the end of Sec.~\ref{sec:RPA}
(in particular, within RPA  the energy scale $v_F |q |$ of the single-pair
continuum  erroneously involves the bare Fermi velocity),
we should self-consistently dress the Gaussian propagator
$f_{\rm RPA} ( Q )$ in Eq.~(\ref{eq:Pipert}) by self-energy corrections.
Formally, this amounts to replacing the RPA interaction by the
exact effective interaction, 
\begin{equation}
 f_{\rm RPA} ( Q ) \rightarrow
 f_{\ast} ( Q ) = \frac{f_q}{1 + f_q \Pi_{\ast} ( Q )}.
\end{equation}
In appendix~\ref{sec:FRG} we justify this procedure using a 
functional renormalization group approach \cite{Schuetz05,Schuetz06}. 
With this substitution, Eq.~(\ref{eq:Pipert}) becomes a complicated 
integral equation for the irreducible polarization, which cannot be 
solved analytically. Fortunately, this integral equation can again 
be simplified by noting that on the right-hand side it is not necessary
to retain the full $Q$-dependence of $\Pi_{\ast} ( Q )$, but
to keep only those terms which contribute to the self-consistent
renormalization of the ZS velocity.
To explain this, let us  introduce again the dimensionless variables
$ i y = i \omega /(v_F q )$ and  $p = q /(2 k_F)$ and define
the dimensionless irreducible polarization
\begin{equation}
 \Pi_{\ast} ( i \omega , q ) = \nu_0 \tilde{\Pi}_{\ast} ( iy , p ).
\end{equation}
The  corresponding dimensionless effective interaction is then 
\begin{equation}
 \tilde{f}_{\ast} ( iy , p ) = \frac{g_p}{1 + g_p \tilde{\Pi}_\ast ( iy , p )},
\end{equation}
where $g_p = \nu_0 f_{ q = p k_F }$,
see also Eqs.~(\ref{eq:ypdef}) and (\ref{eq:tildePidef}).
The dynamic structure factor can then be written as
\begin{eqnarray}
 S ( \omega , q ) & = & \frac{1}{\pi} {\rm Im} \left[ \frac{1}{ f_q + \Pi_{\ast}^{-1} ( 
 \omega + i 0 , q ) } \right]
 \nonumber
 \\
 & = &
 \frac{\nu_0}{\pi}  {\rm Im} \left[ \frac{1}{ g_p + \tilde{\Pi}_{\ast}^{-1} ( 
 x + i 0 , p ) } \right],
 \label{eq:Sfull}
\end{eqnarray}
where $x = \omega /(v_F q )$. 
For our purpose it is now sufficient to approximate
the dimensionless inverse irreducible polarization  by
\begin{equation}
 \tilde{\Pi}_{\ast}^{-1}  ( i y , p ) = Z_1 +  Z_2 y^2,
 \label{eq:Piinvren}
\end{equation}
where the dimensionless renormalization factors $Z_1$ and $Z_2$
should be determined as a function of the interaction such that
the approximation (\ref{eq:Piinvren}) yields the true ZS velocity
$v$. Within RPA, where the non-linear terms in the energy dispersion do
not renormalize the ZS velocity, the irreducible
polarization is approximated by the non-interacting one, so that $Z_1 = Z_2 = 1$.
If we approximate the inverse
polarization  in Eq.~(\ref{eq:Sfull}) by Eq.~(\ref{eq:Piinvren}) 
we obtain for $ \omega > 0$ and $q \rightarrow 0$,
 \begin{equation}
 S ( \omega , q ) \approx \frac{v_F |q |}{2 \pi v Z_2 } \delta ( \omega - v | q | ),
 \label{eq:Sfix} 
\end{equation}
 where the renormalized ZS velocity is
 \begin{equation}
 \frac{v }{ v_F}  = \sqrt{ \frac{Z_1 + g_0}{Z_2} } \equiv x_0 \equiv \sqrt{ 1 + g },
 \label{eq:velren} 
 \end{equation}
with renormalized coupling constant
 \begin{equation}
g = \frac{ g_0 + Z_1 }{Z_2} -1 .
 \label{eq:tildeg0def}
 \end{equation}
In order to avoid the unphysical
splitting of the spectral weight in $S ( \omega , q )$
(as discussed at the end of Sec.~\ref{sec:RPA}, this is an artefact of the RPA)
it is crucial that the true ZS velocity $v$ appears in the
bosonic propagators. Therefore, a naive expansion in powers of the
RPA interaction is not sufficient. However, we may further reduce the
complexity of the calculation by noting that Eq.~(\ref{eq:Sfix})
still contains the correct velocity if we set $Z_{2} \rightarrow 1$ in the
prefactor. Within this approximation, the velocity renormalization
implied by Eq.~(\ref{eq:Piinvren}) can be simply taken into account
via  a re-definition of the coupling constant, $g_0 \rightarrow g$.
It is therefore sufficient to replace the RPA interaction in Eq.~(\ref{eq:Pipert})
by  an effective interaction of the same form
but with a  renormalized effective coupling $g$ 
instead of $g_0$, which should be chosen such that
all interaction corrections to the ZS velocity are  self-consistently taken into account.
Note  that Sch\"{o}nhammer \cite{Schoenhammer07} has recently shown that
within the so-called RPAE approximation (which amounts to solving the
Bethe-Salpeter equation with the bare interaction as irreducible vertex)
the relative position of the collective mode energy and the energy of the
single-pair particle-hole continuum is different from the RPA prediction for the FSM:
In RPAE the ZS mode lies above the non-interacting
single particle-hole continuum which (erroneously) appears in RPA, but
below the Hartree-Fock particle-hole continuum.
This suggests that in order to obtain a correct estimate of the
ZS damping, it is necessary to calculate the location of the ZS energy 
self-consistently. 

In field-theoretical language the constants $Z_1$ and $Z_2$ are counter-terms
which guarantee that our Gaussian propagator depends on  the true ZS velocity. 
In Sec.~\ref{sec:sharp} we shall explicitly calculate the factors $Z_1$, 
$Z_2$ and the corresponding renormalized ZS velocity $v$ to second order in 
our small parameter $p_0$. A similar procedure 
is necessary to self-consistently calculate the true Fermi surface of an interacting
Fermi system \cite{Neumayr03,Ledowski03}. 
The expansion of the modified dimensionless interaction $\tilde{g}_p$
for small $p$ is then
\begin{equation}
 \tilde{g}_p = g + \frac{1}{2} g_0^{\prime \prime} p^2 + O ( p^4 ),
\end{equation}
where
\begin{equation}
 g_0^{\prime \prime} = ( 2 k_F)^2 \nu_0 f_0^{\prime \prime} 
 = {\rm sign} f_0^{\prime \prime}
 \frac{2}{\pi p_c}.
 \label{eq:g0pp}
\end{equation}
In this approximation, our dimensionless effective interaction is
 \begin{equation}
 \tilde{f}_{\ast} ( iy , p ) \approx \tilde{f}_{g} ( iy , p ) =
 \frac{\tilde{g}_p}{1 + \tilde{g}_p \tilde{\Pi}_0 ( iy , p )},
 \end{equation}
which differs from the RPA interaction, because the 
function $\tilde{g}_p$ includes the renormalization of the ZS velocity
due to fluctuations beyond the RPA.

Collecting all terms, our final result for the dimensionless irreducible polarization
to one bosonic loop can be written as 
\begin{equation}
 \tilde{\Pi}_{\ast} ( i y , p ) \approx 
 \tilde{\Pi}_{0} ( i y , p ) + \tilde{\Pi}_1 ( iy , p ) +  \tilde{\Pi}_2 ( iy , p ),
 \label{eq:pol12}
\end{equation}
where the non-interacting polarization is given in Eq.~(\ref{eq:Pi0quad}), and the 
subscripts indicate the powers of $\tilde{g}_p$. The term
$\tilde{\Pi}_1 ( iy , p )$  corresponding to diagram (a) in 
Fig.~\ref{fig:pertdiagrams} can be written as
\begin{widetext}
 \begin{eqnarray}
 \tilde{\Pi}_1 ( iy , p ) & = & -   \int_{- \infty}^{\infty} dp^{\prime} | p^{\prime} |
 \int_{- \infty}^{\infty}  \frac{d y^{\prime}}{2 \pi}
 \tilde{f}_{g} ( iy^{\prime} , p^{\prime} )
 \tilde{L}^{(4)}_S ( iy, p, i y^{\prime} , p^{\prime}  )  ,
 \label{eq:Pi1} 
 \end{eqnarray} 
where the dimensionless symmetrized four-loop
$\tilde{L}^{(4)}_S ( iy, p, i y^{\prime} , p^{\prime}  )$
is defined in Eq.~(\ref{eq:fourloop2}).
The term $\tilde{\Pi}_2 ( iy , p )$ involving two powers
of the effective interaction is of the form
 \begin{equation}
 \tilde{\Pi}_2 ( iy , p ) = \tilde{\Pi}_{2}^{\rm AL} ( iy , p ) + \tilde{\Pi}_{2}^{\rm H} ( iy , p ),
 \end{equation}
where the contribution from the Aslamasov-Larkin (AL) diagram 
in Fig.~\ref{fig:pertdiagrams} (b) is
\begin{eqnarray}
 \tilde{\Pi}_{2}^{\rm AL} ( iy , p ) & = &  -
 \int_{- \infty}^{\infty} dp^{\prime} | p^{\prime} |
 \int_{- \infty}^{\infty}  \frac{d y^{\prime}}{2 \pi}
 \tilde{f}_{g}  ( i y^{\prime} , p^{\prime} )
\tilde{f}_{g} \left(  \frac{  i yp + i y^{\prime} p^{\prime} }{p+p^{\prime} } , 
 p + p^{\prime} \right) 
[ \tilde{L}^{(3)}_S ( iy, p, iy^{\prime}, 
  p^{\prime} ) ]^2 ,
 \label{eq:Pi2}
  \end{eqnarray}
and the contribution from the Hartree  diagram 
in Fig.~\ref{fig:pertdiagrams} (c)  can be written as
\begin{eqnarray}
 \tilde{\Pi}_{2}^{\rm H} ( iy , p ) & = &  - \frac{ g}{1 + g}
  \tilde{L}^{(3)}_S ( iy, p, iy ,-p) 
 \int_{- \infty}^{\infty} dp^{\prime} | p^{\prime} |
 \int_{- \infty}^{\infty}  \frac{d y^{\prime}}{2 \pi}
 \tilde{f}_{g}  ( i y^{\prime} , p^{\prime} )
  \tilde{L}^{(3)}_S ( iy^{\prime}, 
  p^{\prime}, i y^{\prime} , -p^{\prime}).
 \label{eq:Pi2tadpole}
  \end{eqnarray}
 \end{widetext}
Here, the dimensionless symmetrized three-loop $\tilde{L}^{(3)}_S ( iy, p, iy^{\prime}, 
  p^{\prime} )$ is defined in Eq.~(\ref{eq:threeloop2}).
The parameters $Z_1$ and $Z_2$ hidden in the effective
interaction $\tilde{f}_g ( iy, p  )$ should  be determined self-consistently
by evaluating Eqs.(\ref{eq:pol12}--\ref{eq:Pi2tadpole}) and
demanding that the resulting renormalized ZS velocity
is consistent with the result obtained from Eq.~(\ref{eq:Piinvren}). 
We emphasize again that
the above expression for $\Pi_{\ast} ( Q )$  is not based on an expansion in powers 
of $1/m$: all functions appearing in Eqs.~(\ref{eq:Pi1}--\ref{eq:Pi2tadpole}) 
depend on $1/m$ in a rather complicated non-linear manner.

\subsection{Approximation A: neglecting $1/m$-corrections to $\Pi_0 ( Q )$
in  loop integrations} 
\label{subsec:simplification}

Eqs.~(\ref{eq:Pi1}--\ref{eq:Pi2tadpole}) are still too complicated 
to admit an analytic evaluation. In order to
explicitly calculate the dynamic structure factor
without resorting to elaborate numerics, we shall further
simplify the above expressions
by making the following {\it{approximation A}}:
 {\it{We
replace the non-interacting polarization $\Pi_0 ( Q )$
appearing in the effective interaction and
the symmetrized closed fermion loops on the right-hand sides
of Eqs.~(\ref{eq:Pi1}--\ref{eq:Pi2tadpole})
by its asymptotic limit  for small momenta given in
Eq.~(\ref{eq:Pilambda0})}}. 
Keeping in mind that in one dimension the closed fermion loops with $n > 2$ external 
legs can all be expressed in terms of   $\Pi_0 ( Q )$,
the  symmetrized three- and four-loops
 are then approximated by Eqs.~(\ref{eq:Pi3zerofinal}) and (\ref{eq:loop4Peyman2}).
For  consistency, we should also expand the dimensionless
free polarization $\tilde{\Pi}_0 ( i y , p)$  
on the right-hand side of Eq.~(\ref{eq:pol12})
to second order in $p$, see Eq.~(\ref{eq:tildePi0expansion}).
We shall argue below that the above approximation A is {\it{not}} sufficient to calculate
the line-shape of the dynamic structure factor for momenta  $q \lesssim
q_{c} = 1 / ( m | f_0^{\prime \prime} | )$ 
[see Eq.~(\ref{eq:qcdef})], because in this regime the spectral line-shape is
dominated by
the terms neglected in approximation A. On the other hand, for
$q \gtrsim q_{c}$ the line-shape of $S ( \omega , q )$ is essentially determined 
by the quadratic term in the expansion of $f_q$ for small $q$, so that
in this regime A is justified.

It turns out that with this simplification the $y^{\prime}$-integrations
in Eqs.~(\ref{eq:Pi1}), (\ref{eq:Pi2}) and (\ref{eq:Pi2tadpole}) can be done 
analytically for general $\tilde{g}_p$ using  the method of residues.
The form of Eq.~(\ref{eq:Sfull}) suggests that it is natural to expand
the {\it{inverse}} irreducible polarization in powers of $p$ and $p_0$.
This procedure can be formally justified within functional 
bosonization \cite{Kopietz97,Pirooznia07}, where 
the interaction corrections to the {\it{inverse}} irreducible polarization
play the role of the self-energy corrections in the effective bosonized theory.
But it is usually  better to
expand the self-energy 
rather than the Green function in powers of the relevant small parameter, because
the direct expansion of the Green function usually 
leads to unphysical singularities.
Using Eqs.~(\ref{eq:tildePi0expansion}) and (\ref{eq:Pi1}--\ref{eq:Pi2tadpole})
we obtain for the expansion of the inverse irreducible polarization to order
$p_0^2$,
\begin{eqnarray}
 \tilde{\Pi}_{\ast}^{-1}  (iy , p ) & = & 1 + y^2 - \frac{p^2}{3} \frac{1 - 3 y^2}{1+y^2}
 \nonumber
 \\ 
 & - &
  (1 + y^2)^2  \tilde{\Pi}_1 ( iy , p ) 
 \nonumber
 \\ 
 & - &
  (1 + y^2)^2 \tilde{\Pi}_2 ( iy , p )   +O ( p_0^3),
 \label{eq:Piinvexpansion} 
\end{eqnarray}  
where $p$ is assumed to be smaller than the dimensionless momentum-transfer
cutoff $p_0 =q_0 /(2 k_F)$.
It is convenient to introduce the notation
\begin{subequations}
 \begin{eqnarray}
 x_p & = & \sqrt{ 1 + \tilde{g}_p},
 \label{eq:xpdef}
 \\
 a_p & = & x_p +  1 = \sqrt{ 1 + \tilde{g}_p} + 1,
 \label{eq:apdef}
 \\
 b_p & = & x_p -  1 = \sqrt{ 1 + \tilde{g}_p} - 1,
 \label{eq:bpdef}
 \end{eqnarray}
\end{subequations}
so that $a_p b_p = \tilde{g}_p$.
The contribution involving the symmetrized four-loop can then be written as
 \begin{widetext} 
\begin{eqnarray}
 - ( 1 + y^2)^2 \tilde{\Pi}_1 ( iy , p ) & = & 
 {\rm Re} \int_0^{\infty} d p^{\prime} 
\Biggl\{
\frac{  | p^{\prime} | }{x_{p^{\prime}}}
 \frac{ p^{\prime 4} F_1 (  iy, p^{\prime} ) +  p^{\prime 2} p^2 F_2  (  iy, p^{\prime} ) 
 + p^4  F_3  (  iy, p^{\prime} ) }{ 
[  a_{p^{\prime}}^2 p^{\prime 2}  - ( 1 + i y )^2 p^2 ]
[   b_ {p^{\prime}}^2 p^{\prime 2}  - ( 1 - i y )^2 p^2 ] }
\nonumber
 \\
 &   &  \hspace{12mm} -   p^{\prime 2} \tilde{g}_{p^{\prime}}  (1+iy)^2  
 \Bigl[  \frac{ 2 p^{\prime}}{p ( 1-iy)} + 1 \Bigr]
 \frac{| p + p^{\prime} |    }{ x_{p^{\prime }}^2  p^{\prime 2} - 
 [(1-iy) p + p^{\prime} ]^2 }  + ( p^{\prime} \rightarrow - p^{\prime} ) \Biggr\},
 \label{eq:Pi1new}
\end{eqnarray}
with
 \begin{eqnarray}
 F_1 ( iy, p )  
 & = & 4 \tilde{g}_p ( x_p + iy )^2 + \tilde{g}_p^2 
 \left[ \frac{8x_p}{1 - iy} - 4 x_p - \tilde{g}_p - (2 + x_p - \frac{\tilde{g}_p}{2} ) ( 1 + y^2) \right],
 \label{eq:F1def} 
\\
 F_2 ( iy, p ) 
& = & \tilde{g}_p \left[ - ( 1 + iy)^4   + \tilde{g}_p ( 2 - y^2 + y^4 ) - 4 b_p  iy( 1 - y^2)  \right]
 - 2 b_p^2 x_p \frac{1+iy}{1-iy} ( 3 - 6 y^2 - y^4),
 \label{eq:F2def}
 \\
 F_3 ( iy, p ) 
 & = &  - 4  b_p^2 (1+y^2) \left[ 
 1  - \frac{ 1+ y^2}{2} -  \frac{ ( 1 + y^2)^2}{8} \right].
 \end{eqnarray}
\end{widetext} 
Both functions $F_1 ( iy , p )$ and $F_2 ( iy , p )$
contain a singular term proportional to $( 1 - i y )^{-1}$, which after analytic
continuation give rise to a  mass-shell singularity at the energies
$\pm v_F q $ associated with the bare Fermi velocity.
Fortunately, these singularities  cancel  when Eq.~(\ref{eq:Pi1new}) is combined with the
corresponding contributions from 
the expansion of $\tilde{\Pi}_0^{-1} ( iy, p )$
in Eq.~(\ref{eq:Piinvexpansion})
and from the AL diagram given in Eq.~(\ref{eq:Pi2new}) below. 
To show this explicitly, it is useful to 
isolate the singular term in  Eqs.~(\ref{eq:F1def}) and (\ref{eq:F2def}) by setting
\begin{eqnarray}
  F_1 ( iy, p ) & = & \frac{8  \tilde{g}_p^2 x_p}{1-iy} +  \tilde{F}_1 ( iy, p ),
 \\
  F_2 ( iy, p ) & = & - 8 b_p^2 x_p  \frac{(1+iy)^2}{1-iy} +  \tilde{F}_2 ( iy, p ).
\end{eqnarray}
$\tilde{F}_1 ( iy, p )$ and $ \tilde{F}_2 ( iy , p)$ are now analytic functions of $y$,
\begin{widetext}
\begin{eqnarray}
 \tilde{F}_1 ( iy, p ) 
 & = & 4 \tilde{g}_p ( x_p + iy )^2 - \tilde{g}_p^2 
 \left[   4 x_p + \tilde{g}_p + (2 + x_p - \frac{\tilde{g}_p}{2} ) ( 1 + y^2) \right],
 \label{eq:tildeF1def} 
 \\
 \tilde{F}_2 ( iy, p ) 
 & = &
 \tilde{g}_p \Bigl[ - ( 1 + iy)^4   + \tilde{g}_p ( 2 - y^2 + y^4 ) + 4   b_p  iy( 1 +2 iy + y^2)  \Bigr]
 \nonumber
 \\
 & + &    2 b_p^2   (1 + iy) \left[ x_p (1+iy ) (1 + y^2)  - 4 iy \right].
\end{eqnarray}
Eq.~(\ref{eq:Pi1new}) can then be written as
\begin{eqnarray}
 - ( 1 + y^2)^2 \tilde{\Pi}_1 ( iy , p ) & = & 
 {\rm Re} \int_0^{\infty} d p^{\prime} 
 \Biggl\{
 \frac{  | p^{\prime} | }{x_{p^{\prime}}}
 \frac{ p^{\prime 4} \tilde{F}_1 (  iy, p^{\prime} ) +  p^{\prime 2} p^2 \tilde{F}_2  (  iy, p^{\prime} ) 
 + p^4  F_3  (  iy, p^{\prime} ) }{ 
 [  a_{p^{\prime}}^2 p^{\prime 2}  - ( 1 + i y )^2 p^2 ]
 [  b_ {p^{\prime}}^2 p^{\prime 2}  - ( 1 - i y )^2 p^2 ] }
 +   \frac{  8 | p^{\prime} |    }{1-iy} 
 + \frac{ 8  | p^{\prime} |   p^2  (1-iy) }{
 b_ {p^{\prime}}^2 p^{\prime 2}  - ( 1 - i y )^2 p^2 }
 \nonumber
 \\
 &   &  \hspace{16mm} -   p^{\prime 2} \tilde{g}_{p^{\prime}}  (1+iy)^2  
 \Bigl[  \frac{ 2 p^{\prime}}{p ( 1-iy)} + 1 \Bigr]
 \frac{|  p^{\prime} + p |    }{ x_{p^{\prime }}^2  p^{\prime 2} - 
 [ p^{\prime}  +  (1-iy) p    ]^2 }  + ( p^{\prime} \rightarrow - p^{\prime} ) \Biggr\}.
 \label{eq:Pi1new2}
\end{eqnarray}

Next, consider the contribution $\tilde{\Pi}_{2}^{\rm AL} ( iy, p )$
from the Aslamasov-Larkin  diagram in Eq.~(\ref{eq:Pi2}). 
Adopting again approximation A, the symmetrized 
three-loop $\tilde{L}^{(3)}_S (iy, p , iy^{\prime}, p^{\prime} )$
is replaced
by its limit $\tilde{L}^{(3)}_{S,0} (iy, iy^{\prime}, p/ p^{\prime} )$
for $1/m \rightarrow 0$ given in Eq.~(\ref{eq:Pi3zerofinal}). Then we obtain
\begin{eqnarray}
 - ( 1 + y^2)^2 \tilde{\Pi}_{2}^{\rm AL} ( iy , p ) & = & 
 \int_{-\infty }^{\infty} d p^{\prime}  
 | p^{\prime  } |   \tilde{g}_{p^{\prime}}     \tilde{g}_{p^{\prime} + p}
 \int_{ - \infty}^{\infty} 
 \frac{ d y^{\prime}}{2 \pi }
 \frac{ \left[ 1 - y y^{\prime} - ( y + y^{\prime} ) \frac{ p y + p^{\prime} y^{\prime}}{ 
 p + p^{\prime} } \right]^2}{
 \Bigl[ 1 + y^{\prime 2} \Bigr] \Bigl[ x_{p^{\prime}}^2 + y^{\prime 2} \Bigr]
 \Bigl[ 1 +  \left( \frac{ p y + p^{\prime} y^{\prime}}{  p + p^{\prime} } \right)^2 \Bigr]
 \Bigl[ x_{p^{\prime} + p}^2   
 +  \left( \frac{ p y + p^{\prime} y^{\prime}}{  p + p^{\prime} } \right)^2 \Bigr] }.
 \nonumber
 \\
 & & 
\end{eqnarray}
The $y^{\prime}$-integration can now be carried out using the
theorem of residues.  The result can be cast into the following form,
\begin{eqnarray}
 - ( 1 + y^2)^2 \tilde{\Pi}_{2}^{\rm AL} ( iy , p ) & = & 
 {\rm Re} \int_{0}^{\infty} d p^{\prime}  
 p^{\prime  } \frac{| p^{\prime} + p |}{2}  
 \Biggl\{
 \frac{  \tilde{g}_{p^{\prime} + p }    | p^{\prime} + p |   \Bigl[  (p^{\prime} + p ) 
 ( 1 + 2iy x_{ p^{\prime} } + x^2_{ p^{\prime} } ) - p 
 ( y^2 + x^2_{  p^{\prime} })  \Bigr]^2    }{  x_{ p^{\prime}}
 \bigr[   (p^{\prime } + p )^2   - \bigr(  x_{ p^{\prime}}  p^{\prime}   + i y p  \bigr)^2  \bigl]
 \bigr[   x_{p^{\prime} + p }^2 ( p^{\prime } + p )^2  - \bigl(  
 x_{ p^{\prime}}   p^{\prime}   + i y p  \bigr)^2  
 \bigl] } 
 \nonumber
 \\
 &  & \hspace{32mm} 
 +  \frac{  \tilde{g}_{p^{\prime}}    p^{\prime}    
 \Bigl[  p^{\prime}  ( 1 + 2iy x_{ p^{\prime} + p } 
 + x^2_{ p^{\prime} + p } ) + p 
 ( y^2 + x^2_{  p^{\prime} + p })  \Bigr]^2    }{  x_{ p^{\prime} + p }
 \bigr[   p^{\prime 2}  - \bigr(  x_{ p^{\prime} + p }   (  p^{\prime} + p )  - i y p  \bigr)^2  \bigl]
 \bigr[   x_{p^{\prime}}^2 p^{\prime 2}  - \bigl(  
 x_{p + p^{\prime}}   ( p + p^{\prime} )  - i y p  \bigr)^2  
 \bigl] }
 \nonumber
 \\
 &  &  \hspace{32mm} 
 - \Bigl[   \frac{ 2 ( p^{\prime} + p )}{p ( 1-iy)} - 1\Bigr]
 \frac{ \tilde{g}_{p^{\prime} +  p } | p^{\prime} + p |  (1+iy)^2     }{ 
 x_{p^{\prime } + p }^2  ( p^{\prime } + p )^2  - 
 [     p^{\prime} + p -  (1-iy) p  ]^2 } 
 \nonumber
 \\
 &  &  \hspace{32mm} 
 +   \Bigl[  \frac{ 2 p^{\prime}}{p ( 1-iy)} + 1 \Bigr]
 \frac{ \tilde{g}_{p^{\prime}} p^{\prime}  (1+iy)^2     }{ x_{p^{\prime }}^2  p^{\prime 2} - 
 [     p^{\prime} + (1-iy) p  ]^2 }
 \Biggr\} 
 + ( p \rightarrow - p ).
 \label{eq:Pi2new}
\end{eqnarray}
\end{widetext} 

Finally, the contribution (\ref{eq:Pi2tadpole}) of the Hartree-type-of
diagram (c) in  Fig.~\ref{fig:pertdiagrams} is~\cite{footnotehartree}
\begin{equation}
 - ( 1 + y^2)^2 \tilde{\Pi}_{2}^{\rm H} ( iy , p ) =  
  I_{H} (1 - y^2),
 \label{eq:Pi2c}
\end{equation}
with
 \begin{eqnarray}
 I_{H} & = & - \frac{2 g}{1+g}  \int_0^{\infty} dp p 
 \left[ \frac{ 1 + \frac{\tilde{g}_p}{2}}{  \sqrt{ 1 + \tilde{g}_p} } -1 \right]
 \nonumber
 \\
 & = &  - \frac{ g}{1+g}  \int_0^{\infty} dp p 
\frac{ (x_p -1 )^2}{x_p}  .
 \label{eq:gHgeneral}
 \end{eqnarray}
%
%
For $\Theta$-function cutoff this reduces to
\begin{equation}
 I_{H} = - \frac{p_0^2 g}{1+g}
 \left[ \frac{ 1 + \frac{g}{2} }{  \sqrt{ 1 + g} } -1 \right],
 \end{equation}
while for Lorentzian cutoff,
\begin{equation}
 I_{H} = - \frac{p_0^2 g}{1+g}
  \left[ 1 + \frac{g}{2} -   \sqrt{ 1 +g} \right] .
 \end{equation}

Combining all terms we  obtain
the following expansion of the inverse irreducible polarization to second order 
in $p_0^2$,
\begin{eqnarray}
 \tilde{\Pi}_{\ast}^{-1}  (iy , p ) & = & 1 + y^2 
 +p^2  - \frac{2p^2}{3} \left[ \frac{1}{1-iy} + \frac{1}{1+iy} \right]
 \nonumber
 \\ 
 &   &  \hspace{-5mm} +   I_{H} (  1 -y^2  ) +  I ( iy, p )   + O ( p_0^3 ),
 \label{eq:Piinvexpansion2} 
\end{eqnarray}  
where we have used Eqs.~(\ref{eq:tildePi0expansion}) and (\ref{eq:laterref})
to clearly exhibit the mass-shell singularity
generated by the expansion of the inverse free polarization.
The dimensionless integral $I ( iy, p )$ can be written as
\begin{equation}
 I ( iy , p )  =  
 \frac{1}{2}  \int_0^{\infty} d p^{\prime} p^{\prime}
 \left[ J ( iy, p , p^{\prime})   +   J (- iy, p , p^{\prime}) \right]  ,
 \label{eq:Idef}
\end{equation}
where the complex function $J ( iy , p , p^{\prime} )$ is given by
\begin{widetext} 
\begin{eqnarray}
 J ( iy , p , p^{\prime} ) & = & 
 \frac{ p^{\prime 4} \tilde{F}_1 (  iy, p^{\prime} ) +  p^{\prime 2} p^2 
 \tilde{F}_2  (  iy, p^{\prime} ) 
  + p^4  F_3  (  iy, p^{\prime} ) }{  x_{p^{\prime}}
 [  a_{p^{\prime}}^2 p^{\prime 2}  - ( 1 + i y )^2 p^2 ]
 [  b_ {p^{\prime}}^2 p^{\prime 2}  - ( 1 - i y )^2 p^2 ] }
 +  \frac{8}{ 1-iy} + \frac{ 8 p^2 ( 1-iy)}{  
 b_ {p^{\prime}}^2 p^{\prime 2}  - ( 1 - i y )^2 p^2}
 \nonumber
 \\
 & +   &  
  \frac{ |  p^{\prime} + p | }{2 } \biggl\{
 \frac{ \tilde{g}_{ p^{\prime} + p  }  | p^{\prime} + p | 
 \Bigl[  (p^{\prime} + p ) 
 ( 1 + 2iy x_{ p^{\prime} } + x^2_{ p^{\prime} } ) - p 
 ( y^2 + x^2_{  p^{\prime} })  \Bigr]^2 
 }{   x_{ p^{\prime}}
 \bigl[ ( p^{\prime} + p )^2 - ( x_{p^{\prime}} p^{\prime} + iy p )^2 \bigr] 
 \bigl[ x^2_{ p^{\prime} + p} ( p^{\prime} + p )^2 - ( x_{p^{\prime}} p^{\prime} + iy p )^2 \bigr] } 
 \nonumber
 \\
 &     & \hspace{12mm} + 
 \frac{  \tilde{g}_{p^{\prime}}    p^{\prime}   \Bigl[  p^{\prime}  ( 1 + 2iy x_{ p^{\prime} + p } 
 + x^2_{ p^{\prime} + p } ) + p 
 ( y^2 + x^2_{  p^{\prime} + p })  \Bigr]^2    }{  x_{ p^{\prime} + p }
 \bigr[   p^{\prime 2}  - \bigr(  x_{ p^{\prime} + p }   (  p^{\prime} + p )  - i y p  \bigr)^2  \bigl]
 \bigr[   x_{p^{\prime}}^2 p^{\prime 2}  - \bigl(  
 x_{p + p^{\prime}}   ( p + p^{\prime} )  - i y p  \bigr)^2  
 \bigl] }
 \nonumber
 \\
 &     &   \hspace{12mm} -    
 \Bigl[   \frac{ 2 ( p^{\prime} + p )}{p ( 1-iy)} - 1\Bigr]
 \frac{ \tilde{g}_{p^{\prime} +  p } | p^{\prime} + p |   (1+iy)^2     }{ 
 x_{p^{\prime } + p }^2  ( p^{\prime } + p )^2  - 
 [     p^{\prime} + p -  (1-iy) p  ]^2 } 
 \nonumber
 \\
 &  & 
  \hspace{12mm} -    \Bigl[  \frac{ 2 p^{\prime}}{p ( 1-iy)} + 1 \Bigr]
 \frac{ \tilde{g}_{p^{\prime}} p^{\prime}   (1+iy)^2    }{ 
 x_{p^{\prime }}^2  p^{\prime 2} -  [     p^{\prime} + (1-iy) p  ]^2 }
 \biggr\} + ( p \rightarrow -p ).
 \label{eq:Jdef}
\end{eqnarray}
\end{widetext}
Although it is not obvious from Eq.~(\ref{eq:Jdef}),
 the function
 $J ( iy , p , p^{\prime} ) $ vanishes as $ \tilde{g}_{ p^{\prime}}^2$ for
$ p^{\prime} \gg p_0 $, so that
the integral (\ref{eq:Idef}) is ultraviolet convergent as long as
$\tilde{g}_p$ vanishes faster than $1/p$ for $p \rightarrow \infty$.

\subsection{Cancellation of  the mass-shell singularities at $\omega = \pm v_F q$}
\label{subsec:cancellation}

We now show that the mass-shell singularities
at $iy \rightarrow x = \pm 1$ (corresponding to
frequencies $\omega = \pm v_F q $)
arising from the expansion of the non-interacting polarization
in Eq.~(\ref{eq:Piinvexpansion2}) are {\it{exactly cancelled}} by 
corresponding singularities in $I ( x , p )$, because
for $x \rightarrow \pm 1$ the integral $I ( x , p )$ diverges as
\begin{equation}
  I ( x , p )  \sim 
 \frac{ 2 p^2}{3} \frac{1}{ 1  \mp  x }  \; , \; \mbox{ $ x \rightarrow \pm 1$.} 
 \label{eq:massshellcancel}
\end{equation}
To proof this, it is sufficient to calculate the residues
\begin{eqnarray}
 R_{\pm} (p) &   = &  \lim_{ x \rightarrow \pm 1} \left[ ( 1  \mp x ) I ( x , p ) \right]
 \hspace{30mm}
 \nonumber
 \\
  &  = & 
 \frac{1}{2} \int_0^{\infty} d p^{\prime} p^{\prime}  \lim_{ x \rightarrow \pm 1} \left[ 
 ( 1 \mp x ) J (  \pm x , p , p^{\prime} ) \right]. \hspace{10mm} 
\end{eqnarray}
Using $x_p^2 -1 = \tilde{g}_p$ we find from
Eq.~(\ref{eq:Jdef}),
\begin{eqnarray} 
 \lim_{ x \rightarrow \pm 1} \left[ 
 ( 1 \mp x ) J (  \pm x , p , p^{\prime} ) \right]
 & = & 8 - 4 \frac{ | p^{\prime} + p | - | p^{\prime} - p | }{p}
 \nonumber
 \\
 &   & \hspace{-20mm} =  8 \Theta ( | p | - p^{\prime} ) \left( 1 -  p^{\prime} /  | p |  \right).
\end{eqnarray}
Hence,
\begin{equation}
 R_{\pm} (p) = 4 \int_0^{ | p | } d p^{\prime} p^{\prime} \left( 1 -  p^{\prime} / | p |  \right)
 = \frac{ 2 p^2}{3},
\end{equation}
which proofs Eq.~(\ref{eq:massshellcancel}). We conclude 
that the expansion  (\ref{eq:Piinvexpansion2})
of  the inverse irreducible polarization
to second order in $p_0^2$
does not exhibit any mass-shell singularities 
at frequencies $ \omega = \pm v_F  q  $
corresponding to the excitation energy of  non-interacting particle-hole pairs.
This cancellation also corrects the  unphysical feature of the RPA
that  the single particle-hole pair continuum 
is centered at the energy $ v_F | q |$ involving the bare
Fermi velocity $v_F$.  

It is convenient to explicitly cancel the mass-shell 
singularities arising from the expansion of the free polarization in 
Eq.~(\ref{eq:Piinvexpansion2}) against the corresponding singularities in 
$I ( iy, p )$. Therefore we use the identity
\begin{eqnarray}
    \frac{2p^2}{3} \left[ \frac{1}{1-iy} + \frac{1}{1+iy} \right] &  &
 \nonumber 
\\
& & \hspace{-40mm} =
 \frac{1}{2} \int_0^{\infty} d p^{\prime} p^{\prime}  [ J_0 ( iy, p , p^{\prime} ) +
J_0 ( - iy, p , p^{\prime} ) ],
 \end{eqnarray} 
where
 \begin{eqnarray}
 J_0 ( iy, p , p^{\prime} ) & =  & \frac{8}{1-iy}
 \Bigl[ 1 -  \frac{ p^{\prime} + p  +    | p^{\prime} + p  | }{2 p} 
 \nonumber
 \\
 & & \hspace{10mm}  
 + ( p \rightarrow - p ) \Bigr],
 \end{eqnarray}
to write Eq.~(\ref{eq:Piinvexpansion2}) as follows,
\begin{eqnarray}
 \tilde{\Pi}_{\ast}^{-1}  (iy , p ) & = & 1 + y^2 
+p^2  
  +  I_{H} (  1 -y^2  ) +  \tilde{I} ( iy, p ) 
 \nonumber
 \\
 & &
  + O ( p_0^3 ).
 \label{eq:Piinvbetter} 
\end{eqnarray}  
The integral $\tilde{I} ( iy, p ) $ can again be written as
 \begin{equation}
 \tilde{I} ( iy , p )  =  
 \frac{1}{2}  \int_0^{\infty} d p^{\prime} p^{\prime}
\left[ \tilde{J} ( iy, p , p^{\prime})   +   \tilde{J} (- iy, p , p^{\prime}) \right]  ,
 \label{eq:tildeIdef}
 \end{equation}
with
 \begin{equation}
  \tilde{J} ( iy, p , p^{\prime}) = {J} ( iy, p , p^{\prime})
-  {J}_0 ( iy, p , p^{\prime}).
 \end{equation}
We may now explicitly cancel the mass-shell singularities 
in the regularized integrand $\tilde{J} ( iy, p , p^{\prime})$
and obtain after some algebra,
 \begin{widetext} 
\begin{eqnarray}
 \tilde{J} ( iy , p , p^{\prime} ) & = & 
 \frac{ p^{\prime 4} \tilde{F}_1 (  iy, p^{\prime} ) +  p^{\prime 2} p^2 \tilde{F}_2  (  iy, p^{\prime} ) 
 + p^4  F_3  (  iy, p^{\prime} ) }{  x_{p^{\prime}}
[  a_{p^{\prime}}^2 p^{\prime 2}  - ( 1 + i y )^2 p^2 ]
[  b_ {p^{\prime}}^2 p^{\prime 2}  - ( 1 - i y )^2 p^2 ] }
  + \frac{ 8 p^2 ( 1-iy)}{  
 b_ {p^{\prime}}^2 p^{\prime 2}  - ( 1 - i y )^2 p^2}
\nonumber
 \\
 & +   &   
 \frac{ \tilde{g}_{ p^{\prime} + p  }  ( p^{\prime} + p )^2 
 \Bigl[  (p^{\prime} + p ) 
( 1 + 2iy x_{ p^{\prime} } + x^2_{ p^{\prime} } ) - p 
 ( y^2 + x^2_{  p^{\prime} })  \Bigr]^2 
 }{   2 x_{ p^{\prime}}
\bigl[ ( p^{\prime} + p )^2 - ( x_{p^{\prime}} p^{\prime} + iy p )^2 \bigr] 
\bigl[ x^2_{ p^{\prime} + p} ( p^{\prime} + p )^2 - ( x_{p^{\prime}} p^{\prime} + iy p )^2 \bigr] } 
\nonumber
 \\
 &   +   & 
 \frac{  \tilde{g}_{p^{\prime}}     | p^{\prime} + p | p^{\prime}   \Bigl[  p^{\prime}  ( 1 + 2iy x_{ p^{\prime} + p } 
 + x^2_{ p^{\prime} + p } ) + p 
 ( y^2 + x^2_{  p^{\prime} + p })  \Bigr]^2    }{  2 x_{ p^{\prime} + p }
\bigr[   p^{\prime 2}  - \bigr(  x_{ p^{\prime} + p }   (  p^{\prime} + p )  - i y p  \bigr)^2  \bigl]
\bigr[   x_{p^{\prime}}^2 p^{\prime 2}  - \bigl(  
 x_{p + p^{\prime}}   ( p + p^{\prime} )  - i y p  \bigr)^2  
 \bigl] }
\nonumber
 \\
 &   +  &     
 \frac{    ( p^{\prime} + p ) \Bigl[ 8 (p^{\prime} + p ) - 4 (1-iy)p
 + \tilde{g}_{ p^{\prime} + p } ( p^{\prime} + p ) 
\bigl[ \frac{ p^{\prime} + p}{p} ( 3 +iy) + \frac{1}{2} ( 1 + iy)^2 \bigr] \Bigr]
   }{ 
 x_{p^{\prime } + p }^2  ( p^{\prime } + p )^2  - 
 [     p^{\prime} + p -  (1-iy) p  ]^2 } 
 \nonumber
 \\
 & + & 
 \frac{ 
  | p^{\prime} + p |  \Bigl[ - 8 p^{\prime} - 4 (1-iy)p
 + \tilde{g}_{ p^{\prime}  }  p^{\prime} 
\bigl[ \frac{ p^{\prime} }{p} ( 3 +iy) - \frac{1}{2} ( 1 + iy)^2 \bigr] \Bigr]
    }{ 
x_{p^{\prime }}^2  p^{\prime 2} -  [     p^{\prime} + (1-iy) p  ]^2 } + ( p \rightarrow -p ).
 \label{eq:tildeJdef}
\end{eqnarray}
 \end{widetext}

\section{Interaction with sharp momentum-transfer cutoff}
\label{sec:sharp}

\subsection{Explicit evaluation of the irreducible polarization}

In this section we assume that the dimensionless interaction $g_p$ is of the form
 \begin{equation}
 {g}_p = {g}_0 \Theta ( p_0 -  | p| ).
 \label{eq:gtheta}
\end{equation}
In this case the $p^{\prime}$-integration
in  Eq.~(\ref{eq:Idef}) is elementary and can be carried out exactly.
Note that all derivatives of the interaction (\ref{eq:gtheta})
vanish at $p=0$ so that $f_0^{\prime \prime} =0$,  which is certainly
an unphysical feature of the $\Theta$-function cutoff.
The length $q_c$ defined in Eq.~(\ref{eq:qcdef}) is then formally infinite, so that
the regime (\ref{eq:qckF}) does not exist. 
Although for such an interaction
approximation A discussed in Sec.~\ref{subsec:simplification}
(i.e., replacing $\tilde{\Pi}_0 ( iy , p ) \approx \tilde{\Pi}_0 ( iy, 0 ) = [ 1 + y^2]^{-1}$
in  loop integrations) is never justified,
it is still instructive to evaluate Eq.~(\ref{eq:Piinvbetter}), because 
it allows us to explicitly see the partial cancellation between
contributions arising from the first-order diagram 
in Fig.~\ref{fig:pertdiagrams} (a) and the
AL diagram in Fig.~\ref{fig:pertdiagrams} (b).
To clearly exhibit this cancellation,
it is instructive to evaluate the contributions $\tilde{\Pi}_1 ( iy, p )$ 
(first-order in the effective interaction)
and $\tilde{\Pi}_2 ( iy , p )$ (second order in the effective interaction) separately.
Therefore, we specify $\tilde{g}_p = g  \Theta ( p_0 -    | p | )$
in Eqs.~(\ref{eq:Idef},\ref{eq:Jdef}) and perform  
the $p^{\prime}$-integration exactly.
Recall that the effective coupling constant $g$
is defined as a function of the bare coupling $g_0$
via Eq.~(\ref{eq:tildeg0def}).
The $p \rightarrow 0$ limits of the
coefficients $x_p$, $a_p$ and $b_p$ 
given in Eqs.~(\ref{eq:xpdef}--\ref{eq:bpdef}) 
are now denoted by
 \begin{subequations} 
\begin{eqnarray}
x_0 & = & \sqrt{1+g},
 \\
a & = & x_0 +  1 ,
 \\
 b & = & x_0-1.
 \end{eqnarray}
 \end{subequations}
Note that for small $g$,
 \begin{eqnarray}
a & = & 2 + \frac{g}{2} - \frac{g^2}{8} + \frac{g^3}{16} + O ( g^4 ),
 \\
 b & = & a -2 = \frac{g}{2} - \frac{g^2}{8} + \frac{g^3}{16} + O ( g^4 ).
 \end{eqnarray}
After some tedious algebra  
we find that the contribution  
from   the diagram (a) in Fig.~\ref{fig:pertdiagrams}   
to the expansion (\ref{eq:Piinvexpansion})   can be written as
\begin{widetext}
\begin{eqnarray}
 - ( 1 + y^2)^2 \tilde{\Pi}_1 ( iy , p ) & = & - p_0^2 \frac{ b^2 ( 3 + x_0)}{2 a x_0}
  (  2 + g - \Delta  ) 
 \nonumber
 \\
 &  +  &  p^2 \Biggl\{    \frac{2}{3} \frac{1 - 3 y^2}{1+y^2} +     
   \frac{(2 + g)}{g}  ( 4 - g)   - \frac{4 \Delta}{g^2}  \left[  4 + g - \frac{g^2}{4} \right]
 \nonumber
 \\
 &  & \hspace{5mm} +  \frac{    g - \Delta }{   x_0} 
 {\rm Re} \Biggl[ - \frac{ b^2}{a^3}
 (1 -iy)  (x_0 - iy)
 \ln \left( \frac{  p_0^2 a^2  - p^2 ( 1 + iy )^2}{ p^2 (1 + iy )(x_0 - iy ) } \right)
 \nonumber
 \\
 & &        
  \hspace{23mm} +  \frac{ a^2}{ b^3}
 (1-iy)  (x_0 + iy)
 \ln \left( \frac{  1 + iy }{ x_0 + iy  } \right)  
 \Biggl]  \Biggl\},
 \label{eq:Pi1result}
\end{eqnarray}
\end{widetext}
where we have defined
\begin{equation}
 \Delta = 1 + g + y^2 = x_0^2  + y^2 .
 \label{eq:Deltadef}
\end{equation}
If we neglect at this point the contribution $ \tilde{\Pi}_2 ( iy , p )$ involving
two powers of the effective interaction, we recover from the imaginary part 
of Eq.~(\ref{eq:Pi1result}) our previous estimate \cite{Pirooznia07}
for the damping of the ZS mode for $q \rightarrow 0$
\begin{eqnarray}
 \gamma_q \approx  
 \frac{\pi}{8} \frac{g^3}{ x_0 a^4} 
 \frac{ | q |^3}{ v_F m^2}. 
 \label{eq:dampres}
\end{eqnarray}
In view of the discussion at the end of Sec.~\ref{sec:RPA}  
this result should not be surprising: within our approximation the ZS mode
is located at higher energy than the single-pair continuum
and is immersed in the  multi-pair continuum, whose spectral weight
is generated by the  logarithmic terms in Eq.~(\ref{eq:Pi1result}).
The overlap of the multi-pair 
continuum with the ZS mode leads to the $q^3$-damping,
in agreement with the arguments by Teber \cite{Teber06}.

Unfortunately, the term in Eq.~(\ref{eq:Pi1result}) which is responsible
for the result (\ref{eq:dampres}) is exactly cancelled by a similar
term in  $- ( 1 + y^2) \tilde{\Pi}_2^{\rm AL} ( iy , p )$.  
Explicitly carrying out the $p^{\prime}$-integration in Eq.~(\ref{eq:Pi2new}) 
and adding the contribution  (\ref{eq:Pi2c})   from the  Hartree-type-of term,
we obtain for $ | p | < p_0$,
\begin{widetext}
\begin{eqnarray}
 - ( 1 + y^2)^2 \tilde{\Pi}_2 ( iy , p ) & = &  p_0^2 \frac{  b^2}{  2 x_0^3}  g (2 + g - \Delta ) 
 \nonumber
 \\
 &  & \hspace{-20mm} + p_0 ( p_0 -  | p | )  \frac{ b^2}{  a x_0^3} 
 \Bigl[  g ( 2 + g)  - b \bigl( 1 + \frac{g}{4} \bigr)  -    \Delta  x_0^2  \Bigr]
 \nonumber
 \\
 & & \hspace{-20mm} + p^2 \Biggl\{  -  \frac{ 1 - 3 y^2}{3(1 + y^2)}  + 
 \frac{g}{2 x_0} 
 - \frac{(2+g)}{2 g} \Bigl[ 4 - g + \frac{4 }{ x_0 } \Bigr]
 + \frac{2 \Delta}{g^2} \Bigl[ 4 + g - \frac{g^2}{ 4} + \frac{3 g^2}{4 x_0} + x_0 ( 4-g) \Bigr]
 \nonumber
 \\
 & & \hspace{-12mm}
 - \frac{  (4 + g)^2 + 8 g ( 2 + g - \Delta )}{12 x_0 \Delta}  + 
 \frac{ g^2 \Delta }{ 16  x_0^5}
 \ln \left( \frac{ 4  p_0 (  p_0 - | p | ) x_0^2  +  p^2 \Delta}{ p^2 \Delta} \right)
 \nonumber
 \\
 & & \hspace{-12mm} +  \frac{  g - \Delta  }{  x_0  }
 {\rm Re} \Biggl[
 \frac{ b^2}{a^3}
 (1-iy )  (x_0 - iy)
 \ln \left( \frac{  p_0 (p_0 - | p|) a^2  + p^2  (1+iy) (x_0 -iy)  }{ 
 p^2 (1 + iy )(x_0 - iy ) } \right)
 \nonumber
 \\
 & &        
    \hspace{7mm} - \frac{ a^2  }{ b^3   }
 (1-iy)  (x_0 + iy)
 \ln \left( \frac{  1 + iy }{ x_0 + iy  } \right)   \Biggr]                             
 \Biggr\}.
 \label{eq:Pi2result}
\end{eqnarray}
Adding Eqs.~(\ref{eq:Pi1result}) and (\ref{eq:Pi2result}) and rearranging  terms, 
we obtain for the expansion (\ref{eq:Piinvexpansion})
of the inverse irreducible polarization for sharp momentum-transfer cutoff
\begin{eqnarray}
 \tilde{\Pi}_{\ast}^{-1} ( iy , p ) & = & 1 + p_0^2 g_1  + (1+ p_0^2 g_2 ) y^2  +  
 p_0  | p | [ g_3  + g_4   y^2 ]    
 \nonumber
 \\
 & + & \frac{p^2}{2}  \Biggl\{   
 \frac{ 4 g}{ 3 x_0} -2   + \frac{b}{g x_0} [ 8 + 4 g - g^2] +
 \frac{\Delta}{g^2}   \bigl[   16b -   4 g a  + g^2 ( 1 + 3/x_0 ) \bigr]
 \nonumber
 \\
 &  & \hspace{5mm} -
  \frac{ (4 + 3 g)^2  }{6 x_0 \Delta}  + 
 \frac{ g^2 \Delta }{ 8 x_0^5}
 \ln \left( \frac{ 4  p_0 (  p_0 - | p | ) x_0^2  +  p^2 \Delta}{ p^2 \Delta} \right)
 \nonumber
 \\  
 & &  \hspace{5mm} -
 (  1+ y^2  )  \frac{  b^2}{a^3 x_0}
 2 {\rm Re} \Biggl[
 (1-iy )  (x_0 - iy)
 \ln \left( 
 \frac{ p_0 a  - | p| (x_0 - iy )  }{ 
   p_0 a  +  | p | ( 1 + iy )  } \right)
 \Biggr]
 \Biggr\},
 \label{eq:Piastfinal}
\end{eqnarray}
\end{widetext}
where
\begin{subequations}
 \begin{eqnarray}
 g_1 & = &  - \frac{ b^2  }{ 2 x_0^3 } \left[ 3 
 + \frac{g}{2} \frac{  x_0 + 3 }{x_0 +1} \right] 
 \nonumber
 \\
 & = & - \frac{3 }{8} g^2 + \frac{5 }{8} g^3
 + O ( g^4 ) ,
 \\
 g_2 & = &   \frac{ b^2  }{2 x_0^3 } = \frac{1}{8} g^2 - \frac{1}{4} g^3 + O ( g^4 ),
 \label{eq:g2def}
 \\
 g_3 & = &    \frac{ b^2  }{a x_0^3 } \left[ x_0 + \frac{g}{4}  b \right]
 = \frac{1}{8} g^2 - \frac{7}{32} g^3 + O ( g^4), \hspace{10mm}
 \\
  g_4 & = &   \frac{ b^2  }{a x_0 } = \frac{1}{8} g^2 - \frac{5}{32} g^3 + O ( g^4).
 \label{eq:g4def}
\end{eqnarray}
\end{subequations}
Eq.~(\ref{eq:Piastfinal}) has three important properties:
\begin{itemize}
\item 
The logarithmic term 
in Eq.~(\ref{eq:Pi1result}) which is responsible
for the $q^3$-dependence of $\gamma_q$ in
Eq.~(\ref{eq:dampres}), is exactly cancelled
by a similar term with opposite sign arising from the AL diagram.

\item The mass-shell singularity at $\omega = \pm v_F q $
associated with the expansion of the free polarization $\Pi_0 ( \omega , q )$
in Eq.~(\ref{eq:Piinvexpansion2})
has disappeared in Eq.~(\ref{eq:Pi1result}), in agreement with our general considerations
in Sec.~\ref{subsec:cancellation}.

\item 
Eq.~(\ref{eq:Piastfinal}) contains a term proportional to
$1/ \Delta$, which after analytic continuation gives rise to a mass-shell
singularity at the physical energy $\omega = \pm v q $ of the ZS mode.
\end{itemize}
The mass-shell singularity
at $\omega = \pm v q$ is an artefact 
of the sharp momentum-transfer cutoff used in this section
in combination with approximation A discussed in Sec.~\ref{subsec:simplification}.
In fact, we shall show in Sec.~\ref{sec:smooth} that a more realistic
interaction $f_q$ with finite  $f_0^{\prime \prime}$
does not lead to any mass-shell singularities, even if we still
use  approximation A to evaluate Eqs.~(\ref{eq:Pi1}--\ref{eq:Pi2tadpole}).

\subsection{Renormalized ZS velocity}

\begin{figure}[tb]    
\centering   
  \vspace{4mm}
\includegraphics[scale=0.5]{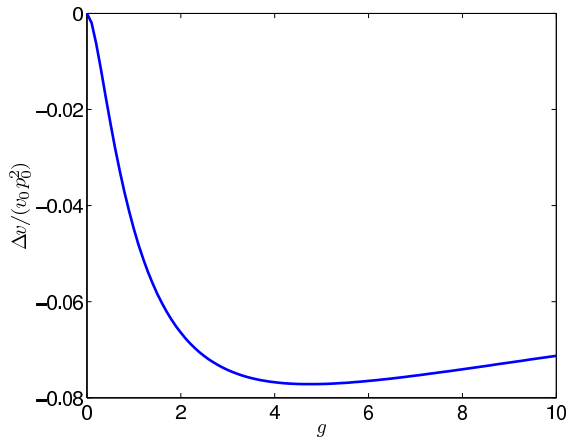} 
\caption{(Color online) Relative renormalization $ \Delta v / (v_0 p_0^2) \equiv
 ( v  -v_0)  / (v_0  p_0^2)  = - g_5/(2 x_0^2)$
of the ZS velocity in units of $p_0^2$ as a function of the interaction strength
$g$, see Eq.~(\ref{eq:AZSren}).
}
    \label{fig:Aplot}
\end{figure}

To calculate the renormalized ZS velocity it is sufficient to set $p=0$
in Eq.~(\ref{eq:Piastfinal}), so that the problems related
to the mass-shell singularity do not arise.
Comparing Eq.~(\ref{eq:Piastfinal}) at $p=0$ with
the defining equation (\ref{eq:Piinvren}) of the renormalization constants
$Z_1$ and $Z_2$, we find to order $p_0^2$,
 \begin{eqnarray}
 Z_i & = & 1 + p_0^2 g_i , \; \; i=1,2, 
 \label{eq:Z0res}
 \end{eqnarray}
which are non-linear self-consistency equations for $Z_1$ and $Z_2$, because
$g_1$ and $g_2$ are defined in terms of the renormalized coupling
$g =  (g_0 + Z_1 -  Z_2 )/Z_2$, see Eq.~(\ref{eq:tildeg0def}).
However, keeping in mind that the difference $g-g_0$ is proportional to $p_0^2$ and
that Eq.~(\ref{eq:Z0res})
is  only valid to order $p_0^2$, we may ignore the self-consistency condition and
set  $Z_1 = Z_2=1$ in the expressions for $g_1$ and $g_2$
on the right-hand side of Eq.~(\ref{eq:Z0res}).
From Eq.~(\ref{eq:velren}) we then obtain for the renormalized 
ZS velocity, 
\begin{eqnarray}
 \frac{v }{ v_F} &  = & \sqrt{ \frac{Z_1 + g_0}{Z_2} }  = \sqrt{ 1 + g },
 \label{eq:velren2} 
 \end{eqnarray}
where
 \begin{equation}
 g = g_0 - p_0^2 g_5,
 \end{equation}
with
 \begin{eqnarray}
 g_5 & = & x_0^2 g_2 - g_1  
 =    \frac{ b^2}{x_0^3} \left[
 2 + \frac{g}{4} \left( 3 + \frac{2}{a} \right) \right]
 \nonumber
 \\
 & = & \frac{1}{2} g^2  - \frac{3}{4} g^3 + O ( g^4 ).
 \label{eq:g5def}
 \end{eqnarray}
To order $p_0^2$ we thus obtain for the energy of the ZS mode
 \begin{equation}
{\omega}_q \approx v | q |,
 \label{eq:tildeomegaq} 
 \end{equation}
with renormalized ZS velocity,
\begin{eqnarray}
 v  & = &     v_F  \sqrt{ 1 + g_0 - p_0^2 g_5} 
 \nonumber
 \\
& = &  v_0
 \left[ 1 - p_0^2 \frac{g_5}{2 x_0^2} +  O ( p_0^4 ) \right],
 \label{eq:AZSren}
 \end{eqnarray}
where $v_0 =  = v_F \sqrt{ 1 + g_0}$ is the RPA result for the ZS velocity.
A graph of the relative change of the ZS velocity 
as a function of the interaction strength $g$
is shown in Fig.~\ref{fig:Aplot}.
Obviously, even for large $g$ and $p_0^2 = O (1)$
the correction to the RPA result $v_0$  never exceeds more than a few percent.

\subsection{Ad hoc regularization of the  mass-shell singularity and spectral line-shape}

Although for sharp momentum-transfer cutoff the dynamic structure factor
exhibits (within approximation A discussed in Sec.~\ref{subsec:simplification}) 
a  mass-shell singularity at 
the ZS energy $v | q |$, it is nevertheless instructive to follow Samokhin \cite{Samokhin98} 
and regularize the singularity by hand using the procedure outlined in Sec.~\ref{sec:RPA}.
Because the natural scale for the momentum dependence is not $2k_F$ but the
scale $q_0$ set by the momentum-transfer cutoff, it is convenient to express 
the momentum dependence via $\tilde{q} = q / q_0$. Setting $p = p_0 \tilde{q}$ 
and writing
\begin{equation}
 S ( \omega , q ) = \frac{\nu_0}{\pi} {\rm Im} 
 \left[ \frac{1}{g_0 + \tilde{\Pi}_{\ast}^{-1} ( x + i 0, \tilde{q} ) }
 \right],
 \label{eq:dynfix}
 \end{equation}
we obtain on the imaginary frequency axis
\begin{widetext}
\begin{eqnarray}
  g_0 + \tilde{\Pi}_{\ast}^{-1} ( iy , \tilde{q} ) & = & \Delta    \left[ 1 + p_0^2 ( g_2 +  
g_4 | \tilde{q} | ) \right] 
- p_0^2  g_6  | \tilde{q} | 
 \nonumber    
 \\
& + & p_0^2 \tilde{q}^2 \Biggl\{ h_0
  - \frac{ h_1  }{\Delta}    +     \Delta  \biggl[  g_7 + 
 g_8 
 \ln \Bigl(  1 +  \frac{4 x_0^2   (1 - | \tilde{q} | ) }{ \tilde{q}^2  \Delta} \Bigr)
 \biggr]
 \nonumber
 \\  
 & &  \hspace{10mm} +
(  g - \Delta )  \frac{  b^2}{  a^3 x_0}
  {\rm Re} \biggl[
 (1-iy )  (x_0 - iy)
 \ln \Bigl( \frac{ a  - | \tilde{q} |  (x_0 -iy )   }{ 
   a   +  | \tilde{q} |  ( 1 + iy)  } \Bigr)
 \biggr]
\Biggr\},
 \label{eq:polfinal}
 \end{eqnarray}
where
 \begin{subequations}
 \begin{eqnarray}
 g_6 & = & x_0^2 g_4 - g_3  
 =    \frac{ b^2}{a x_0^3} g \left[
 2 + g   -  \frac{b}{ g} \left( 1 + \frac{g}{4}  \right) \right]
 =  \frac{3}{16} g^3  + O ( g^4),
 \label{eq:g6def}
  \\
 g_7 & = & \frac{1}{2}  + \frac{3}{2 x_0} - 
 \frac{8}{g}  \Bigl( \frac{a}{4} -  \frac{ b}{ g} \Bigr)
   =   \frac{1}{8} g^2  -  \frac{11}{64}  g^3 + O ( g^4),
 \label{eq:g7def}
 \\
 g_8 & = & \frac{ g^2}{16 x_0^5}  = \frac{1}{16} g^2 - \frac{5}{32} g^3 + O ( g^4),
 \label{eq:g8def}
 \\
 h_0 & = & -1 + 
\frac{ 2 g}{ 3 x_0 }     + \frac{b}{2 g x_0} [ 8 + 4 g - g^2]
  =   1 + \frac{1}{6} g - \frac{1}{12} g^2  + O ( g^3 ),
 \label{eq:h0def}
 \\
 h_1 & = &  \frac{ ( 1 + 3 x_0^2)^2}{12 x_0} =
\frac{ (4 + 3 g)^2  }{12 x_0} = \frac{4}{3} + \frac{4}{3} g + \frac{1}{4}
 g^2 + O ( g^3).
  \label{eq:h1def}
 \end{eqnarray}
 \end{subequations}
\end{widetext}
From Eq.~(\ref{eq:polfinal}) it is obvious that
our functional bosonization approach yields
a systematic expansion of the inverse irreducible
polarization in powers of the small parameter
$p_0 = q_0 /(2 k_F)$.
Note that only  $h_0$ and $h_1$ have finite limits for $g \rightarrow 0$,
whereas  the other couplings $g_1, \ldots , g_8$  vanish at least as
 $g^2$ (the coupling $g_6$ vanishes even as $g^3$).

\begin{figure}[tb]    
\centering   
\includegraphics[scale=0.5]{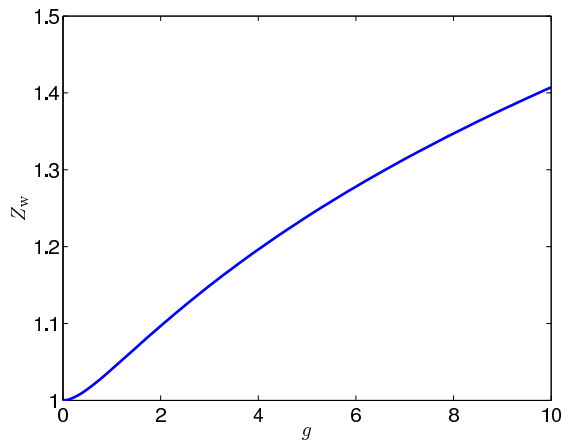} 
\caption{
(Color online) Graph of the factor $Z_{\rm w}$ defined in 
Eq.~(\ref{eq:Zwdef}), which estimates the interaction-induced
relative change of the width of ZS resonance for $q \ll q_c$,
see Eqs.~(\ref{eq:wqres},\ref{eq:Zwdef}).
}
    \label{fig:Zm}
\end{figure}

\begin{figure}[tb]    
\centering   
\includegraphics[scale=0.5]{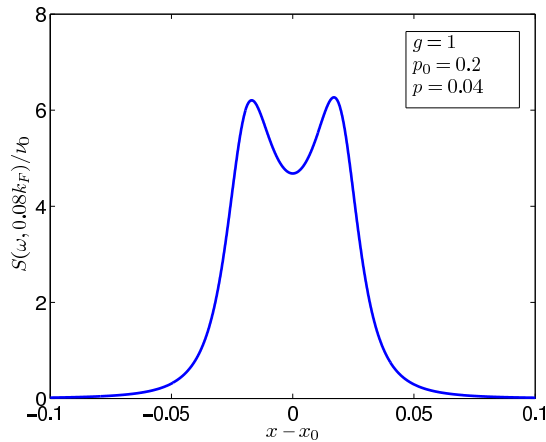} 
\caption{(Color online)
Graph of  the dynamic structure factor  $S ( \omega , q )$ as a function of 
$ x - x_0 = (\omega - v q)/(v_F q )$ 
for fixed  $q = 0.08 k_F $. The line-shape has been calculated 
from Eqs.~(\ref{eq:polfinal}, \ref{eq:wqres}) and (\ref{eq:h1Del}).
The distance between the local maxima is proportional to $\text{w}_q \propto q^2/m$.
}
\label{fig:Sfinalplot}
\end{figure}

In the limit $g \rightarrow 0$
Eq.~(\ref{eq:polfinal}) correctly reduces to the expansion
of the non-interacting inverse polarization
given in Eq.~(\ref{eq:tildePi0expansion}).
However, the term $h_1 / \Delta$ generates
a mass-shell singularity at the true
collective mode energy $\omega = \pm v  q  $.
Fortunately, this singularity 
can be avoided if we use a more physical interaction
whose Fourier transform $f_q$ is analytic
for small $q$, as will be shown explicitly in Sec.~\ref{sec:smooth}.
In this subsection we shall simply regularize 
the mass-shell singularity by hand
using the self-consistent regularization procedure
proposed by Samokhin \cite{Samokhin98}, which we have
already described in detail in Sec.~\ref{sec:RPA}.
Repeating the steps leading from Eq.~(\ref{eq:Imreg}) to Eq.~(\ref{eq:wq0res}),
we obtain from the self-consistent regularization
of the singular term proportional to $h_1 / \Delta$ in
Eq.~(\ref{eq:polfinal}) the following estimate for the width
of the ZS mode,
\begin{equation} 
 \text{w}_q =  \frac{ \sqrt{h_1}}{2 x_0} \frac{q^2}{2m} =
 Z_{\rm w}    \frac{q^2 }{  2 \sqrt{3} m },
 \label{eq:wqres}
\end{equation}
where we have factored out the corresponding estimate 
in the absence of interactions given in Eq.~(\ref{eq:wq0res}), and the
dimensionless factor $Z_{\rm w}$ is given by
\begin{equation}
 Z_{\rm w} = \sqrt{   \frac{ 3 h_1}{4 x_0^2}} = \frac{ 1 + \frac{3}{4} g}{ [1 + g ]^{3/4} }.
 \label{eq:Zwdef}
\end{equation}
Note that $Z_{\rm w} \sim 1 + \frac{3}{32} g^2 + O (g^3)$ for $g \rightarrow 0$, and
$Z_{\rm w} \sim \frac{3}{4} g^{1/4}$ for $g \rightarrow \infty$.
A graph of $Z_{\rm w}$ as a function of the interaction strength 
$g$ is shown in Fig.~\ref{fig:Zm}.
The estimate (\ref{eq:wqres}) for the width of the ZS resonance on the
frequency axis scales as $q^2$, which is 
for small $q$ much larger than our previous estimate
(\ref{eq:dampres}) based on the evaluation of only the first-order diagram (a)
in Fig.~\ref{fig:feynman}. 
The $q^2$-scaling of the width of the ZS resonance
has already been found by Samokhin \cite{Samokhin98}
and has been confirmed later in Refs.[\onlinecite{Pustilnik03,Pustilnik06,Pereira06}].
However, the derivation of Eq.~(\ref{eq:wqres}) is based on a rather ad hoc 
regularization prescription of the mass-shell singularity
in Eq.~(\ref{eq:polfinal}), which ignores in particular the divergent real part of the
term $h_1 / \Delta$. Let us nevertheless proceed and calculate the corresponding dynamic 
structure factor, which can be obtained by replacing the term
$h_1 / \Delta = h_1 / ( x_0^2 + y^2)$  on the right-hand side of Eq.~(\ref{eq:polfinal}) 
by
\begin{equation}
 \frac{h_1}{\Delta}  \rightarrow \frac{h_1}{x_0^2 - 
 \frac{ ( \omega + i \text{w}_q )^2}{ (v_F q )^2 }}.
 \label{eq:h1Del}
\end{equation}
The finite imaginary part $\text{w}_q$ 
in this expression
is a rough estimate of the
modification of the spectral line-shape 
due to the terms which have been neglected by making the
approximation A discussed in Sec.~\ref{subsec:simplification}. 
The typical form  of the dynamic structure factor
in the regime $ p \ll p_0$ 
implied by Eqs.~(\ref{eq:polfinal}, \ref{eq:wqres}) and (\ref{eq:h1Del})
is shown in Fig.~\ref{fig:Sfinalplot}.
Obviously, within our approximation the dynamic structure 
factor does not exhibit any threshold singularities, which
according to Refs.~[\onlinecite{Pustilnik06,Pereira06}]
are a generic feature of the dynamic structure factor of Luttinger liquids.
It turns out that the absence of threshold singularities
in Fig.~\ref{fig:Sfinalplot} is an artefact of the rather simple regularization
prescription (\ref{eq:h1Del}) of the unphysical mass-shell singularity
in Eq.~(\ref{eq:polfinal}). In the following section we shall show how to 
recover the threshold singularities within our functional bosonization approach.

\section{Interaction with regular momentum dependence}
\label{sec:smooth}

In this section we shall show that for a more realistic interaction
whose Fourier transform is for small momenta of the form  
$f_q  = f_0 + \frac{1}{2} f_0^{\prime \prime} q^2 + O ( q^4 )$
with $f_0^{\prime \prime} \neq 0$ we 
do not encounter any mass-shell singularities.
In fact, we believe that even for  sharp momentum-transfer cutoff,
$f_q = f_0 \Theta ( q_0 - q )$, our perturbative result (\ref{eq:Pipert})
does not suffer from mass-shell singularities
as long as we do  not rely on the approximation A discussed in 
Sec.~\ref{subsec:simplification};
in other words, the mass-shell singularity 
$h_1 / \Delta$ in  Eq.~(\ref{eq:polfinal}) 
is an artefact of the sharp momentum-transfer cutoff in combination
with our  neglect of 
curvature corrections to the free polarization in loop integrations.
While
we are not able to evaluate Eqs.~(\ref{eq:Pi1}--\ref{eq:Pi2tadpole})
analytically without relying on approximation A, we shall in this section 
abandon the sharp momentum-transfer cutoff and
assume
that the interaction $f_q$ can be expanded for small $q$ as in Eq.~(\ref{eq:fsmall}).
Later we shall argue that as long as we rely on approximation A,
our result for $S ( \omega , q )$ can  only be trusted 
for $q \gtrsim q_c = 1/( m | f_0^{\prime \prime} | )$, see Eq.~(\ref{eq:qcdef}).
But if $f_0^{\prime \prime}$ is sufficiently large, then
there exists a parametrically large regime $q_c \ll q \ll k_F$ of wave-vectors where our
calculation is valid. 

\subsection{Imaginary part of $\Pi_{\ast}^{-1} ( \omega , q )$}

\begin{figure}[tb]    
\centering   
\includegraphics[scale=0.5]{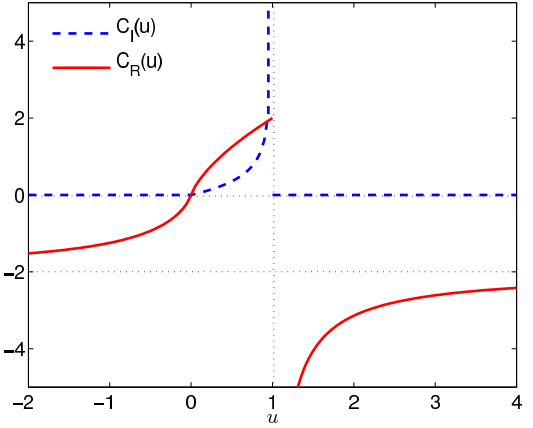} 
\caption{
(Color online) Graph of  the functions $C_I ( u )$ and $C_R ( u )$
 defined in Eqs.~(\ref{eq:YIdef}) and  (\ref{eq:CRdef}).
The dotted lines indicate asymptotic limits.
}
\label{fig:Cplot}
\end{figure}

Let us first calculate the imaginary part of the dimensionless
inverse polarization  $ \tilde{\Pi}^{-1}_{\ast} ( x + i 0 , p )$ given in
Eq.~(\ref{eq:Piinvbetter}) assuming for simplicity $ p > 0$.
From Eqs.~(\ref{eq:tildeIdef}) and (\ref{eq:tildeJdef}) we obtain
\begin{eqnarray}
 {\rm Im}  \tilde{\Pi}^{-1}_{\ast} ( x + i 0 , p ) & = &  {\rm Im} \tilde{I} ( x + i 0 , p )
 \nonumber
 \\
&  & \hspace{-33mm} = \frac{1}{2} \int_0^{\infty} d p^{\prime} p^{\prime} 
 {\rm Im} \left[ \tilde{J} ( x + i 0 , p , p^{\prime} ) +
 \tilde{J} (-  x - i 0 , p , p^{\prime} ) \right].
 \nonumber 
 \\
 & &
 \label{eq:ImPi}
 \end{eqnarray}
In order to calculate the imaginary part of $ \tilde{J} ( x + i 0 , p , p^{\prime} )$,
we first perform a partial fraction decomposition of Eq.~(\ref{eq:tildeJdef}), then carry out
the analytic continuation to the real frequency axis $ iy \rightarrow x + i 0$, and finally
take the imaginary part using ${\rm Im} [ a - x - i 0 ]^{-1} = \pi \delta (a -x )$.
After some lengthy algebra we obtain
 \begin{widetext}
 \begin{eqnarray}
  {\rm Im} \tilde{J} ( x + i 0 , p , p^{\prime} ) & = & - \frac{ \pi |  p^\prime + p | }{2 
x_{p^{\prime}} x_{p^{\prime} + p } }
 \Biggl\{ 
\Theta ( p^{\prime} + p )  \Bigl[
 1 - x_{p^{\prime}}  x_{p^{\prime} + p} - x ( x_{p^{\prime}  } - x_{p^{\prime} + p} ) \Bigr]^2
 \delta \Bigl( p^{\prime} ( x_{p^{\prime}}  + x_{ p^{\prime} + p } ) - p ( x - x_{p^{\prime} + 
 p} ) \Bigr) 
 \nonumber
 \\
& & 
 + \Theta ( - p^{\prime} -  p )  \Bigl[
 1 + x_{p^{\prime}}  x_{p^{\prime} + p} - x ( x_{p^{\prime}  } + x_{p^{\prime} + p} ) \Bigr]^2
 \delta \Bigl( p^{\prime} ( x_{p^{\prime}}  - x_{ p^{\prime} + p } ) - p ( x + x_{p^{\prime} + 
 p} ) \Bigr) \Biggr\} - ( p \rightarrow -p )
 \nonumber
 \\
 &   = & 
 - \frac{ \pi |  p^\prime + p | }{2 
x_{p^{\prime}} x_{p^{\prime} + p } }
 \Bigl[
 1 - x_{p^{\prime}}  \tilde{x}_{p^{\prime} + p} - x ( x_{p^{\prime}  } - \tilde{x}_{p^{\prime} + p} ) \Bigr]^2
 \delta \Bigl( p^{\prime} ( x_{p^{\prime}}  + \tilde{x}_{ p^{\prime} + p } ) - 
p ( x - \tilde{x}_{p^{\prime} +   p} ) \Bigr)  - ( p \rightarrow -p ),
 \nonumber
 \\
 \label{eq:ImJ}
 \end{eqnarray}
\end{widetext}
where we have defined $\tilde{x}_{ p + p^{\prime}} = {\rm sign} ( p + p^{\prime})
{x}_{ p + p^{\prime}}$.
In order to perform the $p^{\prime}$-integration in Eq.~(\ref{eq:ImPi}), we use the fact that
by assumption both $p$ and $p^{\prime}$ are small compared with unity so that we may 
expand $x_p$ to first order in $p^2$,
 \begin{eqnarray}
 x_p & = & \sqrt{ 1 + \tilde{g}_p} 
 \nonumber
 \\
 & = & \sqrt{ 1 + g + \frac{g_0^{\prime \prime}}{2} p^2
 + O ( p^4 ) }
 \nonumber
 \\
 & = &  x_0 +   \frac{ x_0^{\prime \prime}}{2} p^2 + O ( p^4 ),
 \end{eqnarray}
where from Eq.~(\ref{eq:g0pp}),
 \begin{equation}
 x_0^{\prime \prime} =  \frac{  {\rm sign} f_0^{\prime \prime}  }{  \pi x_0 p_c}.
 \end{equation}
Note that for small $p_c$ the coefficient  $x_0^{\prime \prime}$
is large compared with unity.
The $\delta$-functions in Eq.~(\ref{eq:ImJ}) can then be approximated by
 \begin{eqnarray}
 \delta \Bigl( p^{\prime} ( x_{p^{\prime}}  + x_{ p^{\prime} + p } ) - p ( x - x_{p^{\prime} + 
 p} ) \Bigr) 
 \nonumber
 \\
 & & \hspace{-60mm} \approx
\frac{1}{2 x_p} \delta  \Bigl( p^{\prime}
 - p \frac{x - x_p}{2 x_p} 
 \Bigr),
 \label{eq:Dirac1}
 \\
 \delta \Bigl( p^{\prime} ( x_{p^{\prime}}  - x_{ p^{\prime} + p } ) - p ( x + x_{p^{\prime} + 
 p} ) \Bigr)
\nonumber
 \\
 & & \hspace{-60mm} \approx \frac{2}{3 | x_0^{\prime \prime} p | }
 \delta \Bigl( p^{\prime 2} + p^{\prime } p  +   \frac{2 ( x + x_p )}{3 x_0^{\prime \prime}}   \Bigr).
 \label{eq:Dirac2}
 \end{eqnarray} 
In Eq.~(\ref{eq:Dirac1}) we have expanded the argument of the $\delta$-function
to linear order in $p$ and $p^{\prime}$, assuming that both dimensionless
momenta are small.  On the other hand, due to the cancellation
of the leading term in the difference  $x_{p^{\prime}}  - x_{ p^{\prime} + p }$
in the $\delta$-function of Eq.~(\ref{eq:Dirac2}), the corresponding expansion 
has to be carried out to cubic order in the momenta.
The integration in Eq.~(\ref{eq:ImJ}) can now be carried out analytically and
we obtain for small $ p > 0$, 
 \begin{eqnarray}
 {\rm Im}  \tilde{\Pi}^{-1}_{\ast} ( x + i 0 , p )  & = & - \pi^2 p_c 
  \Biggl[ \Theta ( x - x_p ) \tilde{g}_p^2 \tilde{\gamma}_p \frac{ x^2 - x_p^2}{ 12 x_0^4}
 \nonumber
 \\
 &  & \hspace{10mm} + h_1  C_I \Bigl( \frac{ x - x_p}{ \tilde{\gamma}_p} 
 \Bigr) \Biggr] ,
 \label{eq:ImPih1}
 \end{eqnarray}
where 
 \begin{equation}
 \tilde{\gamma}_p = \frac{ 3 p^2}{ 8 \pi x_0 p_c},
 \label{eq:tildegammapdef}
 \end{equation}
and the function $C_I ( u )$ is given by
 \begin{equation}
 C_I  (u )  = \Theta ( u ) \Theta (1-u ) \frac{u}{\sqrt{1-u}}.
 \label{eq:YIdef}
 \end{equation}
Note that the coefficient  $h_1   = ( 1 + 3 x_0^2)^2/(12 x_0)  $ 
on the right-hand side of Eq.~(\ref{eq:ImPih1})
has also appeared for sharp
momentum-transfer cutoff [see Eq.~(\ref{eq:h1def})]
in form of the residue of the mass-shell singularity $h_1/\Delta$
in our expression (\ref{eq:polfinal}) for the irreducible polarization.
A graph of $C_I ( u )$ is shown as the dashed 
line in  Fig.~\ref{fig:Cplot}.
Mathematically, the square-root singularity of $C_I ( u )$ for $u \rightarrow 1$
originates from the  special point  $ x - x_p = \tilde{\gamma}_p$ where
the argument of the Dirac $\delta$-function on the right-hand side
of Eq.~(\ref{eq:Dirac1}) has a double root. 
We believe that the divergence of $C_I ( u )$ for $u \rightarrow 1$ is unphysical
and indicates that the approximations leading to Eq.~(\ref{eq:Dirac1}) are not sufficient
in this regime.
Hence, within our approximations we can only obtain reliable results for the spectral 
line-shape as long as the ratio $( x - x_p )/ \tilde{\gamma}_p$
is not too close to unity.

\subsection{Real part of $\Pi_{\ast}^{-1} ( \omega , q )$}

For  $p_c \ll 1$ and $p \ll 1$ we can 
obtain the contribution from $ {\rm Re} \tilde{I} ( x + i 0 , p )$ analytically from
Eqs.~(\ref{eq:tildeIdef})  and (\ref{eq:tildeJdef})
using the fact that among the corrections of order $p^2$ only terms proportional to
$p^2/p_c$ should be retained.
We obtain for $x > 0$ and $p > 0$,
 \begin{eqnarray}
 {\rm Re} \tilde{I} ( x  + i 0 , p ) & = & I_1 - x^2 I_2
 \nonumber
 \\
 &+  &   \pi p_c h_1
{\rm sign} f_0^{\prime \prime}  
C_R \Bigl( \frac{ x - x_p}{ \tilde{\gamma}_p} \Bigr),
 \hspace{10mm}
 \end{eqnarray}
 where
 \begin{eqnarray}
 I_1 & = &    - \int_0^{\infty} d p p \frac{(x_p -1)^2}{2 x_p^3}  ( 3x_p^2 + 2 x_p +1)
 \nonumber
 \\
 &  & +  2 \pi p_c h_1  {\rm{sign}} f_0^{\prime \prime} ,
 \label{eq:I1def}
 \\
 I_2 & = & \int_0^{\infty} d p p \frac{ (  x_p -1)^2}{x_p},
  \label{eq:I2def}
 \end{eqnarray}
and the function $C_R ( u )$ is given by
 \begin{eqnarray}
 C_R ( u ) & = & \frac{u}{\sqrt{ | 1 - u | }}
 \Biggl[ \Theta ( 1 - u ) \ln \left| \frac{ 1 + \sqrt{ 1-u }}{ 1 - \sqrt{1-u}} \right|
 \nonumber
 \\
 & & - 2 \Theta ( u-1) \arctan \left( \frac{1}{\sqrt{u-1}} \right) \Biggr].
 \label{eq:CRdef}
 \end{eqnarray}
A graph of $C_R ( u )$ is shown in Fig.~\ref{fig:Cplot} (solid line).
Note that $C_R ( u )$ and $C_I ( u )$ can be written as 
$C_R ( u ) = {\rm Re} C ( u + i 0 )$ and
$C_I ( u ) = {\rm Im} C ( u + i 0 )$,
where the complex function $ C ( z )$ is
 \begin{equation}
 C ( z ) = \frac{ z}{  i \sqrt{1-z}} \ln \left( \frac{ \sqrt{ 1-z} +1 }{\sqrt{1-z} -1 } \right).
 \end{equation} 
The real part of our dimensionless inverse polarization can  be written as
   \begin{eqnarray}
 {\rm Re}  \tilde{\Pi}^{-1}_{\ast} ( x + i 0 , p )  & = & Z_1 - Z_2 x^2
 \nonumber 
 \\
  &  +  &    \pi p_c h_1  {\rm{sign}} f_0^{\prime \prime}
 C_R \Bigl( \frac{ x - x_p}{ \tilde{\gamma}_p} \Bigr)  ,
 \hspace{10mm}
 \label{eq:RePiinv}
 \end{eqnarray}
with
 \begin{eqnarray}
 Z_1 = 1 + I_1 + I_H, \; \; \; Z_2 = 1 + I_2 - I_H.
 \end{eqnarray}
By assumption, the bare interaction $f_q$ 
is negligibly small for momentum-transfers exceeding 
$q _0 \ll k_F$,  so that the integrals $I_1$, $I_2$ and $I_H$ are proportional
to $p_0^2 = [ q_0 / (2 k_F )]^2 \ll 1$ and hence $Z_i = 1 + O ( p_0^2)$.
Keeping in mind the self-consistent definition (\ref{eq:velren}) of $x_0$,
we finally obtain for positive  $x $ and $p$,
 \begin{equation}
 g_p + {\rm Re}  \tilde{\Pi}^{-1}_{\ast} ( x + i 0 , p ) = Z_2 \Bigl[  
 x_p^2 - x^2  +  R ( x , p ) \Bigr] ,
 \end{equation}
where
 \begin{equation}
 R ( x , p ) =
    \frac{ \pi p_c h_1 }{ Z_2}  {\rm{sign}} f_0^{\prime \prime} 
 C_R \Bigl( \frac{ x - x_p}{ \tilde{\gamma}_p} \Bigr). 
 \label{eq:sigmaR}
 \end{equation}

\subsection{Spectral line-shape of $S ( \omega , q )$}

\begin{figure}[tb]    
\centering   
\includegraphics[scale=0.5]{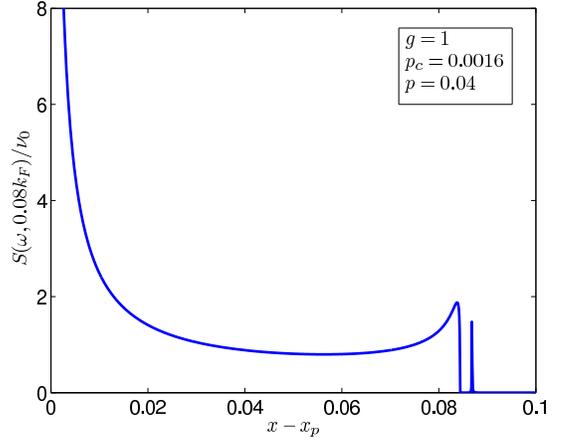} 
\caption{
(Color online) Graph
of  the dynamic structure factor  $S ( \omega , q )$
given in Eq. (\ref{eq:Snewright})
as a function of $x - x_p$ for $p = 0.04 = 25  p_c$ and
${g}=1$.  For simplicity we have set $Z_2 \approx 1 $, which is accurate
for $ p_0 \ll 1$.
For $ p \gg p_c$ most of the spectral weight is carried by the
main shoulder whose lower edge $x \rightarrow  x_p$ 
is bounded by a threshold singularity. The  width of the main shoulder
on the $x$ axis scales as $\tilde{\gamma}_p \propto p^2/p_c$.
Recall that $x = \omega/(v_F q )$, so that the corresponding width 
on the frequency axis scales as $\gamma_q = v_F q \tilde{\gamma}_p \propto q^3/(m q_c)$.
For $ p \gg p_c$ the small ``satellite peak'' emerging above the upper edge of the main shoulder
carries negligible spectral weight and is probably an artefact of our approximations.
}
\label{fig:Sright}
\end{figure}

\begin{figure}[tb]    
\centering   
\includegraphics[scale=0.5]{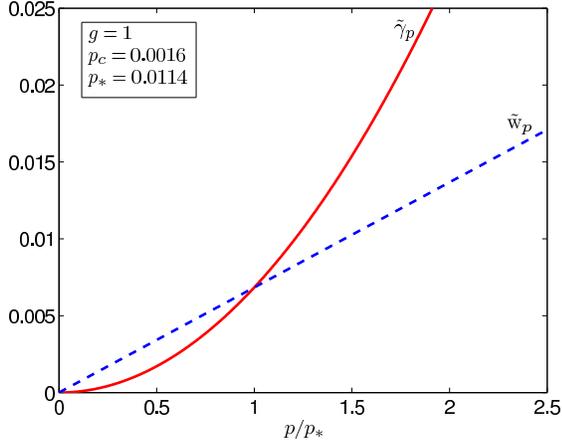} 
\caption{
(Color online) Solid line: dimensionless ZS damping 
$   \tilde{\gamma}_p =  \gamma_q / ( v_F q )  $ 
defined in Eq.~(\ref{eq:gammaq3}) as a function of $p/p_{\ast}$.
Dashed line: estimate of the width $\tilde{\text{w}}_p = \text{w}_q / ( v_F q ) = (Z_{\rm w}/ \sqrt{3}) p$
of the ZS resonance given in Eq.~(\ref{eq:wqres}).
}
\label{fig:gammaw}
\end{figure}

To discuss the line-shape of the dynamic structure factor, it is  convenient to
introduce also the imaginary part of the effective self-energy via
\begin{equation}
 {\rm Im}  \tilde{\Pi}^{-1}_{\ast} ( x + i 0 , p ) = - Z_2
 \Gamma ( x , p  )  ,
\end{equation}
or explicitly,
\begin{eqnarray} 
 \Gamma ( x , p  ) 
   & = &  \frac{ \pi^2 p_c }{ Z_2 }
  \Bigl[ \Theta ( x - x_p ) \tilde{g}_p^2 \tilde{\gamma}_p \frac{ x^2 - x_p^2}{ 12 x_0^4}
 \nonumber
 \\
 &  & \hspace{10mm} + h_1  C_I \Bigl( \frac{ x - x_p}{ \tilde{\gamma}_p} 
 \Bigr) \Bigr] .
 \label{eq:Gammaexplicitres} 
\end{eqnarray}
The dynamic structure factor can then we written as
\begin{equation}
 S ( \omega , q ) = \frac{\nu_0}{\pi Z_2} 
 \frac{ \Gamma ( x , p )}{ [ x^2 - x_p^2 - R ( x , p ) ]^2 + \Gamma^2 (x,p) }.
 \label{eq:Snewright}
\end{equation}
The resulting line-shape
for $ p \gg p_c$ is shown in Fig.~\ref{fig:Sright}.
Obviously, $S ( \omega , q )$ exhibits a threshold singularity at $x  = 
x_p$, corresponding to the threshold frequency
\begin{equation} 
   \omega_q^{-} \equiv v_F q x_p = v q +  \frac{ {\rm sign} f_0^{\prime \prime}}{ 2 \pi x_0 }
   \frac{q^3}{2m q_c }.
\end{equation}
Moreover, most of the spectral weight is smeared out over the interval
$ 0 < x -x_p < \tilde{\gamma}_p$, or  equivalently
 $  \omega_q^{-} < \omega < \omega_q^{-} + \gamma_q$, where
the energy scale $\gamma_q$ is defined by
\begin{equation}
 \gamma_q = v_F q \tilde{\gamma}_p = \frac{3}{8 \pi x_0} \frac{ q^3}{2 m q_c}
 \label{eq:gammaq3}.
\end{equation}
The energy $\gamma_q$ can be identified with the width of the ZS resonance
on the frequency axis. The crucial point is now that for $ q \gg q_c$
Eq.~(\ref{eq:gammaq3})  is much larger than the
estimated broadening 
 $\text{w}_q \propto q^2/m$
of the ZS resonance due to the terms which we have neglected 
by making the approximation A discussed in Sec.~\ref{subsec:simplification} (which 
amounts to ignoring non-linear terms in the energy dispersion
in bosonic loop integrations).
Our  approximation A is therefore only justified 
in the regime where  the broadening
$\gamma_q$ due to the $q$-dependence  of the interaction $f_q$ is large compared
with the broadening $\text{w}_q$ due to the non-linear energy dispersion
in bosonic loop integrations.
We thus conclude that the calculations in this section
are  only valid as long as $\gamma_q \gtrsim \text{w}_q$. A comparison of $\gamma_q$ and 
$\text{w}_q$ is shown in Fig.~\ref{fig:gammaw}.
Obviously, the condition $\text{w}_q = \gamma_q$ defines a characteristic crossover scale
$q_{\ast}$ where the $q$-dependence of the width of the ZS resonance changes from
$q^2$ to $q^3$.  Using Eqs.~(\ref{eq:wqres}) and (\ref{eq:gammaq3}) we obtain the 
following estimate for the crossover momentum scale,
 \begin{equation}
  q_{\ast} = \frac{8 \pi Z_{\rm w} x_0}{ 3 \sqrt{3}}  q_c,
 \end{equation}
which has the same order of magnitude as $q_c = 1 /( m | f_0^{\prime \prime} |)$.
We conclude that the results for $S ( \omega , q )$ presented in this section
are only valid for $q \gtrsim q_\ast$, and hence do not describe
the asymptotic $q \rightarrow 0$ regime.
But the scale $q_\ast$ can be quite small for some interactions.
For example,  if the interaction $f_q$ can be approximated
by a Lorentzian (\ref{eq:Lorentzcut}) with screening wave-vector $q_0 \ll k_F$,
then $q_c  = q_0^2 /( 2 m f_0)$ is quadratic in $q_0$.
For long-range interactions the regime $ q_\ast \lesssim q \ll q_0$ where our calculation is valid can
therefore be quite large and physically more relevant than the asymptotic 
long-wavelength regime $ q \ll q_\ast$.

The small  ``satellite peak'' slightly above the main shoulder in Fig.~\ref{fig:Sright}
is probably an artefact of our approximations, in particular of 
approximation A discussed in Sec.~\ref{subsec:simplification}.
It is easy to show that the satellite peak is located a distance 
$\delta x \propto p_c^3/ p^2 \ll \tilde{\gamma}_p$ above the
upper edge $ x_p + \tilde{\gamma}_p$ of the main shoulder and its width is
proportional to $p^2 \tilde{\gamma}_p \propto p^4/p_c \ll \delta x \ll \tilde{\gamma}_p$.
Note that in the regime $ q \gg q_c$ where our calculation
is valid the threshold singularity
is located at $\omega_q^{-} \approx v q - \gamma_q$ 
(up to corrections of the
order $q^2/m \ll \gamma_q$),
while the energy scale of the satellite peak is  $v q + O ( q^2/m)$.
However, as discussed after Eq.~(\ref{eq:YIdef}),
in the regime
 $ |( x - x_p )/\tilde{\gamma}_p -1| \ll 1$
our approximation A is not reliable, so that the detailed
line-shape in the vicinity of the satellite peak is probably incorrect. 
Fortunately, for $ p \gg p_c$ the satellite peak carries negligible weight,
so that our calculation reproduces the main features of the spectral line-shape.
We speculate that a more accurate evaluation of our self-consistency equation
for $\Pi_{\ast} ( \omega , q )$ derived in Sec.~\ref{sec:selfcon}, which
does not rely on approximation A in Sec.~\ref{subsec:simplification},
will generate additional weight in the dip between
the upper edge of the main shoulder and the
satellite peak, resulting in  a single local maximum at the 
upper edge of the main shoulder. The spectral line-shape
looks then qualitatively similar to the line-shape
proposed in Refs.~[\onlinecite{Pustilnik06,Pereira06}].

\begin{figure}[tb]    
\centering   
\includegraphics[scale=0.5]{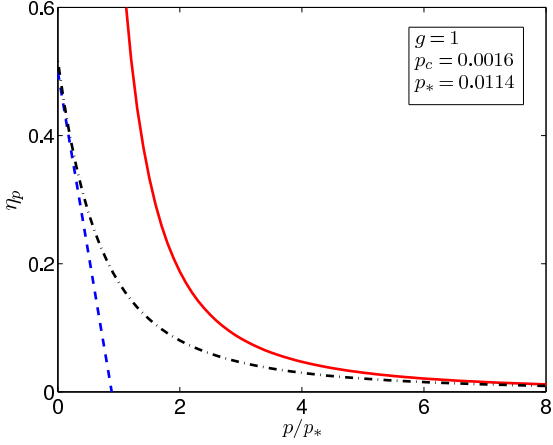} 
\caption{
(Color online)
Solid line:
graph of $ \eta_p $ defined in 
Eq.~(\ref{eq:etapdef})
as a function of $p/ p_{\ast}$ for $f_0^{\prime \prime} < 0$.
 The dashed line is the weak coupling result 
$\eta_p \approx 1/2 - p/(4 \pi p_c)$ 
obtained by Pustilnik {\it{et al.}} in Ref.~[\onlinecite{Pustilnik06}].
The dashed dotted curve is a simple parabolic interpolation.
}
\label{fig:eta}
\end{figure}

Let us next consider the line-shape in the vicinity of the threshold 
singularity $ x \rightarrow x_p$.
For $ 0 < ( x - x_p) / \tilde{\gamma}_p \ll 1$ we may approximate
 \begin{eqnarray}
 \Gamma ( x , p ) & \approx & \frac{ \pi^2 p_c h_1 }{ Z_2 } \frac{ x - x_p}{
 \tilde{\gamma}_p }
 \nonumber
 \\
 & = & 2 \pi x_0 | \eta_p | ( x - x_p) ,
 \label{eq:Gammasmall}
 \\
 R ( x , p ) & \approx & \frac{ \pi p_c  h_1}{ Z_2} \frac{ x - x_p}{
 \tilde{\gamma}_p } \ln \left[ \frac{ 4 \tilde{\gamma}_p}{ x - x_p} \right]
 \nonumber
 \\
 & = &  - 2 x_0 \eta_p (x - x_p)  \ln \left[ \frac{ 4 \tilde{\gamma}_p}{ x - x_p} \right]
,
 \end{eqnarray}
 where we have defined
 \begin{eqnarray}
 \eta_p & = & -  {\rm sign} f_0^{\prime \prime} 
\frac{ \pi p_c   h_1}{2 Z_2 x_0 \tilde{\gamma}_p}
 \nonumber
 \\
 & = &  -  {\rm sign} f_0^{\prime \prime} \frac{ 4 \pi^2 h_1}{3 Z_2 } 
 \frac{ p_c^2}{p^2}
 \nonumber
 \\
 & = & -  {\rm sign} f_0^{\prime \prime} \frac{3 p_{\ast}^2}{4 p^2}
.
 \label{eq:etapdef}
\end{eqnarray}
In the last line we have approximated $Z_2 \approx 1$.
From the above discussion it is clear that this expression can only
be trusted for $p \gtrsim p_\ast$.  A graph of $\eta_p$ as a function of $p/ p_{\ast}$ 
is shown in Fig.~\ref{fig:eta}.
In the regime
$| \eta_p  | \ln [ 4 \tilde{\gamma}_p / ( x - x_p ) ] \gg 1$,
which is equivalent with
 \begin{equation}
  0 < x - x_p \ll 4 \tilde{\gamma}_p \exp \left[ - 1 / | \eta_p |  \right],
 \end{equation}
the dynamic structure factor can thus be approximated by
 \begin{equation}
 S (  \omega , q ) \sim \frac{ \nu_0}{2 x_0 Z_2 | \eta_p | } \frac{1}{ 
( x - x_p) \ln^2 \left[ \frac{ 4 \tilde{\gamma}_p }{ x - x_p } \right]}.
 \label{eq:Slog} 
\end{equation}
According to Pustilnik {\it{et al.}}~\cite{Pustilnik06}, the logarithmic
singularity can be re-summed to all orders, so that it is transformed into an algebraic
one. Assuming that this is indeed correct, we can replace
 \begin{eqnarray}
 x^2 - x_p^2 - R ( x , p ) & \approx &  2 x_0 ( x - x_p ) \left\{
 1 + \eta_p  \ln \left[ \frac{ 4 \tilde{\gamma}_p}{ x - x_p} \right] \right\}
 \nonumber
 \\
 & \rightarrow & 2 x_0 (x - x_p ) \left[ \frac{ 4 \tilde{\gamma}_p}{ x - x_p} \right]^{\eta_p}.
 \label{eq:resum} 
\end{eqnarray}
For $x \rightarrow x_p$ the dynamic structure factor 
then diverges as
  \begin{equation}
 S (  \omega , q ) \sim \frac{ \nu_0}{2 x_0 Z_2 }  \frac{  | \eta_p | }{ 
(4 \tilde{\gamma}_p )^{ 2 \eta_p }} 
\frac{1}{  [  x - x_p]^{ \mu_p }  },
 \label{eq:algebra}
 \end{equation}
with the threshold exponent
 \begin{equation}
  \mu_p = 1 - 2 \eta_p = 1 + {\rm sign} f_0^{\prime \prime} \frac{ 3 p_{\ast}^2}{2 p^2}.
 \label{eq:mupdef}
 \end{equation}
Note that for $f_0^{\prime \prime} < 0$ and $p \ll 1$ 
the weak coupling estimate for $\mu_p$
given by Pustilnik {\it{et al.}}\cite{Pustilnik06}
is in our notation
 \begin{equation}
 \mu_p \approx \frac{p}{ 2 \pi p_c},
 \label{eq:etapust}
 \end{equation}
implying
 \begin{equation}
 \eta_p = \frac{1}{2}[ 1 - \mu_p ] =
\frac{1}{2} \left[ 1 - \frac{p}{ 2 \pi p_c} \right].
 \end{equation}
As shown in Fig.~\ref{fig:eta}, this is consistent with a smooth crossover
to  our result (\ref{eq:etapdef}) at  $p/ p_{\ast} = O (1)$.
Qualitatively, we expect that
the behavior of $\eta_p$ in the crossover regime    
resembles the dashed-dotted interpolation curve in Fig.~\ref{fig:eta}.
Note that $\eta_p \leq 1/2$ for all $p$,  so that $\mu_p \geq 0$.
For some integrable models where
$\eta_p$ has recently been calculated exactly~\cite{Pustilnik06b,Cheianov07} 
the momentum-dependence of $\eta_p$ 
looks  different from our result for the FSM.
For example, in  the Calogero-Sutherland model 
$\eta_p$ is independent of $p$, see  Ref.~[\onlinecite{Pustilnik06b}].
However, the  Fourier transform $f_q$ of the interaction in the Calogero-Sutherland model
vanishes for $q=0$,  while
in the integrable XXZ-chain
considered in Refs.~[\onlinecite{Pereira06,Pereira07,Pereira07b,Cheianov07}] 
the effective interaction of the equivalent one-dimensional fermion system
involves also momentum-transfers of the order of $k_F$.
Moreover, in the XXZ-chain there exists no
crossover scale $q_c$ satisfying 
$q_c = ( m |f_0^{\prime \prime}| )^{-1} \ll k_F$, so that 
the intermediate regime $ q_c \ll q \ll k_F$ where $\gamma_q \propto q^3/q_c$
simply does not exist.
The existence of such an intermediate regime seems to be a special
feature of the FSM considered here, where
$f_{q} $ involves only small momentum-transfers
and has a finite limit for $q=0$.

Within our perturbative approach we cannot justify
the re-summation procedure (\ref{eq:resum}).
Possibly a careful analysis of the
functional renormalization group flow equation for the irreducible
polarization discussed in appendix \ref{sec:FRG} will shed some light onto this
difficult problem. This seems to require extensive numerics, which is beyond the 
scope of this work. Note that for $f_0^{\prime \prime} > 0$
the exponent $\eta_p$ in Eq.~(\ref{eq:etapdef}) is negative, 
so that the singularity in Eq.~(\ref{eq:algebra}) 
is not integrable and exact sum rules \cite{Pines89} cannot be satisfied.
In contrast,  the  original logarithmic
singularity in Eq.~(\ref{eq:Slog}) is  integrable (the integral
$\int_0 dt/ [ t \ln^2 t ]$ is finite), so that at least for $f_0^{\prime \prime} > 0$
the logarithm found in perturbation theory cannot be exponentiated.
On the other hand, an interaction with
$f_0^{\prime \prime} >0$ seems to be unphysical and does not describe
a stable Luttinger liquid~\cite{footnoteessler}.

Finally, consider the tails of the spectral function. 
For $ x \gg x_p$  we obtain from Eqs.~(\ref{eq:Gammaexplicitres}) and
(\ref{eq:Snewright}),
 \begin{equation}
 S ( \omega , q ) \sim \frac{ \nu_0}{\pi Z_2} \frac{ \Gamma ( x , p )}{x^4},
 \label{eq:Sxlarge}
 \end{equation}
 \begin{equation}
 \Gamma ( x , p ) \sim \frac{ \pi^2 p_c}{12 Z_2 x_0^4} \tilde{g}_p^2 \tilde{\gamma}_p x^2.
 \label{eq:Gammaxlarge}
 \end{equation}
Inserting our result (\ref{eq:tildegammapdef}) for $\tilde{\gamma}_p$
we obtain
 \begin{equation}
 S ( \omega , q ) \sim \frac{ \nu_0 \tilde{g}_p^2}{32  Z_2^2 x_0^5} 
\left[ \frac{q^2}{2m \omega } \right]^2,
 \label{eq:Sxlarge2}
 \end{equation}
in agreement with Refs.~[\onlinecite{Teber06,Pereira06,Pereira07,Aristov07}].
Note that the tail of $S ( \omega , q )$ is determined by the first
term on the right-hand side of the damping function
$\Gamma ( x , p )$ given in Eq.~(\ref{eq:Gammaexplicitres}), whereas the
regime close to the ZS resonance is determined by the second term
involving the complex function $C ( z )$.
This is the reason why the spectral line-shape close to
the ZS resonance cannot be simply obtained via extrapolation from the tails assuming a
Lorentzian line-shape.

\section{Summary and Conclusions}
\label{sec:summary}
In this work we have used a functional bosonization approach to
calculate the dynamic structure factor $S ( \omega ,q )$ of 
a generalized Tomonaga model (which we have called {\it{forward scattering model}}),
consisting of spinless fermions in one dimension
with quadratic energy dispersion and an effective density-density interaction
involving only momentum-transfers which are small compared with $k_F$.
We have derived in Sec.~\ref{sec:pert} a self-consistency equation
for the irreducible polarization $\Pi_{\ast} ( \omega , q )$ which does
not suffer from the mass-shell singularities encountered in
other perturbative approaches. 
Although  for the explicit evaluation of  $ S ( \omega , q )$
we had to make some drastic approximations (in particular,
in bosonic loop integrations we have  neglected curvature corrections
to the free polarization, see approximation A discussed in Sec.~\ref{subsec:simplification})
we have found a regime of wave-vectors
$q_{c} \ll q \ll k_F$ where an explicit analytic calculation of the
spectral line-shape is possible.  The crossover  scale
$q_{c} = 1/( m | f_0^{\prime \prime} | )$ is determined by the second derivative
$f_0^{\prime \prime}$ of the Fourier transform of interaction at $q=0$.
For interactions whose
Fourier transform can be approximated by a Lorentzian with 
screening wave-vector $q_0 \ll k_F$,
the crossover scale $q_c$ is proportional to $q_0^2$, so that the
regime $q_c \ll q \ll k_F$ is quite large and can be experimentally
more relevant than the asymptotic long-wavelength regime
$ q \ll q_c$. 
We have shown that for $ q_c \ll q \ll k_F$ the
width of the ZS resonance on the frequency axis scales as $\gamma_q \propto q^3/(m q_c)$.
Our result is consistent with a smooth crossover at $q \approx q_c$ to the asymptotic
long-wavelength result $\gamma_q \propto q^2/m$
obtained by other authors \cite{Samokhin98,Pustilnik06,Pereira06}.
The spectral line-shape is non-Lorentzian, with a main hump
whose low-energy side is bounded by a
threshold singularity at $\omega = \omega_q^- = v q - \gamma_q$, 
a small local maximum around $\omega \approx v q$, and
a high-frequency tail which scales as $q^4/\omega^2$.
For $ \omega \rightarrow \omega_q^- + 0$
the  threshold singularity is within our approximation logarithmic,
 $ S ( \omega, q ) \propto [( \omega - \omega_q^- ) \ln^2 ( \omega - \omega_q^-) ]^{-1}$.
Assuming that higher orders in perturbation theory  
exponentiate the logarithm,
we obtain an algebraic threshold singularity  
with exponent $\mu_q = 1 - 2 \eta_q$ and $\eta_q \propto q_c^2/ q^2$
for $q \gg q_c$.

\vspace{7mm}

Finally, let us point out a number of open problems:

1. It is by now established that, at least
in  integrable models, $S ( \omega , q )$ indeed exhibits algebraic threshold 
singularities  \cite{Pustilnik06b,Cheianov07,Pereira06,Pereira07,Pereira07b}.
However, for generic non-integrable models there is no proof that 
the  logarithmic singularities generated in higher orders
of perturbation theory indeed conspire to transform the logarithm 
encountered at the  first order
into an algebraic singularity, as suggested in Ref.~[\onlinecite{Pustilnik06}].
This would require a thorough analysis of the higher order terms in 
perturbative expansion, which so far has not been performed.

2.  For the explicit evaluation of the self-consistency equation for
the irreducible polarization $\Pi_{\ast} ( \omega , q )$ derived in
Sec.~\ref{sec:selfcon} we had to rely in this work on  approximation
A discussed in Sec.~\ref{subsec:simplification}. 
We have argued that this approximation is not sufficient to
calculate the dynamic structure factor for $ q \lesssim q_c$, because
it neglects the dominant damping mechanism in this regime.
Moreover, for sharp momentum-transfer cutoff our  approximation A 
breaks down for frequencies in the vicinity of  the  mass-shell singularity.
It would be interesting to evaluate the self-consistency equation for
the irreducible polarization $\Pi_{\ast} ( \omega , q )$ derived in
Sec.~\ref{sec:selfcon} without relying on approximation A. 
We believe that in this case
our functional bosonization result for $S ( \omega , q )$ 
does not exhibit any mass-shell singularities even 
for sharp cutoff. The explicit  evaluation of the relevant
integrals is quite challenging and probably requires 
considerable  numerical effort (including a numerical analytic continuation), 
which is beyond the scope of this work.

3. By assumption, the interaction of the FSM considered in this work 
is dominated by small momentum-transfers $q \ll k_F$.
On the other hand, the Fourier transform of the effective
interaction in the Jordan-Wigner transformed  XXZ-chain 
studied in Refs.~[\onlinecite{Pereira06,Pereira07,Pereira07b}] 
has also components involving momentum-transfers 
of the order of $k_F$. It should be interesting to 
investigate more thoroughly how the dynamic 
structure factor depends on the properties of the interaction. 
Unfortunately, the FSM discussed in this work is not integrable and 
there seems to be  no integrable model with quadratic energy dispersion
where the interaction involves only small momentum-transfers and has 
a finite limit for $q \rightarrow 0$.

4. In appendix \ref{sec:FRG} we present a functional renormalization
group equation  [see Eq.~(\ref{eq:flowPitransfer2})]  for the irreducible polarization which goes 
beyond the self-consistent perturbation theory 
based on functional bosonization used here.
A thorough analysis of   Eq.~(\ref{eq:flowPitransfer2})  using numerical methods
still remains to be done. Possibly, this equation
will be a good starting point
for addressing some of the open problems mentioned above.

\section*{ACKNOWLEDGMENTS}
We thank F. H. L.  Essler,  K. Sch\"{o}nhammer, and S. Teber for useful discussions.
The work by P. P. and P. K. was financially supported by the DFG via FOR412.
P. K. is grateful to the  Erwin-Schr\"{o}dinger Institute (ESI)
at the University of Vienna for its hospitality during
the workshop on {\it{Renormalization on Quantum Field Theory,
Statistical Mechanics, and Condensed Matter}}.  
F. S. acknowledges funding by a DFG ``Forschungsstipendium''.

\begin{appendix}

\section{Fermion loops for quadratic dispersion in one dimension}
\label{sec:reduction}
In the functional bosonization approach the vertices
of the interaction part $S_{\rm int} [ \delta \phi ]$ 
of the bosonized action (\ref{eq:Sint}) are
\begin{equation}
\Gamma^{(n)}_{0} ( Q_1, \ldots , Q_n ) =
 i^n (n-1)! \; L^{(n)}_S ( - Q_1 , \ldots , - Q_n ),
 \label{eq:initialGamman2}
\end{equation}
where  the symmetrized closed fermion loops
are defined by
 \begin{eqnarray}
 L^{(n)}_S ( Q_1 , \ldots , Q_n ) & & 
 \nonumber 
\\
&  & \hspace{-35mm} = \frac{1}{n!} 
\sum_{ P(1, \ldots , n )}
 \int_K G_0 ( K ) G_0 ( K - Q_{ P(1)} ) 
 \nonumber
 \\
 & & \hspace{-35mm} \times G_0 ( K - Q_{P(1)} - Q_{ P(2)} )
 \cdots G_0 ( K - \sum_{j=1}^{n-1} Q_{ P(j)} )
 \; .
 \label{eq:fermionloop}
 \end{eqnarray}
Here the sum is over all permutations $P ( 1 , \ldots , n )$ of $1, \ldots , n$, and
the fermionic Green functions   $G_0 ( K )$ should be 
calculated within the self-consistent Hartree approximation, see
Eq.~(\ref{eq:G0Hartree}). For fermions with quadratic energy dispersion in one dimension,
the symmetrized fermion loops (\ref{eq:fermionloop}) can be
calculated exactly. Neumayr and Metzner \cite{Neumayr98,Neumayr99}
(see also Ref.~[\onlinecite{Kopper01}]) 
have derived reduction formulas for
quadratic dispersion in $D$ dimensions which allow to express the 
non-symmetrized loops
 \begin{eqnarray}
 \bar{L}^{(n)} ( \bar{Q}_1 , \ldots , \bar{Q}_n ) & =  & \int_K \prod_{i=1}^{n}
 G_0 ( K - \bar{Q}_i ) 
 \nonumber
 \\
 & & \hspace{-35mm} 
= \int_K
 G_0( K- \bar{Q}_1 ) G_0 ( K - \bar{Q}_2 ) \cdots G_0 ( K - \bar{Q}_n ) ,
 \label{eq:barLn}
 \end{eqnarray}
for  $n > D+1$
in terms of linear combinations
of the more elementary loop $\bar{L}^{(D+1)} ( \bar{Q}_1 , \ldots , \bar{Q}_{D+1} )$. 
In particular, in $D=1$ the non-symmetrized loops
$\bar{L}^{(n)} ( \bar{Q}_1 , \ldots , \bar{Q}_n )$ with $n > 2$ can be expressed in terms 
of the two-loop $\bar{L}^{(2)} ( 0 , -{Q} )  = L^{(2)}_S ( -Q , Q )=-\Pi_0(Q)$.
Given explicit expressions
for the non-symmetrized loops $\bar{L}^{(n)} ( \bar{Q}_1 , \ldots , \bar{Q}_n )$ 
we may construct the corresponding
symmetrized 
loops
$L^{(n)}_S ( Q_1 , \ldots , Q_n)$
by  shifting the labels,
 \begin{eqnarray}
 \bar{Q}_1 & = & 0 \; ,
 \nonumber
 \\
 \bar{Q}_2 & = &  Q_1  \; ,
 \nonumber
 \\
 \bar{Q}_3 & = &  Q_1 + Q_2 \; ,
 \nonumber
 \\
 & \ldots &
 \nonumber
 \\
 \bar{Q}_n & = &  \sum_{j=1}^{n-1} Q_j \; ,
 \end{eqnarray}
so that $\bar{Q}_{i+1} -\bar{Q}_i = Q_i$,
and defining
  \begin{equation}
 {L}^{(n)} ( Q_1 , \ldots , Q_n ) = \bar{L}^{(n)} ( \bar{Q}_1 , \ldots , \bar{Q}_n ) .
 \label{eq:LLn}
 \end{equation}
Then the symmetrized loops are
 \begin{equation}
 L^{(n)}_S ( Q_1 , \ldots , Q_n ) = \frac{ 1}{n!} 
 \sum_{ P(1, \ldots , n )}  {L}^{(n)} ( Q_{P(1)} , \ldots , Q_{P(n)} ) .
 \label{eq:LLsym}
 \end{equation}

In one dimension,  the reduction formula  for the
non-symmetrized loop $\bar{L}^{(n)} ( \bar{Q}_1 , \ldots , \bar{Q}_n )$ 
given by Neumayr and Metzner \cite{Neumayr99} 
can be obtained using a straight-forward partial fraction  decomposition.
Performing the frequency integration
in Eq.~(\ref{eq:barLn}) and introducing the notation
 $\bar{Q}_i  =  ( i \bar{\omega}_i , \bar{q}_i )$ we obtain
 \begin{equation}
  \bar{L}^{(n)} ( \bar{Q}_1 , \ldots , \bar{Q}_n ) = \sum_{i=1}^n
 \int_{ - k_F}^{k_F} \frac{ dk}{2 \pi} \prod_{ \stackrel{j=1}{ j \neq i} }^n \frac{1}{\Omega_{ij} (k )},
 \label{eq:Ln1}
 \end{equation}
where
 \begin{equation}
 \Omega_{ij} (k) = i ( \bar{\omega}_i - \bar{\omega}_j ) + \xi_{k} - \xi_{ k + \bar{q}_i - 
\bar{q}_j},
 \end{equation}
and $\xi_k = \frac{k^2}{2 m} + f_0 \rho_0 - \mu  =  ( k^2 - k_F^2)/(2m)$.
Defining
\begin{equation}
 k_{ij} = \frac{ \bar{q}_j - \bar{q}_i}{2} + i m \frac{ \bar{\omega}_j - \bar{\omega}_i}{
 \bar{q}_j - \bar{q}_i },
 \end{equation}
we may
alternatively write Eq.~(\ref{eq:Ln1})  as
\begin{equation}
  \bar{L}^{(n)} ( \bar{Q}_1 , \ldots , \bar{Q}_n ) = \sum_{i=1}^n
 \int_{ - k_F}^{k_F} \frac{ dk}{2 \pi} \prod_{ \stackrel{j=1}{ j \neq i} }^n 
\frac{m}{  ( \bar{q}_j - \bar{q}_i) (k - k_{ij}  )  }\,.
 \label{eq:Ln2}
 \end{equation}
We can now perform another partial fraction expansion
to obtain
\begin{eqnarray}
  \bar{L}^{(n)} ( \bar{Q}_1 , \ldots , \bar{Q}_n ) & = & \sum_{\stackrel{i,j=1}{i \neq j}}^n
 \left[ \prod_{ \stackrel{l =1}{l \neq i,j}}^n H_{ijl} \right]^{-1} 
 \frac{m}{ \bar{q}_j - \bar{q}_i} 
 \nonumber
 \\
 & & \times
 \int_{ - k_F}^{k_F} \frac{ dk}{2 \pi} 
\frac{1}{  k - k_{ij}    },
 \label{eq:Ln3}
 \end{eqnarray}
with
 \begin{equation}
 H_{ijl} = - \frac{ ( \bar{q}_l - \bar{q}_i)  ( \bar{q}_l - \bar{q}_j)   }{2m}
 + i ( \bar{\omega}_i - \bar{\omega}_l )  + i  ( \bar{\omega}_j - \bar{\omega}_i )
 \frac{ \bar{q}_l - \bar{q}_i}{\bar{q}_j - \bar{q}_i }.
 \end{equation}
In the special case $n=2$ this yields
 \begin{equation}
\bar{L}^{(2)} ( \bar{Q}_1 , \bar{Q}_2 ) = \frac{m}{\pi ( \bar{q}_1 - \bar{q}_2 )}
 \ln \left| \frac{ k_F + k_{12}}{k_F - k_{12} } \right|.
 \end{equation}
In order to give an explicit formula for the  function $L^{(n)} (Q_1 , \ldots , Q_n )$
defined in Eq.~(\ref{eq:LLn}), which depends on
the  external momenta and frequencies $Q_i = ( i \omega_i , q_i )$,
we introduce the notation
 \begin{equation}
 q_{ij} = \bar{q}_i - \bar{q}_j  =
 \left\{ \begin{array}{cc} \sum_{ l=j}^{i-1} q_l\,,& i > j \\
                      - \sum_{ l=i}^{j-1} q_l\,, & j > i 
 \end{array} \right. ,
 \end{equation}
and similarly for $\omega_{ij} = \bar{\omega}_i - \bar{\omega}_j$. These quantities
fulfill $q_{ij}= q_{il}+q_{lj}$ and $\omega_{ij}=\omega_{il}+\omega_{lj}$, such that 
$H_{ijl}$ can be reexpressed as
\begin{equation}
	H_{ijl}= \frac1{q_{ij}}\Big[ i(\omega_{il}q_{lj}-q_{il}\omega_{lj}) - \frac{q_{li}q_{lj}q_{ij}}{2m}\Big]\,,
\end{equation}
which is manifestly symmetric under exchange of $i$ and $j$, i.e., $H_{jil}=H_{ijl}$. 
This yields
\begin{equation}
	L^{(n)} (Q_1 , \ldots , Q_n )= - \sum_{\stackrel{i,j=1}{i <j}}^n
 \left[ \prod_{ \stackrel{l =1}{l \neq i,j}}^n H_{ijl} \right]^{-1} \Pi_0(Q_{ij})\,, 
 \label{eq:Lnfinal}
\end{equation}
with $Q_{ij}=(i\omega_{ij},q_{ij})$. This result is equivalent with Eq.~(19) of 
Ref.~[\onlinecite{Neumayr99}].
Finally, in order to obtain the symmetrized loops
$L^{(n)}_S (Q_1 , \ldots , Q_n )$ 
in Eq.~(\ref{eq:LLsym}),
an additional
summation over the $n!$ permutations
is necessary.
Evidently, the resulting expressions are rather complicated. In the following two 
appendices we shall therefore discuss the symmetrized three-loop and the
symmetrized four-loop separately.
However, without explicitly  evaluating the loops the following two
general properties can be established:

\begin{enumerate}

\item 
The symmetrized $n$-loops 
	$L^{(n)}_S (Q_1 , \ldots , Q_n )$
are finite for all values
of their arguments \cite{Neumayr99}. 
This guarantees that in the perturbative expansion
of the irreducible polarization $\Pi_{\ast} ( Q )$ in powers of the RPA
interaction no infrared singularities are encountered.

\item In the limit $1/m \rightarrow 0$ the 
symmetrized $n$-loop is proportional to 
$(1/m)^{n-2}$. More precisely,
the dimensionless symmetrized 
$n$-loops  $\tilde{L}^{(n)}_S (Q_1 , \ldots , Q_n )$,
defined via
 \begin{eqnarray}
 (n-1) ! L^{(n)}_S (Q_1 , \ldots , Q_n ) &  &
\nonumber
 \\
& & \hspace{-40mm} = \frac{ \nu_0}{ ( m v_F^2)^{n-2} }
\tilde{L}^{(n)}_S (Q_1 , \ldots , Q_n ),
 \label{eq:Loopm}
 \end{eqnarray}
have finite limits for $1/m \rightarrow 0$.
For large $m$ the vertices
$\Gamma_0^{(n)} ( Q_1 , \ldots , Q_n )$
in  the
interaction part $S_{\rm int} [ \delta \phi ]$
of our effective action (\ref{eq:Sint})
are therefore proportional
to increasing powers of the small parameter $1/m$,
which  justifies the perturbative treatment of these vertices.

\end{enumerate}

\section{Symmetrized three-loop}
 \label{sec:threeloop}

The explicit expression for the symmetrized three-loop 
can be written as
\begin{eqnarray}
  L_S^{(3)} ( i \omega_1 , q_1 ; i \omega_2 , q_2 ; 
- i \omega_1 - i \omega_2 , - q_1 - q_2 )  
  &  &
 \nonumber
 \\
 & & \hspace{-70mm}
  = - {\rm Re}  \left[
\frac{1}{ i \omega_1 q_2 -i \omega_2 q_1 -  q_1 q_2 \frac{ q_1 + q_2 }{2m}}
 \right]
\nonumber
 \\
 & &  \hspace{-61mm} \times \Bigl[ q_1 \Pi_0 ( i \omega_1 , q_1 )    + q_2  \Pi_0 ( i \omega_2 , q_2 ) 
 \nonumber
 \\
& &  \hspace{-57mm}
- ( q_1 + q_2) \Pi_0 ( i \omega_1 + i \omega_2 , q_1 + q_2 ) \Bigr] 
 \; .
 \label{eq:threeloop}
 \end{eqnarray}
Introducing again the variables $ i y_1 =  i \omega_1  /(v_F q_1)$,
$p_1 = q_1 / (2 k_F)$ (and similarly for $i y_2$ and $p_2$) and the 
dimensionless function $\tilde{\Pi}_0 ( iy , p ) = \nu_0^{-1}
 \Pi_0 (  i \omega , q ) $ [see Eqs.~(\ref{eq:ypdef}) and (\ref{eq:tildePidef})],
we may write the symmetrized three-loop in the dimensionless form (\ref{eq:Loopm}),
 \begin{eqnarray}
 & & 2 L_S^{(3)} ( i \omega_1 , q_1 ; i \omega_2 , q_2 ; 
- i \omega_1 - i \omega_2 , - q_1 - q_2 ) 
 \nonumber
 \\
 &   &    
=  \frac{ \nu_0}{m v_F^2} 
 \tilde{L}^{(3)}_S ( i y_1 , {p}_1 ; i y_2 , {p}_2 )
 \; ,
 \label{eq:loop3phy}
 \end{eqnarray}
with 
 \begin{eqnarray}
 \tilde{L}^{(3)}_S ( i y_1 , {p}_1 ; i y_2 , {p}_2 )
 & = &
 \frac{1}{ (y_1 - y_2)^2 +   ({p}_1 + {p}_2)^2}
  \nonumber
 \\
 & & \hspace{-30mm}  \times
 \Biggl[ 
  \frac{1}{s_2}
 \tilde{\Pi}_0 ( i y_1 , {p}_1 ) 
 +   \frac{1}{s_1}
\tilde{\Pi}_0 ( i y_2 , {p}_2 )
 \nonumber
 \\
 & & \hspace{-28mm} 
- \biggl( \frac{1}{s_1} + \frac{1}{s_2}  \biggr) \tilde{\Pi}_0 ( i y_1 s_1  +
i   y_2 s_2 , {p}_1 + {p}_2 )
  \Biggr],
 \label{eq:threeloop2}
 \end{eqnarray}
where we have defined
 \begin{equation}
 s_1 = \frac{{p}_1}{{p}_1 + {p}_2 }
 = \frac{r}{r+1}
 \; , \;  s_2 = \frac{{p}_2}{{p}_1 + {p}_2 }
 = \frac{1}{r+1}
 \; ,
 \label{eq:ssdef}
 \end{equation}
with 
 \begin{equation}
 r = \frac{p_1}{p_2}.
 \end{equation}
For later convenience we also define
\begin{equation}
 r_1 = \frac{{p}_1}{{p}_1 - {p}_2 }
= \frac{r}{r-1}
 \; , \;  r_2 =  \frac{{p}_2}{{p}_2 - {p}_1 }
 = \frac{-1}{r-1}
 \; .
 \label{eq:rrdef}
 \end{equation}
Note that by construction $s_1 + s_2 = r_1 + r_2 =1$.

At the first sight it seems that 
the symmetrized three-loop diverges for
$| {p}_1 / {p}_2 | \rightarrow 0$ or
$| {p}_2 / {p}_1 | \rightarrow 0$. Moreover,
the prefactor in
 Eq.~(\ref{eq:threeloop}) 
diverges in the special limit ${p}_1 \rightarrow  {p}_2$ and
$y_1 \rightarrow y_2$. 
It turns out, however, that all divergencies cancel and
the symmetrized three-loop is everywhere of the order of unity.
This non-trivial cancellation cannot be obtained by power-counting
and can be viewed to be a consequence of the
asymptotic Ward-identity associated with the separate conservation
of left-and right-moving particles for linearized energy 
dispersion \cite{Dzyaloshinskii74,Metzner98}. 
We shall show shortly that a similar cancellation protects also the 
symmetrized four-loop from divergencies.
The symmetrization of the loops is crucial to cancel the divergencies.
In diagrammatic language, the symmetrization properly takes vertex and 
self-energy corrections into account. 

\begin{figure}[tb]    
\centering   
  \vspace{4mm}
\includegraphics[scale=0.55]{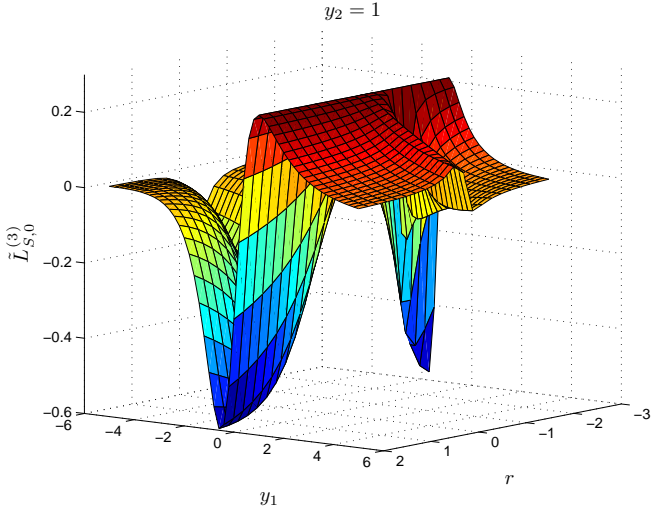} 
\caption{(Color online) Graph of the function 
 $\tilde{L}^{(3)}_{S,0} ( i y_1 , i y_2 , r )$ given in Eq.~(\ref{eq:Pi3zerofinal})
for $y_2 =1$ as a function of $y_1$ and $r = p_1/p_2$. 
}
    \label{fig:threeloop}
\end{figure}

The limiting behavior of the function
 $\tilde{L}^{(3)}_S ( i y_1 , {p}_1 ; i y_2 , {p}_2 )$ 
for $p_1 \rightarrow 0$ and $p_2 \rightarrow 0$ is not unique but
depends on the ratio $r = p_1 / p_2$.
Using Eq.~(\ref{eq:Pilambda0}) we obtain after some algebra,
 \begin{equation}
\lim_{p_i \rightarrow 0, p_1/p_2 =r } \tilde{L}^{(3)}_S ( i y_1 , {p}_1 ; i y_2 , {p}_2 )
 =  \tilde{L}^{(3)}_{S,0} ( i y_1 , i y_2 , r )
 \; ,
 \end{equation}
with 
\begin{eqnarray}
\tilde{L}^{(3)}_{S,0} ( i y_1 , i y_2 , r ) & = & 
 \nonumber
 \\
 &  & \hspace{-30mm} - \frac{ 1 - y_1 y_2 - (y_1 + y_2 )(s_1 y_1 + s_2 y_2)}{  [1 + y_1^2]
[1 + y_2^2] [ 1 + ( s_1 y_1 + s_2 y_2 )^2]}
 \; ,
 \label{eq:Pi3zerofinal}
 \end{eqnarray}
which is manifestly finite for all values of its arguments.
A graph of the function $\tilde{L}^{(3)}_{S,0} ( i y_1 , i y_2 , r )$ 
is shown in Fig.~\ref{fig:threeloop}.
%
%
%
%
Note that a finite limit of the dimensionless function
$\tilde{L}^{(3)}_{S} ( i y_1 , {p}_1 ; i y_2 , {p}_2 )$
for small momenta does not contradict
the loop cancellation 
theorem~\cite{Dzyaloshinskii74,Kopietz95,Kopietz97,Metzner98,Neumayr98,Neumayr99},
because according to
Eq.~(\ref{eq:loop3phy}) the physical symmetrized three-loop
$L_S^{(3)} ( i \omega_1 , q_1 ; i \omega_2 , q_2 ; 
- i \omega_1 - i \omega_2 , - q_1 - q_2 ) $ involves an extra factor of
$1/m$, so that it vanishes for $1/m \rightarrow 0$.

\section{Symmetrized four-loop}
\label{sec:fourloop}

The symmetrized four-loop is more complicated than the three-loop.
However, the four-loop determines the correction
to the irreducible polarization
to first order in the RPA interaction, so that
we need it for our calculation.
Actually, we need the four-loop only for 
the special arguments $Q_3 = - Q_1$ and $Q_4 = - Q_2$.
It is useful to introduce the notation
  \begin{subequations} 
 \begin{eqnarray}
 y_\pm & = & y_1 \pm y_2
 \; ,
 \\
 p_{\pm }  & = & p_1 \pm p_2
 \; , 
\end{eqnarray}
\end{subequations}
and the complex functions
 \begin{eqnarray}
 C_{\pm}  (  i y_- , p_1 , p_2  )
 & = &   \frac{1}{   p_1 p_2 [
  i y_- -  p_{\pm} ]  }
 \; ,
 \end{eqnarray}
 \begin{equation}
 W ( iy, p ) = \frac{ 1}{2 p}
 \left[ \frac{1}{ i y +1 +  p }
 - \frac{1}{ i y +1 -  p } \right]
 \; .
 \end{equation}
We also need
 \begin{eqnarray}
 & & \hspace{-10mm} {\rm Re} W ( iy , p )  =  \frac{ y^2 -1 +  p^2}{
 [ y^2 + ( 1 +  p )^2 ][y^2 + ( 1 -  p )^2 ]}
 \label{eq:ReW}
 \; , 
\\
 & &  \hspace{-10mm} {\rm Im} W ( iy , p )  =  \frac{ 2 y}{
 [ y^2 + ( 1 +  p )^2 ][y^2 + ( 1 -  p )^2 ]}
 \label{eq:ImW}
 \; .
 \end{eqnarray}
The functions $C_\pm  (  i y_- , p_1 , p_2  )$ are
singular for $p_i \rightarrow 0$, while $ W ( iy, p )$ has a finite
limit for $p \rightarrow 0$,
 \begin{eqnarray}
 \lim_{p \rightarrow 0} W ( iy , p ) & = & - \frac{1}{ (1 + iy)^2 }
 \; .
 \label{eq:Wzero}
 \end{eqnarray}
Using our  general result (\ref{eq:Lnfinal})
for the non-symmetrized $n$-loops $L^{(n)} ( Q_1 , \ldots , Q_n )$
and performing the sum (\ref{eq:LLsym})
over all permutations of the external labels,
we obtain for the dimensionless symmetrized four-loop 
(as defined in Eq.~(\ref{eq:Loopm}) for $n=4$) for the special combination
of external labels needed in Eq.~(\ref{eq:Pi1}),
 \begin{eqnarray}
  & & 6 L_S^{(4)} ( i \omega_1 , q_1; -i \omega_1 , - q_1 ; i \omega_2 , q_2 ;
 - i \omega_2 , -q_2 )  
 \nonumber
 \\
 & & =  
 \frac{ \nu_0}{ (m v_F^2)^2} \tilde{L}^{(4)}_S ( i y_1 , p_1 ; 
iy_2 , p_2 )
 \; , 
 \end{eqnarray}
with
 \begin{eqnarray}
  \tilde{L}^{(4)}_S ( i y_1 , p_1 ; iy_2 , p_2 ) &  = &
 \nonumber
 \\
&   &  \hspace{-35mm} + \frac{p_1}{2}{\rm Re} \left[   p_+ {C}_+^2    
+    p_- {C}_-^2      
 + 2 p_1  {C}_+^{\ast}   {C}_-    \right]
  \tilde{\Pi}_0 ( i y_1 , p_1 )
\nonumber
 \\
 &   &  \hspace{-35mm}+ \frac{p_2}{2}     {\rm Re} \left[
     p_{+} {C}_+^2   
 -   p_-   {C}_-^2   
 -  2 p_2   {C}_+      {C}_-  \right]
  \tilde{\Pi}_0 ( i y_2 , p_2 )
\nonumber
 \\
  &  & \hspace{-35mm}
  -   p_+^2  [ {\rm{Re}} {C}_+]^2 
 \tilde{\Pi}_0 ( i y_1 s_1 + iy_2 s_2 , p_+ )   
\nonumber
 \\
  &  & \hspace{-35mm}
-  p_-^2   [ {\rm{Re}} {C}_- ]^2 
\tilde{\Pi}_0 ( i y_1 r_1 + iy_2 r_2 , p_-)
 \nonumber
 \\
 &   &  \hspace{-35mm} + \frac{1}{2} {\rm Im} [ {C}_+   - {C}_- ]
 {\rm Im} [ W ( i y_1, p_1)  - W ( i y_2 , p_2) ]
\nonumber
 \\
 &   & \hspace{-35mm}
 - {\rm Re}[ W ( i y_1, p_1)  W ( i y_2, p_2) ]
 \; ,
 \label{eq:Pi4v1}
 \end{eqnarray}
where we have written
 $C_{\pm} = C_{\pm} ( i y_{-} , p_1 , p_2  )$.
After some algebra Eq.~(\ref{eq:Pi4v1})  can be cast into the  form
 \begin{widetext}
 \begin{eqnarray}
  \tilde{L}^{(4)}_S ( i y_1 , p_1 ; iy_2 , p_2 ) &  = &
 2 \frac{  ( p_1^2 +    3 p_2^2   ) y_-^4
 + 2  ( p_1^2 - p_2^2)^2 
  y_-^2 +    (p_1^2-  p_2^2)^3  }{ p_2^2
 [ y_-^2 +  p_{+}^2 ]^2 
  [ y_-^2 +  p_{-}^2 ]^2} 
   \tilde{\Pi}_0 ( i y_1 , p_1 )
\nonumber
 \\
 &  + & 
 2 \frac{  (   p_2^2 +   3 p_1^2 ) y_-^4
 + 2  (p_2^2 - p_1^2)^2 
  y_-^2 +     ( p_2^2 - p_1^2)^3  }{ p_1^2
 [ y_-^2 +  p_{+}^2 ]^2 
  [ y_-^2 +  p_{-}^2 ]^2} 
   \tilde{\Pi}_0 ( i y_2 , p_2 )
 \nonumber
 \\
 & - &
  \frac{\tilde{\Pi}_0 ( i y_1 s_1 + iy_2 s_2 , p_+ )    }{{s}_1^2 {s}_2^2 
 [ y_-^2 +  p^2_+]^2 } 
 - 
 \frac{\tilde{\Pi}_0 ( i y_1 r_1 + iy_2 r_2 , p_- )    }{{r}_1^2 {r}_2^2 
 [ y_-^2 +  p^2_-]^2 } 
\nonumber
 \\
 &  +  &     \frac{   2 y_- {\rm Im} 
 [ W ( i y_1, p_1)  - W ( i y_2 , p_2) ]   }{  [ y_-^2 +  p_{+}^2 ]
  [ y_-^2 +  p_{-}^2 ]  } 
 - {\rm Re}[ W ( i y_1, p_1)  W ( i y_2, p_2) ]
 \; .
 \label{eq:fourloop2}
 \end{eqnarray}
\end{widetext}
Naive power counting would suggest that
this expression is  singular for
$ y_1 \rightarrow y_2$ or $ | p_1 | \rightarrow | p_2 |$, 
or if $p_1 / p_2$ approaches either zero or infinity. 
However,  similar to the symmetrized three-loop, 
all singularities cancel in Eq.~(\ref{eq:fourloop2}), so that
the symmetrized four-loop
 remains finite and of the order of unity
for all values of its arguments.

\begin{figure}[tb]    
\centering   
  \vspace{4mm}
\includegraphics[scale=0.55]{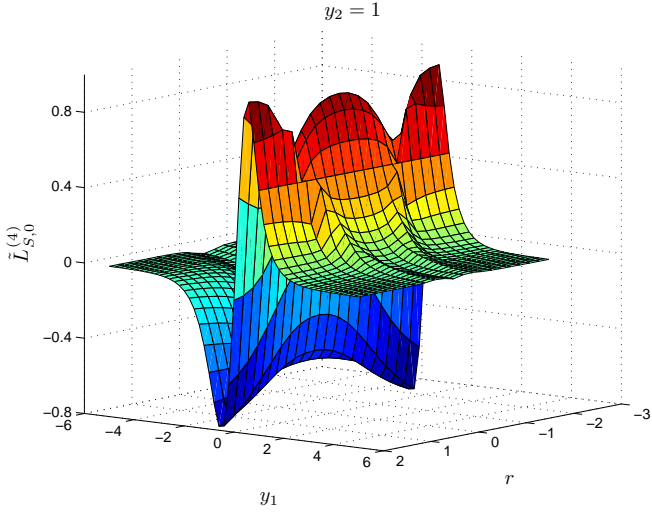} 
\caption{(Color online) Graph of the function 
 $\tilde{L}^{(4)}_{S,0} ( i y_1 , i y_2 , r )$ given in Eq.~(\ref{eq:Pi4zero})
for $y_2 =1$ as a function of $y_1$ and $r = p_1/p_2$. 
}
    \label{fig:fourloop}
\end{figure}

For simplicity, consider again the limit $p_1 \rightarrow 0$ and
$p_2 \rightarrow 0$ with constant $r = p_1 / p_2$. Then
 \begin{equation}
 \lim_{p_i \rightarrow 0, p_1/p_2 = r}
 \tilde{L}^{(4)}_S ( i y_1 , p_1 , i y_2 , p_2 ) 
 = \tilde{L}^{(4)}_{S,0} ( i y_1 , i y_2 ,  r )
 \; ,
 \end{equation}
with
 \begin{widetext}
 \begin{eqnarray}
\tilde{L}^{(4)}_{S,0} ( i y_1 , i y_2 ,  r )
 & = & \frac{1}{(y_1 - y_2)^4 } \Biggl[
 \frac{  2 (r^{2} + 3)}{ 1 + y_1^2}    +  \frac{ 2( r^{-2} + 3)}{ 1 + y_2^2}
 - \frac{1}{  s_1^2 s_2^2 [ 1 + ( s_1 y_1 + s_2 y_2 )^2 ] }
- \frac{1}{  r_1^2 r_2^2 [ 1 + ( r_1 y_1 + r_2 y_2 )^2 ] } 
 \nonumber
 \\
 &  & \hspace{5mm} + 
\frac{ 4 y_-^2 [ 1 - 2 y_1 y_2 - y_1 y_2 ( y_1^2 + y_2^2 + y_1 y_2 )] }{
 [ 1 + y_1^2]^2  [1 + y_2^2 ]^2 } \Biggr]
+ \frac{ 4 y_1 y_2 - ( 1 - y_1^2)(1 - y_2^2) }{
[ 1 + y_1^2]^2  [1 + y_2^2 ]^2 }
 \; .
 \label{eq:Pi4zero}
 \end{eqnarray}
The important point is now that
the singular prefactor $1/(y_1 - y_2)^4$ in Eq.~(\ref{eq:Pi4zero})
is compensated by a factor of 
 $(y_1 - y_2)^4$ arising from the sum of the five terms 
in the square braces.
In fact, we can explicitly cancel this singularity by combining
these terms differently,
\begin{eqnarray}
\tilde{L}^{(4)}_{S,0} ( i y_1 , i y_2 ,  r ) & = &
  - \frac{ [1-y_1^2][1-y_2^2]}{ [ 1 + y_1^2]^2[1+y_2^2]^2}
+ \frac{1}{  [1 + y_1^2]^2 [ 1 + y_2^2]^2
[ 1 + (s_1 y_1 + s_2 y_2)^2][1 + ( r_1 y_1 + r_2 y_2)^2]}  \Biggl\{
\nonumber
 \\
& &
2 r_1^2 s_1^2 [ 1 + y_1^2] [ -1 + 3 y_1^2 + y_2^2  + 8 y_1 y_2 - 3 y_1^2 y_2^2]
+  2 r_2^2 s_2^2 [ 1 + y_2^2] [ -1 + 3 y_2^2 + y_1^2  + 8 y_1 y_2 - 3 y_1^2 y_2^2]
 \nonumber
 \\
& &
\hspace{3mm}  +  (r_1^2 s_2^2 + r_2^2 s_1^2)  \Bigl[
[ 1 + y_1^2][ 1 + y_2^2] [-6 + y_1^2 + y_2^2 + 4 y_1 y_2] + 8 [
1 -2 y_1 y_2 - 3 y_1^2 y_2^2] \Bigr]
 \Biggr\}.
 \label{eq:loop4Peyman}
\end{eqnarray}
Note that the coefficients $r_1$, $r_2$, $s_1$ and $s_2$ are 
not independent but can be
expressed in terms of a single parameter $r = p_1 / p_2$, as 
given in Eqs.~(\ref{eq:ssdef}) and (\ref{eq:rrdef}).
Introducing the notation
 \begin{equation}
 t_1 = s_1 r_1 = \frac{p_1^2}{p_1^2 -p_2^2}  = \frac{ r^2}{ r^2 -1} \; , \;
 t_2 = s_2 r_2 =      \frac{p_2^2}{p_2^2 -p_1^2}  =  \frac{-1}{r^2 -1},
 \end{equation}
so that $t_1 + t_2 =1$, we may alternatively write
 \begin{eqnarray}
\tilde{L}^{(4)}_{S,0} ( i y_1 , i y_2 ,  r ) & = &
  - \frac{ [1-y_1^2][1-y_2^2]}{ [ 1 + y_1^2]^2[1+y_2^2]^2}
+ \frac{1}{  [1 + y_1^2]^2 [ 1 + y_2^2]^2
[ 1 + (s_1 y_1 + s_2 y_2)^2][1 + ( r_1 y_1 + r_2 y_2)^2]}  \Biggl\{
\nonumber
 \\
& &
-1 + 6 y_1 y_2  + t_1 t_2 (y_1 -y_2)^2 [ y_1^2 + y_2^2 +6 y_1 y_2 ]
 \nonumber
 \\
& & 
+ 2 (t_1 y_1 + t_2 y_2 )^2  y_1 y_2 (4 - y_1 y_2 )
 +  2  (t_1 y_1 + t_2 y_2 ) 
\Bigl[  (t_1 y_1 - t_2 y_2 ) (y_1^2 - y_2^2) + (t_1 y_2 + t_2 y_1 ) \Bigr]
 \nonumber
 \\
& &
+ (t_1 y_1^2 + t_2 y_2^2)^2 +  (t_1 y_1^2 + t_2 y_2^2) (2 - y_1^2 y_2^2 )
+ (t_1 y_2^2 + t_2 y_1^2)
 \Biggr\}.
 \label{eq:loop4Peyman2}
\end{eqnarray}
\end{widetext}
A graph  of the  function $\tilde{L}^{(4)}_{S,0} ( i y_1 , i y_2 , r )$ 
is  shown in Fig.~\ref{fig:fourloop}.
%
%
%
%
%

\section{Functional renormalization group equation for the irreducible polarization}
\label{sec:FRG}

\begin{figure}
  \begin{center}
    \epsfig{file=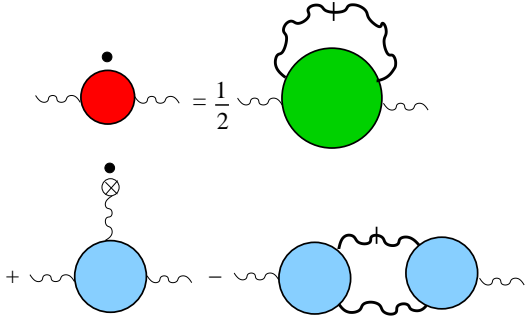,width=0.8\hsize}
  \end{center}
  \caption{
 (Color online) Graphical representation of the exact FRG
flow equation (\ref{eq:flowPitransfer})
for the irreducible polarization in the momentum-transfer cutoff scheme. 
The shaded circles represent the one-interaction-line irreducible
vertices, the
thick wavy lines denote the exact cutoff-dependent
boson propagator (effective interaction)
defined in Eq.~(\ref{eq:Fdef}),
the small crossed circle is the flowing vacuum expectation value of
the bosonic field $\phi$, and
the small black dot denotes
the derivative with respect to the flow parameter $\Lambda$.
The slashed wavy lines represent the single-scale
propagator given in  Eqs.~(\ref{eq:Fdotdef},\ref{eq:singlescaleR}).
}
  \label{fig:flowpi}
\end{figure}

\begin{figure}
  \begin{center}
    \epsfig{file=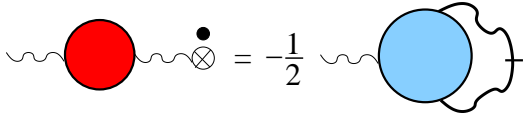,width=0.8\hsize}
  \end{center}
  \caption{
(Color online) Graphical representation of the exact FRG
flow equation (\ref{eq:flowPitransfer})
for the vacuum expectation value of the
bosonic Hubbard-Stratonovich field.
The symbols are explained in the caption of Fig.~\ref{fig:flowpi}.
}
  \label{fig:flowvacuum}
\end{figure}

In this appendix we shall derive a
formally exact functional renormalization group (FRG)
equation for the irreducible polarization which can be used
to generate a perturbative expansion
of the polarization in powers of bosonic loops
with  either the RPA interaction or the true effective interaction as  propagators. 
We use the
momentum-transfer cutoff scheme proposed
in Ref. [\onlinecite{Schuetz05}], where only
the free bosonic part $S_0 [ \phi ]$ is regularized by suppressing
bosonic fluctuations with momenta $q$ smaller than a certain cutoff $\Lambda$.
One possibility is to introduce the cutoff as 
a multiplicative $\Theta$-function \cite{Morris94} by replacing
in Eq.~(\ref{eq:S0phi})
 \begin{equation}
 f_q \rightarrow \Theta ( q_0 > | q | > \Lambda ) f_q,
 \end{equation}
where  $\Theta ( X ) =1$ if the logical expression $X$
is true, and $\Theta ( X ) =0$ if $X$ is wrong.
Alternatively, we may insert an additive cutoff $R_{\Lambda} ( q )$ in the inverse
propagator \cite{Wetterich93},
 \begin{equation}
 f^{-1}_q \rightarrow  f_q^{-1} + R_{\Lambda} ( q ),
 \end{equation}
where $R_{\Lambda} ( q ) = \nu_0 R ( q^2 / \Lambda^2) $
with $R ( 0 ) = 1$ and $R ( \infty ) =0$.
A convenient choice is the Litim
regulator $R (x) = (1-x) \Theta (1-x )$, see Ref.~[\onlinecite{Litim01}]. 
All correlation functions then depend on the cutoff $\Lambda$.
Denoting the flowing irreducible polarization  by $\Pi_{\Lambda } (Q )$,
the true irreducible polarization of our model is recovered  in the limit 
$ \lim_{\Lambda \rightarrow 0} \Pi_{\Lambda} ( Q )
= \Pi_{\ast} ( Q ) $.
An exact hierarchy of FRG flow equations for the
one-line irreducible vertices of our model can then be obtained by
differentiating the corresponding generating functional
$\Gamma_{\Lambda} [ \langle \bar{c} \rangle , \langle c \rangle , 
 \langle \phi \rangle ]$
with respect to $\Lambda$ and  expanding 
$\Gamma_{\Lambda}$ in powers of the expectation values of the fields.
A slight complication arises from the fact that even in the absence of external sources
the bosonic field $\phi_Q$ has a finite expectation value $\phi^0_Q$, so that
for finite external sources
 \begin{equation}
  \langle \phi_Q \rangle = \phi_Q^0 + \delta \phi_Q.
 \end{equation}
In our model, the vacuum expectation value $\phi_Q^0$ in the absence
of sources is related to the exact density $\rho = \int_K \langle
 \bar{c}_K c_K \rangle$ via the Poisson equation \cite{Schuetz06},
  \begin{equation}
  \phi_Q^0 =  \delta_{ Q,0} \bar{\phi} \; \;  , \; \;  \bar{\phi} = -i f_0 \rho,
 \end{equation}
where $\delta_{ Q , 0} = \beta V    \delta_{\bar{\omega} , 0}  \delta_{ q,0}$.
The one-line irreducible vertices  can then be defined by expanding
the generating functional
$\Gamma_{\Lambda} [ \langle \bar{c} \rangle , \langle c \rangle , 
 \langle \phi \rangle ]$ in powers of the expectation values
$\bar{\psi} =  \langle \bar{c} \rangle$, 
$\psi = \langle {c} \rangle$, and $\delta \phi = \langle \phi \rangle - \phi^0$,
 \begin{widetext}
\begin{eqnarray}
     \Gamma_{\Lambda}[\bar{\psi}, \psi,\phi^0 + \delta \phi ] & = & 
 \sum_{n=0}^{\infty}\sum_{m=0}^{\infty}
    \frac{1}{(m!)^2 n!}
    \int_{K_1'}\dots\int_{K_m'}
    \int_{K_1}\dots\int_{K_m}
    \int_{Q_1}\dots\int_{Q_n}
    \delta_{ K'_1 + \ldots + K'_m ,  K_1 + \ldots + K_m  + Q_1 + \ldots + Q_n }
    \nonumber \\
 & \times &
    \Gamma_{\Lambda}^{(2m,n)}(K'_1,\dots,K'_m;
    K_1,\dots, K_m;Q_1,\dots,Q_n) 
    \bar{\psi}_{K'_1} \cdots \bar{\psi}_{K'_m}
    \psi_{K_1}\cdots \psi_{K_m}
    \delta \phi_{Q_1}\cdots  \delta  \phi_{Q_n}\,.
    \label{eq:expansion2}
\end{eqnarray}
Note that the vertices
$\Gamma^{(2m,n)}_{\Lambda}$  implicitly depend
on the vacuum expectation value $\bar{\phi}$.
Following Ref.~[\onlinecite{Schuetz06}],
it is convenient to include to contribution $(2 \beta V f_0)^{-1}  ( \delta \phi_0)^2$
arising from the fluctuation of the zero mode in the Gaussian part of the bosonic
action  (\ref{eq:S0phi})
into the definition of the irreducible vertex
$\Gamma^{(0,2)}_{\Lambda} (-Q,Q)$ with two external bosonic legs, so that
 \begin{equation}
  \Gamma^{(0,2)}_{\Lambda} (- Q , Q ) =  (\beta V)^{-1}  \delta_{Q,0} f_0^{-1}
+    \Pi_{\Lambda} (Q ).
 \end{equation}
The flowing irreducible polarization  $\Pi_{\Lambda} (Q ) $ then satisfies the
exact flow equation \cite{Schuetz05,Schuetz06},
\begin{eqnarray}
  \partial_{\Lambda} \Pi_\Lambda ( Q) &=& \frac12 \int_{{Q^{\prime}}}
  \dot{F}_{\Lambda }({Q}^{\prime})
  \Gamma^{(4)}_{\Lambda} ({Q}^{\prime} ,- {Q}^{\prime} , {Q} ,- Q)
   + \Gamma^{(3)}_{\Lambda} ( Q , - Q , 0 ) 
 \partial_{\Lambda} \bar{\phi}_{\Lambda} 
 \nonumber
 \\
 & - & 
  \int_{Q'}
  \dot{F}_{\Lambda} ({Q}')F_{\Lambda} ( Q+ {Q}') 
  \Gamma^{(3)}_{\Lambda} (- {Q} , {Q}+ {Q}' ,- {Q}' )
  \Gamma^{(3)}_{\Lambda} ( {Q}' ,- {Q}- {Q}' , {Q})  .
  \label{eq:flowPitransfer}
\end{eqnarray}
\end{widetext}
Here for
a sharp momentum-transfer cutoff the bosonic  propagator is
  \begin{equation}
   {F}_{\Lambda }({Q} ) = \Theta ( q_0 > | q |  > \Lambda ) 
   \frac{ f_q }{ 1 + f_q \Pi_{\Lambda} ( Q ) },
 \label{eq:Fdef}
  \end{equation}
and the corresponding
single-scale propagator is
  \begin{equation}
   \dot{F}_{\Lambda }({Q} ) = - \delta ( | q | - \Lambda )
   \frac{ f_q }{ 1 + f_q \Pi_{\Lambda} ( Q ) } .
 \label{eq:Fdotdef}
  \end{equation}
Alternatively, if we work with a smooth
additive cutoff  then
\begin{equation}
   {F}_{\Lambda }({Q} ) = 
   \frac{ f_q }{ 1 + f_q [ \Pi_{\Lambda} ( Q )  + R_\Lambda ( q ) ]},
 \label{eq:FRdef}
  \end{equation}
and
\begin{equation}
   \dot{F}_{\Lambda }({Q} ) = [ - \partial_\Lambda R_{\Lambda} ( q ) ] 
   [ {F}_{\Lambda }({Q} )]^2.
 \label{eq:singlescaleR}
  \end{equation}
The vertices 
 \begin{equation}
\Gamma^{(n)}_{\Lambda} ( Q_1, \ldots , Q_n ) \equiv
 \Gamma^{(0,n)}_{\Lambda}  ( Q_1, \ldots , Q_n ) 
 \end{equation}
 are the totally symmetrized
one-interaction-line irreducible vertices with $n$ external bosonic legs.
A graphical representation of Eq.~(\ref{eq:flowPitransfer}) 
is shown in Fig.~\ref{fig:flowpi}. 
The flow of the vacuum expectation value $\bar{\phi}_{\Lambda}$
of the bosonic field is determined
by the  exact FRG equation \cite{Schuetz06}
 \begin{equation}
  [ f_0^{-1} + \Pi_{\Lambda} ( 0 ) ]  \partial_{\Lambda} \bar{\phi}_{\Lambda}
 = - \frac{1}{2} 
  \int_{Q'} \dot{F}_{\Lambda} ( {Q}') 
  \Gamma^{(3)}_{\Lambda} (  {Q}' ,- {Q}' ,0),
 \label{eq:flowvac}
 \end{equation}
which follows from the requirement 
that the FRG does not generate any tadpole vertices with only one external bosonic leg.
A graphical representation of Eq.~(\ref{eq:flowvac}) is shown
in Fig.~\ref{fig:flowvacuum}.
We may use Eq.~(\ref{eq:flowvac}) to eliminate the derivative of the flowing
vacuum expectation value in the flow equation (\ref{eq:flowPitransfer}) to obtain
the following exact FRG flow equation for the irreducible polarization 
 \begin{widetext}
\begin{eqnarray}
  \partial_{\Lambda} \Pi_\Lambda ( Q) &=& \frac12 \int_{{Q^{\prime}}}
  \dot{F}_{\Lambda }({Q}^{\prime})
  \Gamma^{(4)}_{\Lambda} ({Q}^{\prime} ,- {Q}^{\prime} , {Q} ,- Q)
   -  \frac{1}{2} \frac{f_0}{ 1 + f_0 \Pi_{\Lambda} (0)}  
\Gamma^{(3)}_{\Lambda} ( Q , - Q , 0 )
  \int_{Q'} \dot{F}_{\Lambda} ( {Q}') 
  \Gamma^{(3)}_{\Lambda} (  {Q}' ,- {Q}' ,0)
 \nonumber
 \\
 & - & 
  \int_{Q'}
  \dot{F}_{\Lambda} ({Q}')F_{\Lambda} ( Q+ {Q}') 
  \Gamma^{(3)}_{\Lambda} (- {Q} , {Q}+ {Q}' ,- {Q}' )
  \Gamma^{(3)}_{\Lambda} ( {Q}' ,- {Q}- {Q}' , {Q})  .
  \label{eq:flowPitransfer2}
\end{eqnarray}
\end{widetext}
As shown in Ref.~[\onlinecite{Schuetz05}] (see also the appendix of
the first work cited in Ref.~[\onlinecite{Schuetz06}] for a formal proof), 
in the momentum-transfer cutoff  scheme the 
vertices at the initial scale $\Lambda = \Lambda_0 \equiv q_0$
satisfy non-trivial initial conditions.
The requirement that the vertex $\Gamma^{(0,1)}_{\Lambda_0}$
with one bosonic leg vanishes at  $\Lambda = \Lambda_0 $
implies that the fermionic  self-energy  
 $\Sigma_{\Lambda_0} (K) = 
\Gamma_{\Lambda_0}^{(2,0) } (K , K )$ at the initial scale
is given by the self-consistent Hartree approximation,
 \begin{eqnarray}
 \Sigma_{\Lambda_0} & = & f_0 \rho_{0}, 
 \end{eqnarray}
where the initial density $\rho_{0}$ satisfies
the Hartree self-consistency condition (\ref{eq:Hartreerho}).
The initial  conditions for the purely bosonic vertices
$\Gamma^{(0,n)}_{\Lambda_0} \equiv \Gamma^{(n)}_{\Lambda_0}$ are for $n=2$,
 \begin{equation}
 \Gamma^{(0,2)}_{\Lambda_0} ( -Q , Q ) = 
 (\beta V)^{-1}  \delta_{Q,0} f_0^{-1}
 -  L^{(2)}_S ( Q ,- Q),
 \end{equation}
and for $n >2$,
 \begin{equation}
\Gamma^{(0,n)}_{\Lambda_0} ( Q_1, \ldots , Q_n ) =
 i^n (n-1)! \; L^{(n)}_S ( - Q_1 , \ldots , - Q_n ),
 \label{eq:initialGamman}
 \end{equation}
where  $L^{(n)}_S ( Q_1 , \ldots ,  Q_n )$
are the symmetrized closed fermion loops
with $n$ external legs defined in Eq.~(\ref{eq:fermionloop}).
A graphical representation of 
Eq.~(\ref{eq:initialGamman})
is shown in Fig.~\ref{fig:closedloop}.
By definition, the symmetrized two-loop is
(up to a minus sign)  given by  the non-interacting polarization $\Pi_0 ( Q )$,
 \begin{equation}
 L^{(2)}_S ( - Q,   Q ) = \int_K G_0 ( K ) G_0 ( K + Q ) = - \Pi_0 ( Q ).
 \label{eq:L2pol}
 \end{equation}
However, in contrast to Eq.~(\ref{eq:Pi0}),
the fermionic Green functions in Eq.~(\ref{eq:L2pol}) are self-consistent
Hartree  Green functions as defined in  Eq.~(\ref{eq:G0Hartree}).
Finally, in the momentum-transfer cutoff scheme the initial value of 
the three-legged vertex $\Gamma^{(2,1)}_{\Lambda_0} ( K^{\prime} ;
K ; Q )$ is
 \begin{equation}
\Gamma^{(2,1)}_{\Lambda_0} ( K^{\prime} ; K ; Q ) = i,
 \end{equation}
which follows from Eq.~(\ref{eq:S1}). All other vertices
vanish at the initial scale $\Lambda = \Lambda_0$.

The above RG equation (\ref{eq:flowPitransfer2}) is exact but not closed, 
and should be augmented by FRG flow equations for the vertices
$\Gamma^{(3)}_{\Lambda}$ and $\Gamma^{(4)}_{\Lambda}$, which in turn involve
higher order bosonic vertices.
Only for linearized energy dispersion
the closed loop theorem guarantees that
$\Gamma^{(n)}_{\Lambda}  ( Q_1 , \ldots , Q_n ) =0$, so that we recover
the well-known result that the RPA is exact for the Tomonaga-Luttinger model. 
To motivate a sensible truncation procedure for quadratic energy dispersion,
we note that
the vertices $\Gamma^{(n)}_{\Lambda}$ with $ n \geq 3$ are
irrelevant in the renormalization group sense:
If we assign to the momentum-independent part of the interaction
$f_0 = f_{q=0}$ a vanishing scaling dimension, then in $D$ dimensions
the vertices $\Gamma_{\Lambda}^{(n)}$ have  
scaling dimensions $-( D + z_{\phi} ) ( n/2 -1 )$, 
where $z_{\phi}$ is the
dynamic exponent of the bosonic field mediating the forward scattering 
interaction.  
In one dimension $z_{\phi} =1$ due to the 
linear dispersion of the ZS mode, so that in $D=1$
the vertex $\Gamma^{(3)}$ is irrelevant with scaling dimension $-1$, while
$\Gamma^{(4)}$ is irrelevant with scaling dimension $-2$.
Because the renormalization group flow of irrelevant
couplings is usually not important, it is reasonable to truncate the
infinite hierarchy of flow equations by 
approximating the vertices $\Gamma^{(3)}_{\Lambda}$ and
$\Gamma^{(4)}_{\Lambda}$ in Eq.~(\ref{eq:flowPitransfer})
by their initial values at $\Lambda = \Lambda_0$,
 \begin{eqnarray}
\Gamma^{(n)}_{\Lambda} ( Q_1, \ldots , Q_n ) & \approx &
\Gamma^{(n)}_{\Lambda_0} ( Q_1, \ldots , Q_n ) 
 \nonumber
 \\
 &  & \hspace{-20mm} =
 i^n (n-1)! L^{(n)}_S ( - Q_1 , \ldots , - Q_n ) ,
 \end{eqnarray}
where the symmetrized closed fermion loops are defined 
in Eq.~(\ref{eq:fermionloop}).
In particular, for the vertices appearing in
Eq.~(\ref{eq:flowPitransfer2}) we substitute
 \begin{eqnarray}
 \Gamma^{(3)}_{\Lambda} ( Q_1 , Q_2 , -Q_1 - Q_2 ) & &
 \nonumber
 \\
 & &  \hspace{-40mm} \rightarrow  -2i  L^{(3)}_S ( - Q_1 , - Q_2 , Q_1 + Q_2 ),
\label{eq:Gamma3initial}
 \\
 \Gamma^{(4)}_{\Lambda} ( Q_1 , - Q_1 , Q_2, - Q_2 ) & &
 \nonumber
 \\ 
& & \hspace{-40mm} \rightarrow  6 L^{(4)}_S ( - Q_1 ,  Q_1 , - Q_2 ,  Q_2 ).
 \label{eq:Gamma4initial}
 \end{eqnarray}
Then Eq.~(\ref{eq:flowPitransfer2}) becomes a closed integro-differential
equation for $\Pi_{\Lambda} (Q )$, the solution of which gives for
$\Lambda \rightarrow 0$ a non-perturbative estimate for the
irreducible polarization.
We can easily recover from
 (\ref{eq:flowPitransfer2})
the  perturbative expansion  (\ref{eq:Pipert})
of the irreducible polarization
$\Pi_{\ast} ( Q )$ in powers of the RPA interaction
if  we replace the flowing polarization
on the right-hand side of  Eq.~(\ref{eq:flowPitransfer2})
by the non-interacting one,
$\Pi_{\Lambda} ( Q ) \rightarrow \Pi_0 ( Q )$, which amounts to
replacing the flowing effective interaction in $F_{\Lambda} ( Q )$ by the
RPA interaction defined in Eq.~(\ref{eq:fRPAdef}).
With these substitutions, it is easy to integrate
both sides of Eq.~(\ref{eq:flowPitransfer2}) over the flow parameter
$\Lambda$ and recover Eq.~(\ref{eq:Pipert}).
Alternatively, we may truncate
Eq.~(\ref{eq:flowPitransfer2}) by replacing
the flowing polarization on the right-hand side
by its limiting value without cutoff,
$\Pi_{\ast} ( Q ) = \lim_{\Lambda \rightarrow 0} \Pi_{\Lambda} ( Q )$.
Then we arrive that the one-loop self-consistency equation for $\Pi_{\ast} ( Q )$
given in Eqs.~(\ref{eq:pol12}--\ref{eq:Pi2tadpole}).

Note that even with the truncation (\ref{eq:Gamma3initial},\ref{eq:Gamma4initial})
the FRG flow equation (\ref{eq:flowPitransfer2}) is non-perturbative, because
the renormalization of the polarization is
self-consistently taken into account in the bosonic loop integrations. 
It should be interesting to analyze  Eq.~(\ref{eq:flowPitransfer2}) 
numerically and try to extract the spectral line-shape. Possibly, one can check in this way
whether
the resummation procedure proposed
by Pustilnik {\it{et al.}} \cite{Pustilnik06} is justified also for non-integrable models.

\end{appendix}


\begin{thebibliography}{99}
\bibitem{footnote1}
We focus on neutral spinless fermions in this work.
In charged Fermi systems the ZS mode becomes the plasmon~\cite{Pines89}.
\bibitem{footnoteprint}
There is a printing error in 
the expression for $\omega_q$  given in
Eq.~(6) of Ref.~[\onlinecite{Pirooznia07}];
the correct result for $\omega_q$
is given in Eq.~(\ref{eq:dispersion}) of the present work.
\bibitem{footnotethreedim}
In three dimensions,
the single particle-hole pair continuum 
is within RPA smeared out over a frequency interval
$0 < \omega \lesssim v_F | \bd{q} | $,
while  the collective ZS mode lies (for repulsive interactions) above the upper limit
of the single-pair continuum~\cite{Pines89}.  
More generally, in arbitrary dimensions and for repulsive short-range interactions
the long-wavelength dispersion of the ZS mode
is $ v | \bd{q} |$ with $v > v_F$.
\bibitem{footnotehartree}
The three-loop 
 $\tilde{L}^{(3)}_S ( iy, p, iy ,-p)$ 
appearing in the Hartree-type-of contribution (\ref{eq:Pi2tadpole})
to the irreducible polarization is 
ambiguous for the required combination of
arguments. Because Hartree interactions should be  static, 
we define  $\tilde{L}^{(3)}_S ( iy, p, iy ,-p)$ as the limit of
 $\tilde{L}^{(3)}_S ( iy, p, iy^{\prime} ,p^{\prime})$ 
where we first set
the frequency sum 
$ \omega + \omega^{\prime} = v_F (q y + q^{\prime} y^{\prime})$
equal to zero and then take the limit $  p^{\prime} \rightarrow -p$.
\end{thebibliography}
\end{document}